\documentclass[12pt]{article}
\usepackage[dvips]{graphicx}
\usepackage[]{epsf}
\usepackage{amsmath}

\newcommand{\gtrsim}%
{\mathop{\raisebox{-.4ex}{\rlap{$\sim$}} \raisebox{.4ex}{$>$}}}
\newcommand{\lesssim}%
{\mathop{\raisebox{-.4ex}{\rlap{$\sim$}} \raisebox{.4ex}{$<$}}}

\def\vek#1{\mbox{\protect\boldmath $#1$}}

\newcommand{\sla}[1]{/\!\!\!#1}

\newcommand{\epem}{e^+ e^-}
\newcommand{\mpmm}{\mu^+ \mu^-}
\def\HtoWW{H\to W^+ W^-}
\def\HtoZZ{H\to Z^0 Z^0}
\def\Htobb{H\to b\bar{b}}
\def\Mhiggs{m_H}

\newcommand{\met}{\rlap{\,/}E_T}

\begin{document}


\title{
\vspace*{-1.0in}
\begin{flushright}
{\normalsize FERMILAB-FN-701 \\[-12pt] hep-ex/0107044}
\end{flushright}
\vspace*{1.0in}
Linear Collider Physics}

\author{
	Paul       F. Derwent,
	Bogdan     A. Dobrescu,$^*$
	Andreas    S. Kronfeld, \\ 
	Heather    E. Logan,
	Konstantin T. Matchev,$^\dagger$
	Adam          Para, \\
	David      L. Rainwater,
	S\l awomir    Tkaczyk, and \\
	William    C. Wester~III \\[0.7em]
	{\it Fermi National Accelerator Laboratory, Batavia, Illinois}
}
\date{\today}

\maketitle
\thispagestyle{empty}

\begin{abstract}
We report on a study of the physics potential of linear $e^+e^-$
colliders.
Although a linear collider (LC) would support a broad physics program,
we focus on the contributions that could help elucidate the origin
of electroweak symmetry breaking.
Many extensions of the standard model have a decoupling limit, with a
Higgs boson similar to the standard one and other, higher-mass states.
Mindful of such possibilities, we survey the physics of a (nearly)
standard Higgs boson, as a function of its mass.
We also review how measurements from an LC could help verify several
well-motivated extensions of the standard model.
For supersymmetry, we compare the strengths of an LC with the LHC.
Also, assuming the lightest superpartner explains the missing dark
matter in the universe, we examine other places to search for a signal
of supersymmetry.
We compare the signatures of several scenarios with extra spatial
dimensions.
We also explore the possibility that the Higgs is a composite,
concentrating on models that (unlike technicolor) have a Higgs boson
with mass of a few hundred GeV or less.
Where appropriate, we mention the importance of high luminosity, for
example to measure branching ratios of the Higgs, and the importance of
multi-TeV energies, for example to explore the full spectrum of
superpartners.
\end{abstract}

\newpage

\pagestyle{plain}
\pagenumbering{roman}

\begin{center}
	\includegraphics[width=4.0in]{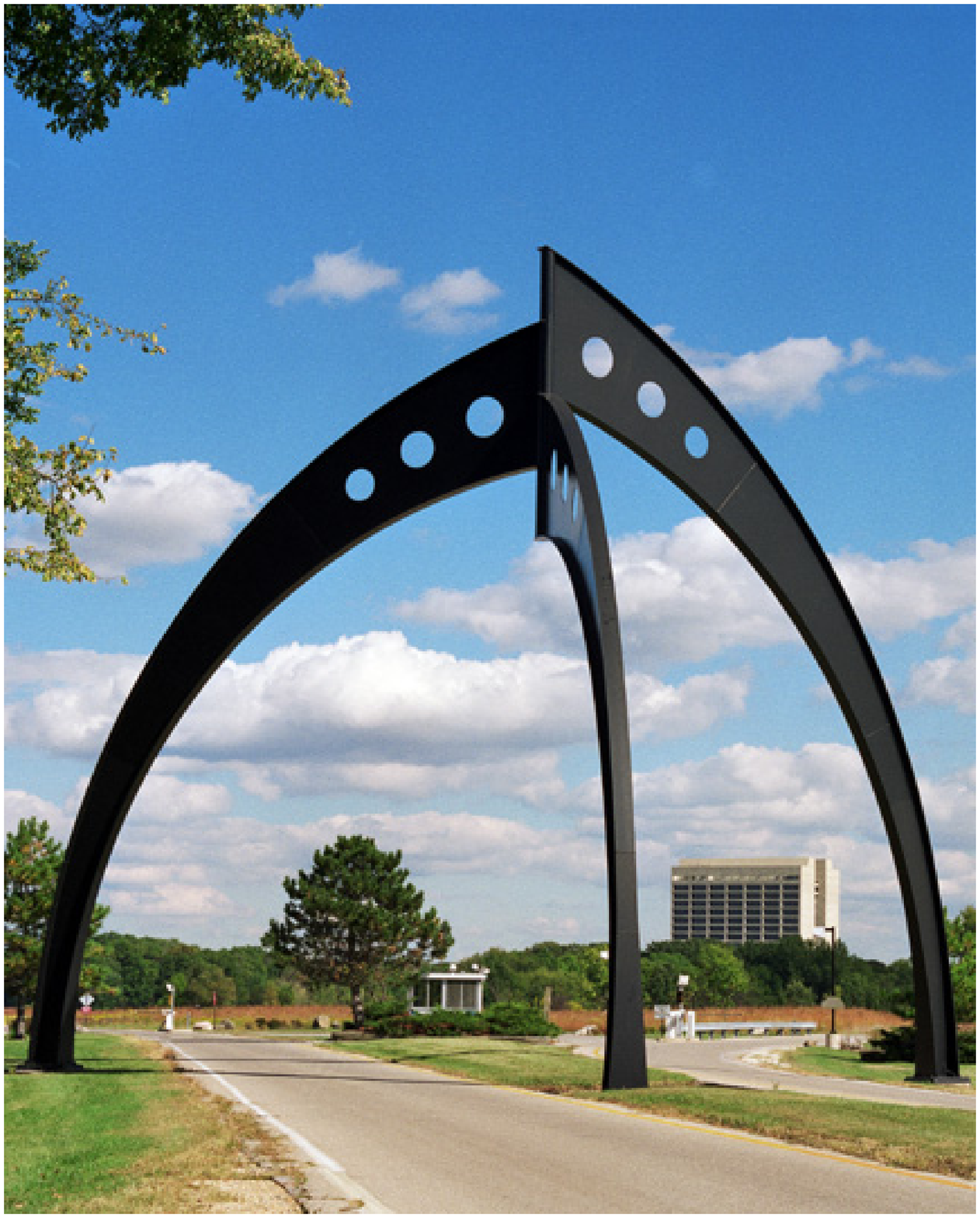}

	\vspace{2em}
	``Broken Symmetry''\\
	Sculpture by R. R. Wilson at Fermilab's western entrance.
\end{center}

\section*{Executive Summary}
About a year ago, the Fermilab Directorate asked us to study the
physics potential of linear $e^+e^-$ colliders with center-of-mass
energy ranging between several hundred~GeV and a few~TeV.
This range covers not only the next energy frontier, but also the
energy scale where we expect to discover the mechanism that breaks
electroweak symmetry.

There are mature designs from DESY, KEK, and SLAC for linear colliders
that start at an energy of 0.5~TeV and can be upgraded to around 1~TeV.
It is therefore pragmatic to summarize first what can be achieved
below 1~TeV, and then the physics that would require a multi-TeV
lepton collider.

The 0.5--1~TeV linear collider (LC) will have a broad program of
studies of the standard
$SU(3)_C \times SU(2)_W \times U(1)_Y$ gauge interactions.
It will almost certainly also produce Higgs bosons.
In many extensions of the standard model the lightest Higgs boson has
properties similar to the Higgs of the standard model.
The physics of a nearly-standard Higgs boson depends on its mass.
\begin{itemize}
	\item If the Higgs boson is light, with mass $m_H<2m_W$, many decay
	modes are accessible, yielding measurements of the couplings
	to vector bosons ($W$ and $Z$), charged leptons ($\tau$),
	up-type quarks ($c$), and down-type quarks ($b$).
	These measurements are vital, because they test how nature
	generates these particles' masses.
	Measurements of loop-induced $\gamma\gamma$ and gluon-gluon
	branching ratios are also possible.
	With the proposed LC designs, the precision would be a few percent.
	At hadron colliders some ratios of couplings can be measured, but
	the $b\bar b$ mode may be difficult, and the $c\bar c$ and $gg$
	modes are probably impossible.

	Branching ratios are also sensitive to the effects of virtual
	contributions of higher-mass states.
	Consequently, high (integrated) luminosity is valuable for
	measuring the couplings as precisely as possible.

	\item If $m_H>2m_W$ the decays to $WW$ and $ZZ$ dominate, and
	decays to quarks and charged leptons are rare, if not very rare.
	More than 1~ab$^{-1}$ integrated luminosity would be needed to
	measure the rare branching ratios.
	If $m_H>2m_t$ the branching ratio to $t\bar{t}$ is large enough
	to measure.
\end{itemize}
Note that, for all masses, the LC precisely measures the Higgs coupling
to the $W$ and $Z$~bosons, which is interesting, because it demonstrates
how much of the known $W$ and $Z$ masses come from the observed Higgs.

The higher mass regions are often disregarded, because fits of the
standard model to precisely measured electroweak observables suggest
that the Higgs boson is light.
We believe this argument is not robust.
The Higgs makes a small contribution to the precisely measured
observables, and could be compensated by similarly small
contributions from TeV-scale particles.
In some extensions of the standard model this cancellation does
take place.
Then, the Higgs phenomenology of a 0.5--1~TeV LC would look much like
the standard model, even with a Higgs mass of a few hundred~GeV.

The decays of the Higgs boson(s) also could be grossly non-standard.
Then, as a rule, experiments at a 1~TeV LC could be the key to
understanding the physics of electroweak symmetry breaking.
For example, the LC is better suited than a hadron collider for
measuring the partial width of the Higgs boson to invisible final
states.

If supersymmetry plays a leading role in breaking electroweak symmetry,
it is likely, but not certain, that the lightest superpartners can be
pair-produced in a 1~TeV~LC.
If so, one could make precise measurements of masses and mixing angles
of the lightest superpartners.
The precision is useful for gaining insight into the mechanism
responsible for breaking supersymmetry.

If the lightest superpartner were to explain the missing (non-baryonic)
dark matter in the universe, indirect signals of supersymmetry may
appear soon.
Such a signal could appear during the next several years, either in
astrophysical searches or in particle physics experiments.

Similarly, there are many models with a composite Higgs boson that
would lead to a rich phenomenology below 1~TeV.
For example, a composite Higgs could couple in a nearly standard way to
the known particles, yet decay to other new particles.
These models also give concrete examples of nearly standard, heavy Higgs
bosons, whose contribution to electroweak observables is compensated by
further new states lying above 1~TeV.

Most extensions of the standard model postulate additional states in
the multi-TeV region.
In supersymmetry, energies above 1~TeV are probably needed, in the long
run, to produce the full spectrum of superpartners and Higgs bosons.
If there are extra dimensions, higher energies would be needed to show
the pattern of Kaluza-Klein excitations.
Composite models often contain additional fermions with multi-TeV
masses.
While indirect, low-energy tests can be helpful in ruling out specific
models, on-shell production of these particles would be more valuable.
Thus, multi-TeV lepton collisions will probably also be needed to
understand fully the mechanism of electroweak symmetry breaking.

The problem of electroweak symmetry breaking is too important to
ignore, especially since the key energy scale might be close at hand.
The scale could be within reach of CDF and D0 in Run~II of the Tevatron,
and is certainly within reach of the LHC experiments.
That being said, a lepton collider facility with
high luminosity and a flexible energy will also be necessary to
comprehend fully the Higgs mechanism and related phenomena.
The linear $e^+e^-$ collider is the most promising candidate to fill
this need.
We recommend that study of physics at a future linear collider
continue, and we encourage more colleagues to get more involved.

\newpage

\tableofcontents

\newpage

\pagenumbering{arabic}
\section{Introduction}
In December 1999, the Directors' Office of Fermilab asked us to 
undertake a Study of the physics possibilities of linear $e^+e^-$ 
colliders at center-of-mass energies between 300~GeV and as high as a 
few~TeV.
The charge from the Associate Director for Research reads
\begin{quote}
\begin{flushright}
December 3, 1999
\end{flushright}
Dear Fermilab Colleague: 

I would like to ask you to participate in a physics study of linear
$e^+e^-$ colliders at Fermilab.  The laboratory is interested in
assessing the physics capabilities of a linear collider and how
they depend on the collider parameters.  Three labs (SLAC, KEK, and
DESY) have advanced designs for linear colliders with an initial
center-of-mass energy of 500~GeV with an upgrade path to an energy
of around 1~TeV.  Given the likely high cost of such facilities,
it is imperative to understand what the LC would contribute to the
worldwide high-energy physics program in the LHC era.

The charge for the group will be to deliver a report by September 18,
2000, which should explicitly include:
\begin{enumerate}
\item	An analysis of the capability for Higgs physics as a function
	of energy and luminosity.  This should include measurement
	of Higgs boson parameters including couplings and indirect
	measurements of virtual effects for very massive Higgs bosons.
\item	A comparison with the physics capability of the LHC experiments
	in some well-defined scenarios for physics beyond the
	standard model.
\end{enumerate}
As you know, SLAC and Fermilab have begun a collaboration on the NLC
design.  The experience gained from the accelerator collaboration and
the physics study should make it possible for the Fermilab community
to develop an informed opinion on the merits of proposed accelerators.

The Fermilab physicists who are being asked to start the physics
study are Paul Derwent, Andreas Kronfeld, Stephan Lammel, Adam Para,
S\l awek Tkaczyk, Rick Van Kooten (Indiana U.), and G. P. Yeh.
Kronfeld and Tkaczyk will be the coordinators.  It is expected and
imperative that other people, both from the Lab and the user community,
join the study as it develops.  The local Fermilab group should also
interact with the Worldwide Study of the Physics and Detectors for
Future Linear $e^+e^-$ Colliders.

Sincerely, 

Mike Shaevitz
\end{quote}
Our study group consisted mostly of physicists who were new to the
linear collider (LC), together with some who have followed its
developments in the past.
Our meetings were open to all, and were often attended by frank 
skeptics of the physics potential of the LC.
Whether pro, con, or neutral we all agreed with the Directors' 
sentiment that the decision whether or not to build an LC must involve 
informed members of the Fermilab community.
Moreover, we agree that it is appropriate to focus our study on the 
Higgs boson and, more generally, extensions of the standard Higgs 
sector that could explain the origin of electroweak symmetry breaking.

Before summarizing our findings, let us point out that the physics 
program of the LC extends well beyond the physics of electroweak 
symmetry breaking.
Near $\sqrt{s}=2m_t$ the LC will be able to trace out the threshold 
for $t\bar{t}$ pairs, yielding a precise determination of~$m_t$.
The precision attained this way, and probably also from $t\bar{t}$
production above threshold, will be far better than that at hadron
colliders.
The program of QCD pursued at LEP and SLC, including two-photon 
physics, will continue at the LC, for example tracing out how the 
coupling~$\alpha_s$ runs with~$\sqrt{s}$.
The LC will also produce $W^\pm$ and $Z^0$ bosons copiously, providing 
interesting measurements of anomalous triple and quartic gauge-boson 
couplings, including energy dependence.
Furthermore, the LC can revisit the $Z^0$-pole and produce 
$10^8$--$10^9$ $Z^0$s from polarized beams.
This would refine further the beautiful measurements performed at LEP 
and SLC, particularly on
the left-right polarization asymmetry at the $Z$ pole,
the $Z$'s line-shape,
and in $B$ physics.
Thus, the program of the LC contributes to nearly all of (experimental) 
high-energy physics.

Nevertheless, unraveling the mechanism of electroweak symmetry 
breaking is the central problem of our time.
Other compelling problems---such as the origin of flavor, the 
mechanism(s) of $CP$ violation, and even the origin of neutrino
masses---seem to be connected to it.
Yet for these problems a fundamental understanding may well require 
experiments at extremely high energies, whereas the electroweak 
symmetry is broken around the TeV scale that will be accessible to the 
CERN Large Hadron Collider (LHC) and a future LC.
Moreover, in the context of the LC, decisions on operating energies, 
and luminosity integrated at each energy, will almost certainly be 
dominated by our desire to understand the Higgs and whatever else 
accompanies it, such as supersymmetric partners of the known 
particles.
Thus, while it is important not to forget that the LC will produce 
excellent results across the board, the most critical information for 
assessing its value concerns electroweak symmetry breaking. 

It is a truism that the ``standard model'', with one Higgs doublet, 
describes all available data.
Most of the success of the standard model comes from its gauge sector: 
low-energy QED, electroweak radiative corrections, and perturbative 
QCD at high energies.
In this report we take for granted many results based on the 
well-tested gauge interactions of the standard model.
We also take seriously the solid theoretical arguments showing that
the scalar sector of the standard model breaks down at some
energy scale.
On the other hand, almost everything that touches the Higgs sector 
(including fermion masses and $CP$ violation) is tested either poorly 
or indirectly.
In particular, there are only rough guides to the scale at which the 
one-doublet description breaks down.
There are many ideas for a more fundamental theory operating at this 
new scale and above, but only experiments can demonstrate which one is 
realized in Nature.
Therefore, in this report we try to treat various possibilities for the 
Higgs sector without unnecessary theoretical prejudice.

Where it has been tested, the one-doublet Higgs sector fits
the experimental data, although many models with richer TeV-scale
physics fit equally well.
Most viable models possess a so-called decoupling limit, in which all 
particles associated with electroweak symmetry breaking are very 
heavy, except for a relatively light, $CP$-even scalar.
In the decoupling limit the one-doublet model is a good effective theory 
up to the mass scale of the heavy particles.
Thus, almost by construction, models with a decoupling limit can
describe the data, particularly the precisely measured electroweak 
observables, just as well as the one-doublet model.

Many of these models, whether based on supersymmetry or on strong 
dynamics, also remain viable away from the decoupling limit.
Frequently, the models predict particles that would be produced in 
$e^+e^-$ collisions with $\sqrt{s}=1$~TeV, or even less.
An intriguing twist is that some of the new particles could be so 
light that the Higgs could decay into them.
More generally, once models stray from the decoupling limit, they 
almost always predict a rich phenomenology that would require many 
complementary measurements to disentangle the underlying physics.

On the basis of many classes of models, it is clear that there are 
grounds to anticipate essential measurements from an LC with 
$\sqrt{s}=0.5$--1~TeV.
Nevertheless, one cannot rule out the decoupling limit with a Higgs 
boson whose properties are close to the one in the standard model and 
with other new particles beyond the reach of a 1-TeV~LC.
Here there are some gaps in the LC literature, so we shape the
discussion of the Higgs boson around the standard model.
The phenomenology is sensitive to the Higgs mass, so, depending on the
mass, there are different tradeoffs between energy and luminosity.
Note, however, that in this report we approach the standard Higgs 
sector not as fundamental, but as an effective field theory, valid up
to some finite energy.

In this report we also discuss some aspects of specific models.
We concentrate our attention on models with supersymmetry, extra 
dimensions, or dynamical electroweak symmetry breaking in four 
dimensions.
In our opinion, these are the best motivated theoretical frameworks,
and are broad enough phenomenologically to cover many other 
possibilities.
In supersymmetry we elaborate on some features that challenge the
conventional wisdom.
With extra dimensions we emphasize especially the properties of the
Kaluza-Klein states and the models with a composite Higgs 
boson made of Kaluza-Klein excitations of the standard gauge
bosons and fermions.

In the next several years, experiments at the Tevatron and the LHC 
will search for the Higgs boson and other new particles, and the 
scientific value of the LC must be weighed against their anticipated 
measurements.
With enough integrated luminosity, experiments at the Tevatron may be 
the first to observe a Higgs boson.
At the LHC an observation of a Higgs boson, with production and decay
properties like the one in the standard model, is certain---no matter 
what its mass.
The LHC should---and the Tevatron could---discover light superpartners,
Kaluza-Klein excitations, or something else to point at the origin
of electroweak symmetry breaking.
As a rule, the LHC experiments will also measure some, but not all,
of the Higgs boson's couplings at the 10\% level.
Thus, they will not be able to test fully whether one field gives mass 
to gauge bosons and fermions, as in the standard model.
The LHC experiments also will not measure the self-coupling of the
Higgs, which is needed to reconstruct the Higgs potential and test 
directly the mechanism of spontaneous symmetry breaking.

To get an idea of how the LC can elucidate discoveries of the hadron 
colliders, it is useful to recall a few basic features of LC 
experimentation.
In $e^+e^-$ collisions, signatures and backgrounds are typically both
electroweak processes.
Therefore, the signal and background cross sections are comparable,
and they are calculable at the percent level.
With a linac the center-of-mass energy can be varied over a wide range,
and it is known precisely, so the LC can home in on any interesting
threshold (below its $\sqrt{s}_{\rm max}$).
The proposed LC designs are several colliders in one:
here we speak not only of the possibility of $e^-e^-$, $e\gamma$, 
and $\gamma\gamma$ collisions---though those are potentially 
interesting---but also of the polarization of the beams.
Above the electroweak scale left- and right-handed fermions are 
fundamentally different, so choosing the right combinations, in 
response to the data, could prove vital.
The cleanliness, flexibility, and versatility of the LC give it several 
advantages that counterbalance the higher and broader reach of the LHC.

In Sec.~2 we review some properties of the linear colliders under
design and R\&D.
Section~3 covers the Higgs physics at an LC, including some background
on hadron colliders.
Here we focus on the properties of a Higgs with couplings similar to 
the standard model.
Section~4 considers additional new physics that we expect to be a part 
of electroweak symmetry breaking, concentrating on supersymmetry, 
extra spatial dimensions, and composite Higgs bosons.
In Sec.~5 we summarize our views, give some recommendations, and
identify some open problems that warrant further scrutiny.

While our local Study was in progress, the American Study of
Physics and Detectors of Linear $e^+e^-$ Colliders, in which some of
us participate, posted a ``whitepaper''~\cite{Bagger:2000iu} on the
physics program of the LC at $\sqrt{s}=500$~GeV.
For the most part, the whitepaper makes it easier for us to write this 
report, because it is clear, up to date, and not too long.
The whitepaper concentrates on the arguments for a 500-GeV stage of 
the~LC.
Much of its analysis pertains to the standard model when viewed as 
valid up to very high energies, or to supersymmetric extensions of the 
standard model.
Under these circumstances there would be a light Higgs boson.
This report, on the other hand, takes a more general view of the 
standard model as an effective theory and examines a broader class of 
models of electroweak symmetry breaking.
In particular, one of the main original contributions of this report 
is to survey the capabilities of the LC for a nearly-standard Higgs
boson, as a function of its mass, including the region $m_H>165$~GeV.

Finally, while we were preparing the final version of this report, two
comprehensive reports on linear collider physics appeared.
One is the physics volume of the TESLA Technical Design Report from the
TESLA Collaboration~\cite{Tesla}, and the other is a resource book from
the American Linear Collider Working Group~\cite{Abe:2001ab}.
The latter includes some of our material on intermediate-mass and heavy
Higgs bosons.
Both cover all aspects of the LC high-energy physics program, including
top quark physics, properties of electroweak bosons, and~QCD.

\section{Accelerator Parameters}

There are several $\epem$ linear collider design efforts currently 
underway, differing most significantly in the choice of RF 
acceleration.
The TESLA design from DESY uses superconducting RF cavities with 
resonant frequency of 1.3~GHz.
The NLC/JLC-X design from SLAC and KEK uses normal conducting X-band
cavities with resonant frequency of 11.4~GHz.
The JLC-C design from KEK uses normal conducting C-band cavities with 
resonant frequency of 5.7~GHz.
These three designs all use klystrons as the RF power source.
The CLIC R\&D program at CERN uses 30~GHz normal conducting 
cavities, coupled to a drive beam linac for power.
The choice of the RF acceleration method causes differences in the 
bunch structure parameters of the various designs.
We will not go into detail on the designs here but just touch briefly 
on the time structure, energy, and luminosity.
Table~\ref{table-lc_params} summarizes these parameters.
\begin{table}[t]
\centering
\caption{Baseline parameters for four linear collider designs.}
\label{table-lc_params}
\vskip 6pt
\begin{tabular}{lcccc}
\hline \hline
  & TESLA & NLC/JLC-X & JLC-C & CLIC \\
\hline 
$\sqrt{s}$ (TeV) & 0.5--0.8 & 0.5--1.0 & 0.5 & 0.5--3 \\ 
${\cal L}$ (10$^{34}$ cm$^{-2}$ s$^{-1}$)
	& 3.4--5.8 & 2.2--3.4 & 0.43 & 1--10 \\
RF cavities & superconducting & normal & normal & normal \\
RF power source & klystrons & klystrons & klystrons & drive beam \\
bunches/train & 2820--4886 & 190 & 72 & 150 \\ 
bunch separation (ns) & 337 & 1.4 & 2.8 & 0.7 \\
repetition rate (Hz) & 5  & 120  & 50 & 200  \\ 
\hline \hline
\end{tabular}
\end{table} 
More detailed descriptions and references can be found in
Refs.~\cite{Kuhlman:1996rc,Toge:2000mm,Brinkmann:1999rz,Assmann:2000hg}.

All designs plan to have polarized $e^-$ beams, with $P_{e^-}=80$\%.
There is also some work on polarized $e^+$ sources, achieving perhaps 
$P_{e^+}=40$\% at full luminosity and 60\% with reduced
luminosity~\cite{Walker}.
With both beams polarized the effective polarization for $e^+e^-$ 
annihilation processes is 
$P_{\rm eff}=(P_{e^-}+P_{e^+})/(1 + P_{e^-}P_{e^+})$.

\subsection{TESLA}

The TESLA design uses superconducting cavities, operated at 2 K,
with a resonant frequency of 1.3~GHz.  The baseline design calls for
$\sqrt{s}=500$ GeV, with an upgrade path to 800~GeV.  For the baseline
design, the beam structure will be long bunch trains of 2820 bunches
separated by 337~ns, for a total bunch train length of 950 $\mu$s
(285 km) and a 5 Hz repetition rate.  The nominal luminosity at 500~GeV
is $3.4\times 10^{34}$~cm$^{-2}$~s$^{-1}$. The higher energy
requires higher gradient from the superconducting cavities~\cite{Tesla:II}.

\subsection{NLC and JLC}

The unified NLC/JLC-X design uses normal conducting cavities with
a resonant frequency of 11.4~GHz.
The baseline design calls for $\sqrt{s}=500$~GeV,
with an upgrade path to 1~TeV~\cite{NLC:Snowmass}.
For the present baseline design, the beam
structure will be bunch trains of 190 bunches separated by 1.4~ns,
for a total bunch train length of
0.27~$\mu$s (81~m) and a repetition rate of 120~Hz.  The nominal
luminosity is $2.2\times 10^{34}$~cm$^{-2}$~s$^{-1}$.  The baseline
option only fills half the linac tunnel with RF cavities, enabling
a straightforward upgrade to 1~TeV.

The JLC collaboration is also pursuing an accelerator based on C-band 
(5.7~GHz) as a backup for the X-band design.
Most of the components of the C-band main linac satisfy the
specifications of the 500~GeV JLC.
The C-band design is considered by some to be a serious option, if one
wants to build a normal-conducting machine as early as possible.

\subsection{CLIC}

The CLIC design uses normal conducting cavities with a resonant
frequency of 30 GHz.  Compared to the other designs, it is earlier
in the R\&D phase.  The novel approach of CLIC is that the
RF power is delivered to the accelerating cavities by a drive beam
with coupled cavities.  The design is being optimized for 3~TeV,
but the concept is being developed for 0.5--5~TeV.  For the baseline
design, the beam structure will be bunch trains of 150 bunches
separated by 0.7~ns, for a total bunch train length of 0.1~$\mu$s
(30 m) and a repetition rate of 200~Hz.  The nominal luminosity is
$10^{34}$--$10^{35}$~cm$^{-2}$~s$^{-1}$.

\subsection{Detector Backgrounds}

\par  Studies have been done of backgrounds in the detectors from
extra $\epem$ pair production.  Since these are generally low momentum
electrons, they could spiral in the central magnetic fields, causing
larger occupancies in the tracking detectors.  
Because the TESLA and
NLC/JLC designs have very different time structures, we have 
investigated the density of
extra hits per bunch crossing (for TESLA) or bunch train (for NLC/JLC).

\par
\begin{figure}[bp]
\centering
	\includegraphics[width=0.7\textwidth]{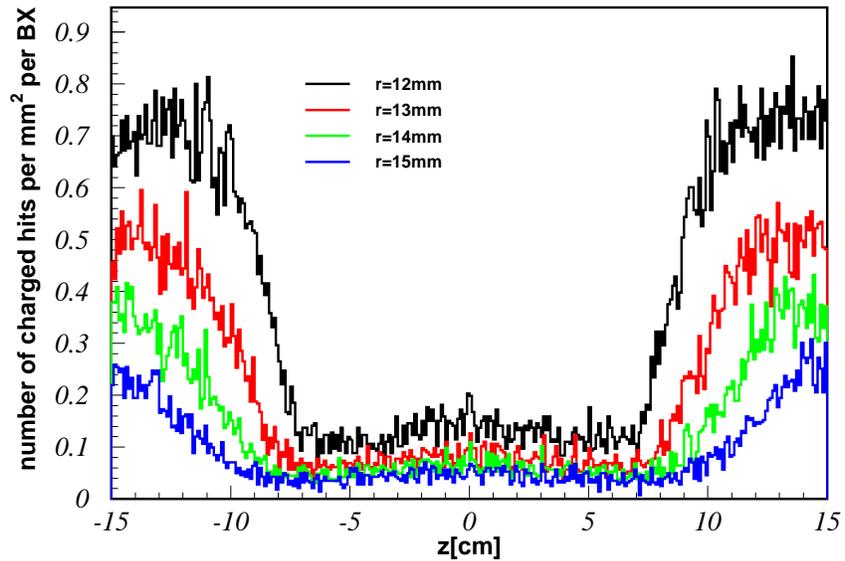}
	\caption[1:fig:MB]{The background hit densities at TESLA in each 
	vertex layer as a function of a distance along the beamlines for 
	3~T magnetic field.  From Ref.~\cite{ref-Eur_density}.}
	\label{1:fig:MB}
\end{figure}

\begin{figure}
\centering
	\includegraphics[width=0.5\textwidth]{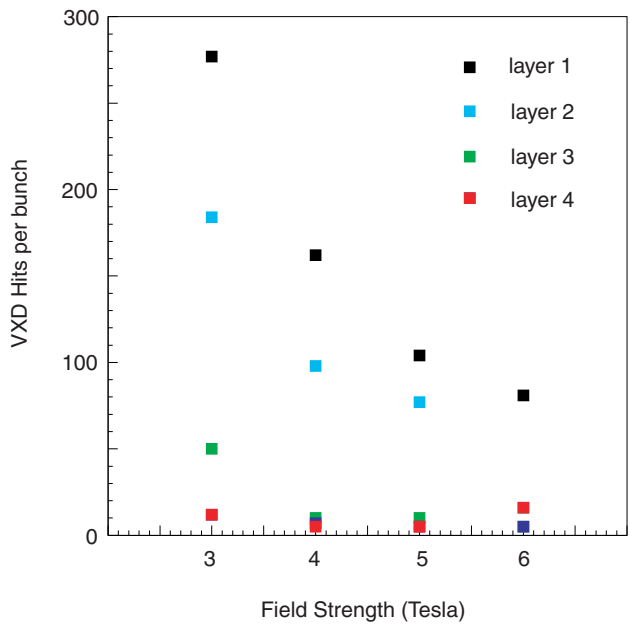}
	\caption[2:fig:TWM]{The raw number of hits at NLC in each layer 
	as a function of magnetic field for the small detector.  (The
	masking layout was not modified for the low magnetic field, so 
	hits from pairs which impacted the outer face of the M1 mask were
	deleted.)  From Ref.~\cite{ref-VXD_density}.}
	\label{2:fig:TWM}
\end{figure}

  With the 337~ns bunch spacing in the TESLA design, individual
bunch crossings can be resolved with the detector readout electronics. 
Therefore, it is appropriate to consider hits per bunch crossing 
as a measure of background hits.
With the bunch spacing of 2.8~ns in the (earlier) NLC/JLC design, 
the detector readout electronics will most likely integrate 
95 bunches, and the relevant unit for background measure is 
number of hits per bunch train.
Simulations done in the ECFA-DESY Study~\cite{ref-Eur_density},
for $\sqrt{s}=500$ GeV and a central magnetic field 3~T,
give 0.2--0.5~hits/mm$^2$/bunch at a radius of 1.2~cm; see
Fig.~\ref{1:fig:MB}.
In contrast, simulations done for the
NLC/JLC VXD~\cite{ref-VXD_density} result in 3 hits/mm$^2$/train
for a 3~T central magnetic field (or 0.03 hits/mm$^2$/bunch).
Figure~\ref{2:fig:TWM} shows the results of these simulations, 
where number of background hits per bunch is plotted 
as a function of the magnetic field at various radii. 
(The 280 hits in Layer 1 for the magnetic field of 3 T, 
shown in Figure~\ref{2:fig:TWM}, correspond to a hit density of 
 ~3 hits/mm$^2$/train.) 

Both groups are working on designs of vertex
detectors with a first active layer placed at a radial distance of
1.2~cm away from the colliding beams.
For both bunch time structures, it is felt
that it is possible to operate such detectors with hit densities
seen in the simulations.

In addition to beam-induced background sources discussed in the 
previous studies, overlapping hadronic $\gamma\gamma$ interactions 
provide a source of background.
Preliminary studies of their effect on reconstruction of 
physics processes of interest, {\it e.g.}, for Higgs decays, 
have been done in the ECFA-DESY Study~\cite{ref-MB_ggbkg}. 
The authors concluded that a combination
of kinematic and vertex topology selections can reduce the effects of 
$\gamma\gamma$ interactions, with moderate losses in reconstruction 
efficiency.  The background events resulted in a very small 
additional number of charged hits in the inner layer of the vertex 
detector in the amount of $3.4 \times 10^{-5}$~hits/mm$^2$/bunch.

Optimization work on the mask designs is still in progress, to reduce
the beam backgrounds even further.
In addition, there are other ideas leading to background reduction,
which remain to be evaluated.
One possibility, currently under study, is to increase the strength of
the magnetic field in the detector.
Another option, relevant for the NLC/JLC beam structure, is to
reduce the number of bunches recorded by the electronics during the
collisions.
Such reduction can be achieved by applying pipelined front-end readout 
with a length of the integration window shorter than the 266~ns 
duration of the NLC/JLC bunch train.
However, the presently achieved reduction factors are very good for 
both machine designs and result in acceptable levels of background.

\section{Higgs Bosons}
\label{sec:higgs}
This section details our study of the physics of the Higgs boson(s).
In this report, we consider a Higgs to be any excitation of a field
whose vacuum expectation value breaks electroweak symmetry.
They arise from the same dynamics as the \emph{longitudinal} $W^\pm$
and $Z^0$ bosons.
It is an experimental fact that the latter exist, because the $W$ and
$Z$ have mass.
At the same time, the measured quantum numbers of quarks and leptons 
unmistakably reveal an $SU(2)_W\times U(1)_Y$ gauge symmetry mediated by 
the transverse $W$ and $Z$.
These two observations can be reconciled only if the symmetry is 
spontaneously broken.
Then the theory of the Higgs mechanism shows how massless gauge bosons
and massless scalars (or Nambu-Goldstone bosons) can interact to form
a massive vector boson, and dictates how the physical Higgs bosons
couple to~$W$ and~$Z$.

For example, the standard model contains a complex doublet of 
fundamental scalar fields.
Three of the degrees of freedom in the doublet become the longitudinal 
modes of the $W^\pm$ and $Z^0$, while the fourth becomes the single 
Higgs boson of this model.
The standard model's doublet also generates masses for quarks and
charged leptons.
The true nature of the Higgs sector remains experimentally obscure,
however.
It could be richer, with several Higgs bosons sharing in the mass 
generation of $W$ and $Z$, with some or all of them generating the 
fermion masses.
And the full mechanism that breaks the electroweak symmetry should
contain additional particles as well, probably at the TeV scale.

At an LC the process $e^+e^-\to Z^*\to ZH$ gives a superb way to 
search for Higgs bosons, without relying on the decay products of 
the~$H$.
If several fields give mass to the $Z$, several Higgs states may be 
found in this way.
Even if the dominant branching ratio is to invisible particles, a 
Higgs still could be observed easily as a bump in the missing mass 
recoiling against the $Z$, perhaps even when broad.
Under these and other circumstances, the LC also could observe Higgs 
bosons that would escape detection at hadron colliders. 

As explained in the introduction, we will focus on a Higgs boson with
couplings similar to the one in the standard model.
We do not necessarily believe that this is the most likely situation,
or the most interesting.
But it is a well-motivated example, because most manifestly viable 
models have a limit, called the decoupling limit, for which sub-TeV 
phenomena
can be described by the standard model, plus small corrections
from higher-mass states, which would be produced only at the LHC or
a multi-TeV~LC.

\begin{figure}[b]
	\centering
	\includegraphics[totalheight=\textwidth,origin=c,angle=90]{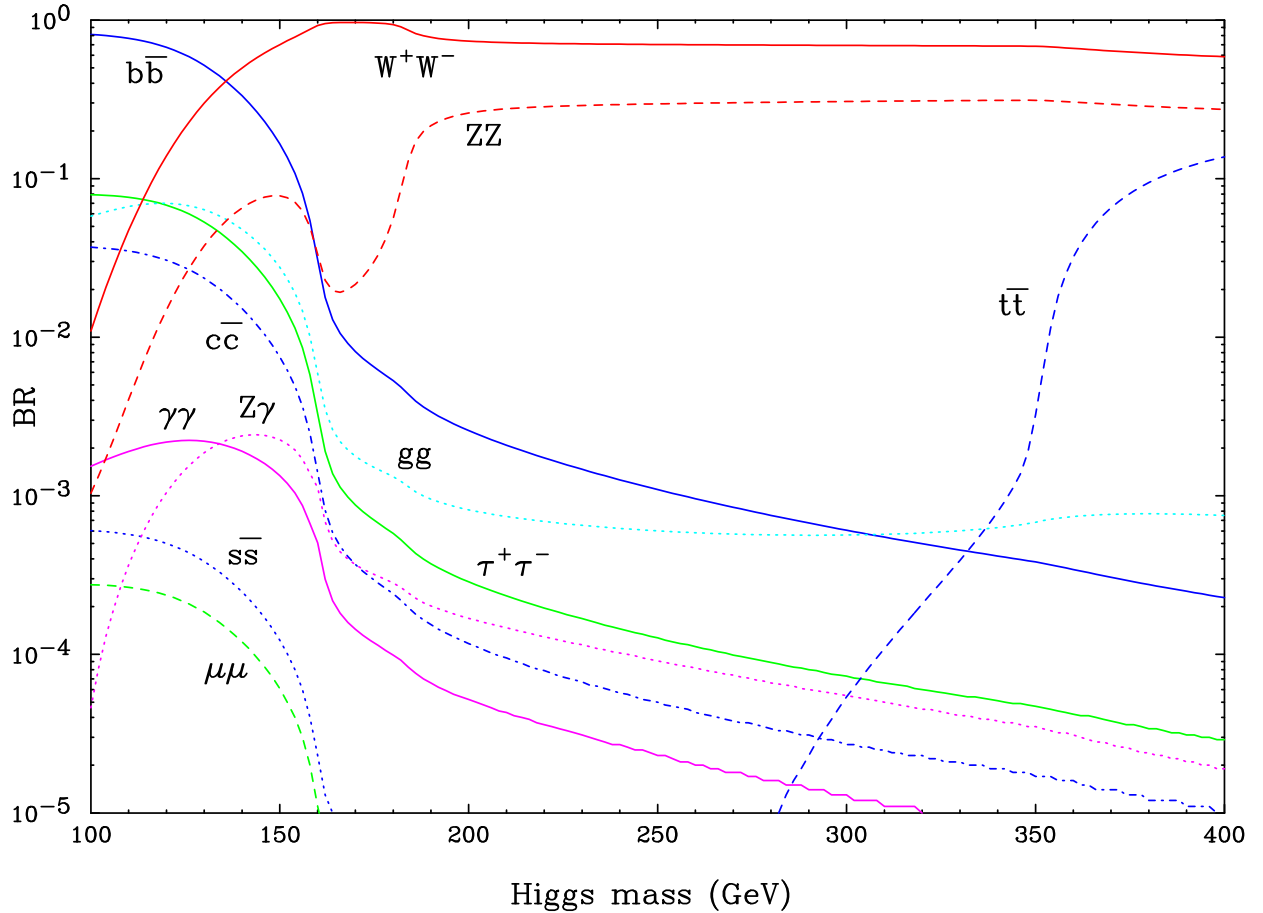}
	\vskip -2.0cm
	\caption[1:fig:HBR]{Branching ratios for the Higgs boson in the 
	standard model, as a function of Higgs mass $m_H$, as computed by
	HDECAY~\cite{Djouadi:1998yw}.}
	\label{1:fig:HBR}
\end{figure}

Figure~\ref{1:fig:HBR} shows the branching ratios for decays of the 
standard model Higgs boson~$H$, as a function of its
mass~\cite{Djouadi:1998yw}.
There are four qualitatively different regions:
\begin{enumerate}
	\item very light Higgs, $m_H\lesssim 113$~GeV.
	The largest standard branching ratios are $b\bar{b}$, 
	$\tau^+\tau^-$, $c\bar{c}$, $gg$, and $\gamma\gamma$.  The last 
	two proceed through loop processes.  At the highest masses in this 
	range $W^\pm W^{\mp*}$ and $ZZ^*$ branching ratios should be 
	observable.

	\item light Higgs, $113~{\rm GeV}\lesssim m_H<2m_W$.
	The branching ratios to $b\bar{b}$, $c\bar{c}$, 
	$\tau^+\tau^-$, $W^\pm W^{\mp*}$, $ZZ^*$ are all measurable, 
	as are the loop-mediated decays $gg$, $\gamma\gamma$, and 
	$Z\gamma$.

	\item intermediate Higgs, $2m_W<m_H<2m_t$.
	The branching ratios to~$WW$ and~$ZZ^{(*)}$ are large and easy 
	to measure.  The decay to~$b\bar{b}$ is rare; to~$gg$, $c\bar{c}$ 
	and~$\tau^+\tau^-$ very rare.

	\item heavy Higgs, $m_H>2m_t$.
	Similar to case~3, but now the branching ratio to $t\bar{t}$ 
	should be measurable.
\end{enumerate}
In cases~1 and~2, the total width of the Higgs is narrow and must be
determined indirectly.
As we explain below, the LC can do so in a model-independent way.
It is broad enough in cases~3 and~4 to allow a direct measurement at
either the LHC or an LC.

The appeal of the LC is extremely strong in cases~1 and~2.
Case~1 is ruled out by non-observation at LEP, unless the Higgs has
a non-standard coupling to $Z$ or decays in a non-standard way.
Some regions of the parameter space in supersymmetric models allow
such behavior, and it is a feature of several other models too.
These scenarios also can pose problems for the LHC, 
either in observation or elucidation.
In case~2 the standard-model-like Higgs is a bonanza for an LC,
already at modest~$\sqrt{s}$.
With the expected luminosity, several tens of thousands of Higgs
bosons should be produced.
The LC would be in a position to check experimentally a remarkable
feature of the standard model, namely that the same Higgs field
gives mass to gauge bosons, charged leptons, up-type quarks, and
down-type quarks.
In fact, the precision should be enough to probe the effects of
virtual corrections from higher-mass particles.

On the other hand, if the decay to real $W^+W^-$ pairs is kinematically
allowed (cases 3 and 4), it and the similar decay to $ZZ^*$ or real
$ZZ$ swamp the rates to quarks and charged leptons.
The integrated luminosity needed to measure the $b\bar{b}$ 
branching ratio, not to mention the $c\bar{c}$ and 
$\tau^+\tau^-$, has not been thoroughly investigated.
We make a first attempt below, showing that ingenuity as well as 
very high integrated luminosity will be needed,
but we have not even started to worry about systematic limitations.

With this background in mind, the rest of this section covers what
hadron colliders at Fermilab and CERN can do (very briefly) and then
reviews the main measurements that an LC can add.
The latter is based mostly on published work and focuses on the light
Higgs.
We add to this a discussion on how to measure the spin and parity of
a light Higgs boson.
Next, we revisit the arguments for a light Higgs.
We note that the data-driven upper bound, currently at 170~GeV 
at~95\% confidence level,
assumes that the standard model is valid up to very high scales.
This assumption is not likely to be right, and if the standard model 
is treated as an effective theory, the bounds are much weaker.
Thus, we also consider properties of a standard-model(-like) Higgs 
boson of intermediate or heavy mass.




\subsection{Higgs Physics at Hadron Colliders}
\subsubsection{Discovery potential}

Now that LEP has ended running, it will be the task of Run~II of the 
Fermilab Tevatron or of the CERN Large Hadron Collider (LHC) to 
determine if a Higgs sector does indeed exist, or some other mechanism 
is responsible for electroweak symmetry breaking and fermion mass 
generation.
Higgs physics in hadronic collisions is complicated by the presence of
several production mechanisms, and the presence of hadronic backgrounds
that obscure some final states.
Nevertheless, it is clear that the LHC especially will provide a wealth
of information on the Higgs boson, and the contribution of the~LC must
be weighed against it.

The Tevatron's $p\bar{p}$ collision energy has been raised to
$\sqrt{s}=2$~TeV, increasing several theoretical Higgs production
cross sections considerably.
The coinciding luminosity upgrade grants it significant
potential to discover a (standard-model-like) Higgs boson,
up to about 180~GeV via a combinatorial analysis of several
channels~\cite{Carena:2000yx}.
The search strategy requires large integrated luminosity, 15~fb$^{-1}$
or more.
In certain scenarios in the minimal supersymmetric standard model (MSSM),
Tevatron's discovery potential is
dramatically better than in the standard model.
There are, however, also large regions of parameter space in which
the Higgs boson would be unobservable.
Furthermore, even if a Higgs-like state is observed at the Tevatron,
it will be a challenge to measure accurately enough its properties
so as to distinguish the underlying model.
Largely this is a function of luminosity: with even twice as much
data as anticipated, one may be able to draw some conclusions about
the nature of the Higgs sector, such as couplings and spin.

If no Higgs boson is observed at the Tevatron, then attention will
shift to the LHC.
Its $pp$ collisions at $\sqrt{s}=14$~TeV will have enough energy
to produce a Higgs of any mass, up to the unitarity constraint of
about 1~TeV.
The LHC is also a good machine for determining the structure
of a Higgs sector.
We outline here both the likely discovery modes of a standard-model Higgs
at the LHC, as a function of the Higgs mass, and measurement prospects
for the quantum numbers which would define the Higgs sector observed.

The dominant production mode over the entire possible mass range of
the Higgs is gluon-gluon fusion, $gg\to H$.
At all Higgs masses above the experimental limit, weak boson fusion
(WBF) is the next largest cross section, about a factor of 8 smaller
than gluon fusion over most of the mass range.
The associated production modes---$WH$, $ZH$, $t\bar{t}H$---have cross
sections that fall off swiftly as the Higgs mass increases.
In each case one must consider several different decay channels,
depending on the mass, as discussed above.
Neither the size of the production cross section nor the dominant
branching ratio is a good indicator of the best discovery channel.
Discovery potential is instead a complicated function of the relative
size of the cross sections, the decay mode under consideration, and the
richness of the final event structure.
The last is important for providing discriminating power against the
enormous QCD backgrounds that will be present at the LHC.
For instance, if $m_H=120$~GeV, the dominant branching ratio is
to $b$ quark pairs, but the QCD background to $gg\to H\to b\bar{b}$
is about five orders of magnitude larger, making observation of this
channel hopeless.
It is more promising to examine WBF events, which naturally yield 
far-forward and far-backward jets of very high energy for tagging, or 
associated $t\bar{t}H$ production, in which complicated event 
structures are found.
Another alternative is to use other final states: although they have
smaller branching ratios, the backgrounds are often much less severe.

With this overview in mind, the following paragraphs sketch the likely
standard-model Higgs discovery channels, as a function of Higgs mass.
\vspace*{1em}


\noindent $\bullet$ {\sl Higgs mass from 110 to 125 GeV}
\nopagebreak

For this mass range, the significant decays are to $b\bar{b}$ and 
$\tau^+\tau^-$.
The rare decay $H\to\gamma\gamma$ is important, however, owing to 
drastically lower backgrounds. 
At present, $t\bar{t}H$ associated production is the only channel in 
which an observation of $H\to b\bar{b}$ has been shown to be feasible.
CMS and ATLAS studies indicate that 
this would require 100~fb$^{-1}$ or more of integrated luminosity to 
reach $5\sigma$~\cite{CMS,ATLAS}.
With 30--50~fb$^{-1}$ both ATLAS and CMS can reach $5\sigma$ in several
other channels: $gg\to\gamma\gamma$, $qq'\to qq'\gamma\gamma$,
$qq'\to qq'\tau^+\tau^-$~\cite{CMS,ATLAS}, and, for $m_H>115$~GeV,
$qq'\to qq'WW$~\cite{Kauer:2001hi}.
For photon pairs in gluon fusion, the backgrounds are extremely large 
and CMS can probably perform better than ATLAS due to its better 
photon mass resolution.
Thus, for CMS, discovery in this mass range is likely to be via gluon 
fusion Higgs production and decay to a pair of photons.
For ATLAS, the situation is somewhat less certain.
With 50~fb$^{-1}$, it would be able to observe all four modes mentioned
above at approximately the same significance, somewhat above $5\sigma$,
but pending full detector simulation one cannot confidently guess
which one will win out.
But this is rather moot, as both experiments would enjoy the 
confidence of confirming discovery in multiple channels, nearly 
simultaneously.
\vspace*{1em}


\noindent $\bullet$ {\sl Higgs mass from 125 to 200 GeV}
\nopagebreak

For $m_H \gtrsim 125$~GeV, the decay $H\to W^{(*)}W^{(*)}$ becomes
very significant.
Due to the uniqueness of the signature $W^{(*)}W^{(*)}\to e\mu\sla{p}_T$,
in both gluon fusion and WBF production modes, discovery
potential shifts from the fermionic or rare decays to these channels.
Although gluon fusion has a much higher signal rate than WBF, it is
not as clean as WBF, and the channels turn out to
be competitive with each other.

For $m_H < 150$~GeV or so, the WBF channel is probably somewhat
better, although this has not yet been explored fully by the detector
collaborations.
In particular, $gg\to H\to W^{(*)}W^{(*)}$ has not been explored in
published form by the collaborations for $m_H < 150$~GeV.
The WBF channel would require approximately 10--15~fb$^{-1}$ at
$m_H=125$~GeV~\cite{Kauer:2001hi}, but the amount of data required
for a $5\sigma$ observation drops rapidly with increasing Higgs mass,
to only 5~fb$^{-1}$ at $m_H = 140$~GeV~\cite{Rainwater:1999sd}.
For $m_H \gtrsim 150$~GeV, both gluon fusion and WBF would require
less than 5~fb$^{-1}$, or half a year of running at design turn-on
luminosity of the LHC.
We note that the studies of $gg\to H\to W^{(*)}W^{(*)}$ have not yet 
taken advantage of the transverse mass distribution of the $W$
pair, as the WBF studies have, so prospects there may improve
somewhat~\cite{Dittmar:1997ss}.
Considering the short amount of time after turn-on that a discovery
could be made, however, a discovery would probably be
announced in both channels simultaneously, again providing additional
confidence.
\vspace*{1em}


\noindent $\bullet$ {\sl Higgs mass above 200 GeV}
\nopagebreak

Above $m_H = 200$~GeV the only decay channels to consider for 
discovery potential are those to weak bosons, $H\to 
W^+W^-,ZZ$~\cite{CMS,ATLAS}.
Above $m_H = 350$~GeV, the decays $H\to t\bar{t}$ require large 
integrated luminosity and, thus, cannot compete.
The channel $gg\to H\to ZZ\to 4\ell$ is clearly extremely powerful, 
requiring less than 5~fb$^{-1}$ of integrated luminosity up to about 
400~GeV in Higgs mass, and about 5~fb$^{-1}$ up to 500~GeV.
Above $m_H = 400$~GeV, it is likely that production by gluon fusion 
would provide the quickest discovery, simply due to the higher rate 
and low backgrounds: $S/B \gg 1/1$ for gluon fusion in this mass 
range.


\subsubsection{Width measurements}

Beyond discovery of a Higgs-like resonance, experiments must measure its 
couplings to standard particles and test whether it behaves like the
one-doublet or some other Higgs sector. 
One may trade these couplings for Higgs partial decay widths, the
sum of which is the Higgs total width.
The Higgs total width grows with $m_H$, and does not exceed
experimental resolution in direct reconstruction for Higgs masses
below about 220~GeV.
For this range, width resolution is several tens of percent, falling off to 
$\sim 10\%$ around $m_H = 300$~GeV, and achieving a best value of
$\sim 3$--$4\%$ for $m_H = 400$~GeV~\cite{CMS,ATLAS}. 

For $120 < m_H < 200$~GeV, the total width can be determined
indirectly for a standard Higgs sector by summing up the observed
decay branching ratios in several different channels.
Phenomenological studies suggest the decay width to $W^+W^-$ in
such a scenario could be measured to about 8--$10\%$, depending
on the Higgs mass.
The width to $ZZ$ cannot be measured quite as well, but one can improve
its resolution by assuming the $SU(2)$ relation between~$\Gamma_{H\to WW}$
and~$\Gamma_{H\to ZZ}$.
For Higgs masses where decays to tau pairs are visible, the partial
width to $\tau^+\tau^-$ could also be measured:
$\Gamma_{H\to\tau\tau}/\Gamma_{H\to WW}$ could be determined to about
10--$20\%$~\cite{Zeppenfeld:2000td}.
Then, because the width to $b\bar{b}$ is not measured at
LHC, one must calculate~$\Gamma_{H\to b\bar{b}}$ from the
measured~$\Gamma_{H\to\tau\tau}$.
In this way one can determine the total width of the Higgs to 10--20\%,
with better accuracy for larger Higgs mass~\cite{Zeppenfeld:2000td}.
Nevertheless, one should keep in mind the theoretical assumptions that
are required to extract the width from the measurements.
In particular, in the light region, where $H\to b\bar{b}$ dominates, one
must assume that the same Higgs boson generates $m_b$ and $m_\tau$.

For a Higgs sector that is not very much like the standard model, these 
indirect measurements can become much more complicated. 
In general one would be able to detect deviations from the standard
model by taking ratios 
of partial widths, but the uncertainties in width extractions are complicated 
functions of the model. 
Recent work suggests that the LHC will have good capability to detect
invisible Higgs branching ratios as small as $15\%$, but this would
require a modification to the current ATLAS and CMS trigger 
designs~\cite{Eboli:2000ze}.


\subsubsection{Spin determination and $CP$ properties}

Any narrow resonance observed at the LHC with Higgs-like couplings would have 
to be confirmed to be spin zero. The LHC can do a fairly good job at this, but 
has difficulty for Higgs masses above 400 GeV. First, if $H\to\gamma\gamma$ is 
observed, the Landau-Yang Theorem implies that the resonance is not a 
vector, cf.\ Sec.~\ref{sec:spin}. 
The more powerful technique, however, which works for 
$130 \lesssim m_H \lesssim 400$~GeV, is to examine the azimuthal distribution 
of the reconstructed pairs of $Z$ bosons in 
$gg\to H\to ZZ\to \ell^+\ell^-\ell^+\ell^-$~\cite{CMS,ATLAS}. 
This measurement does require large statistics, but has good discriminating 
power.
Recent work~\cite{Plehn:2001nj}, examining the azimuthal distribution
of the tagging jets in $qq'\to Hqq'$, demonstrates that the LHC can
determine the $CP$ nature of a light Higgs, as well as the tensor
structure of the $HWW$ vertex.
An LC can carry out such studies with higher precision, however.


\subsubsection{Higgs self-coupling}

Exploration of a Higgs sector would not be complete without also
measuring the Higgs self-coupling~$\lambda$.
In the standard model this is the free parameter that fixes the
Higgs mass.
In models with two Higgs doublets there is more than one self-coupling;
for the MSSM these become gauge couplings and are rigidly determined.
$\lambda$ can be determined only via direct production of two or more
Higgs bosons, the standard-model cross section for which is extremely 
small at the LHC and the backgrounds are large.
Thus, it is probable that the LHC cannot make a measurement
here~\cite{Djouadi:1999rc}.

\subsection{Light Higgs at LC}
\label{sec:lightHiggs}

If the Higgs boson has a mass between LEP's lower limit and $2m_W$, the
LC can verify experimentally the main features of the Higgs mechanism.
It can also measure small deviations from the standard model, which
indicate higher-mass states coupled to the Higgs.
This is a very exciting prospect.
This subsection reviews how the coupling of the Higgs to other
particles and to itself will be attained.
A full description of the light Higgs physics program is outlined in
Refs.~\cite{Battaglia:2001jb,Tesla,Abe:2001ab}.

\begin{figure}[b]
	\includegraphics[width=0.48\textwidth]{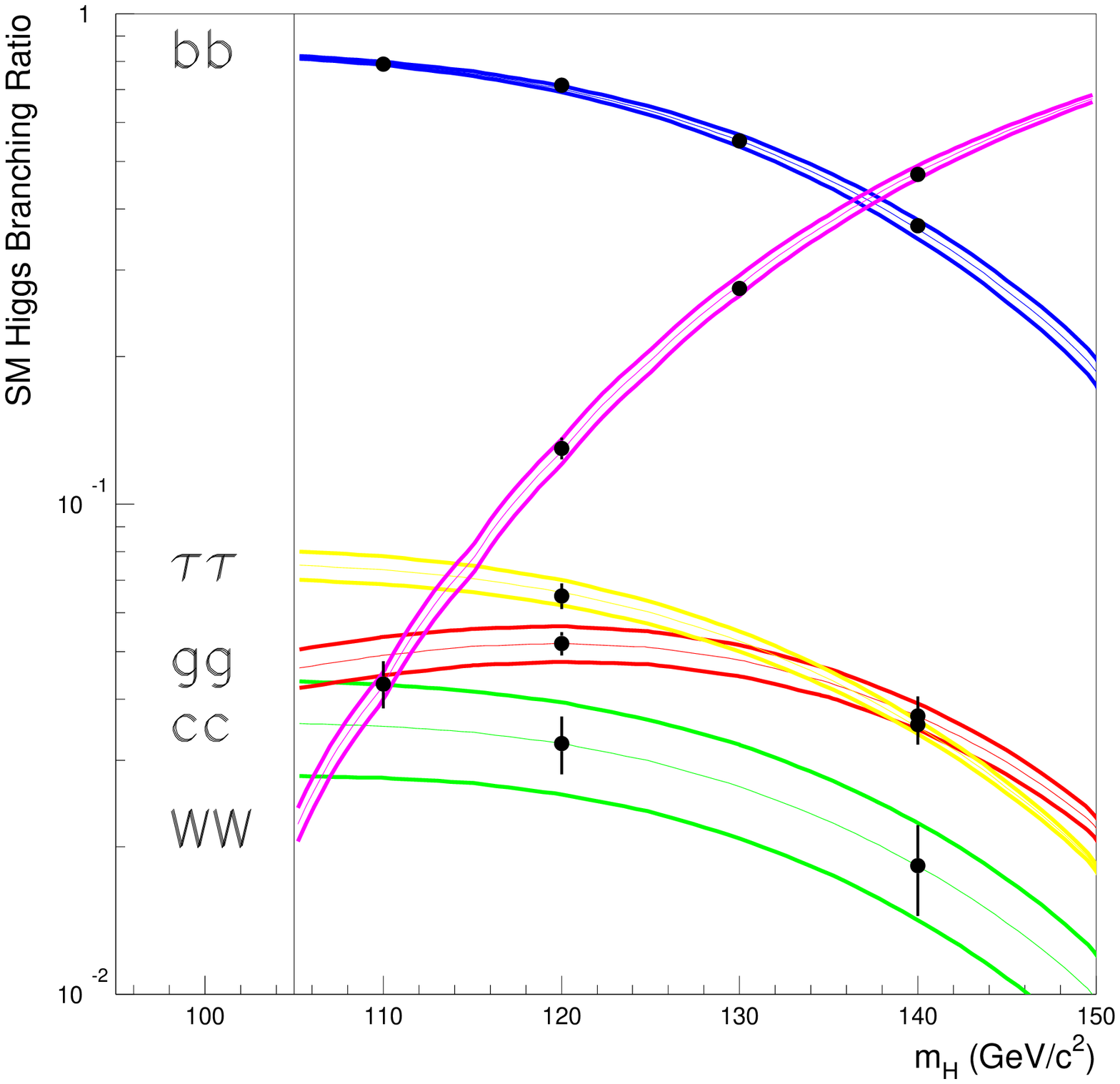}
	\hfill
	\includegraphics[width=0.48\textwidth]{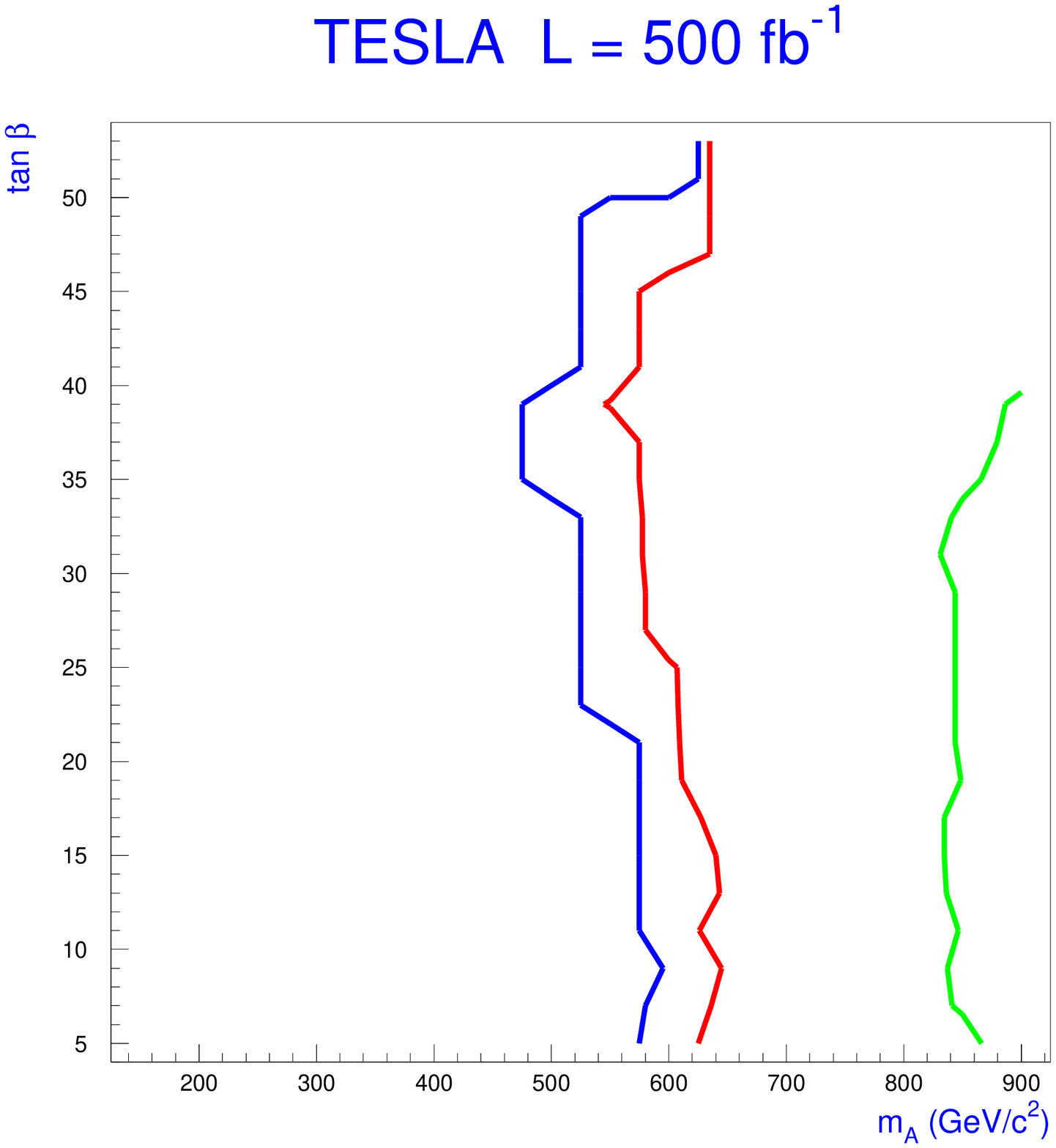}
	\caption[3:fig:hbrsim]{Simulation of 
	a standard-model Higgs, assuming $\int{\cal L}dt=500~{\rm fb}^{-1}$ 
	and $\sqrt{s}=350$~GeV. From Ref.~\cite{Battaglia:1999re}.
	(a) Simulated uncertainties in branching ratios vs.\ $m_H$.
	(b) Reach in the mass $m_A$ of the $CP$-odd state of the MSSM;
	from left to right the curves exclude 95\%, 90\%, and 68\% of
	the parameter space considered in Ref.~\cite{Battaglia:1999re}.}
	\label{3:fig:hbrsim}	
\end{figure}

The two main production channels are $e^+e^-\to Z^*\to ZH$ and 
$e^+e^-\to \bar{\nu}_eW^{+*}W^{-*}\nu_e\to \bar{\nu}_eH\nu_e$.
With standard-model couplings to $Z$ and $W$, the LC will produce
several tens of thousands of Higgs bosons this way, given design
luminosity of around 300~${\rm fb}^{-1}{\rm yr}^{-1}$.
The results~\cite{Battaglia:1999re} for branching ratios of a
recent simulation of Higgs production in $500~{\rm fb}^{-1}$ at
$\sqrt{s}=350$~GeV is shown in Fig.~\ref{3:fig:hbrsim}(a).
This study relies on developments in heavy flavor tagging to separate 
$b$- and $c$-flavored jets from each other and from jets with light 
hadrons.
As one can see, these measurements would be good enough to show that 
the putative Higgs boson generates mass for vector bosons, charged 
leptons, down-type quarks, and up-type quarks.
Through $H\to {\rm loop}\to gg$, it would also probe for non-standard 
colored particles coupling to~$H$.

These measurements would yield a wealth of information.
They test the one-doublet model, in which the ratio of branching 
ratios of particles coupling directly to the Higgs should 
correspond to the ratio of squared masses.
They also test indirectly for more massive states, which produce 
radiative corrections or mixing effects that modify the ratio test.
These deviations diminish for higher mass, so the higher the integrated
luminosity, the higher the reach.
This is illustrated in Fig.~\ref{3:fig:hbrsim}(b), which gives bounds
on the mass~$m_A$ of the $CP$-odd Higgs of the MSSM.
Because this analysis~\cite{Battaglia:1999re} studies only part of the
MSSM parameter space, the numerical reach given here is not definitive.
A more recent study~\cite{Carena:2001bg} finds a similar reach,
except in small regions of MSSM parameter space where the $h^0$
has exceptionally standard properties.
Nevertheless, this kind of analysis indicates that a compelling program
of precision physics is possible.

The LC also gives an indirect, but model-independent, measurement of
the total width.
(A~light Higgs is narrower than the mass resolution.)
The production cross section for $ZH$ ($H\bar{\nu}\nu$) depends on
the partial width $\Gamma_{H\to ZZ}$ ($\Gamma_{H\to WW}$).
Thus, independent measurements of the cross section and the branching
ratio can be combined to obtain the full width without theoretical
assumptions.
A recent study finds an uncertainty on the width of 5--10\% when
$m_H<2m_W$~\cite{Drollinger:2000lc}, somewhat better than at the LHC.
The essential contribution of the~LC would be to determine the width
with no theoretical ingredients.

The LC can also study $t\bar{t}H$ events to measure the top Yukawa 
coupling~\cite{Dawson:1998im}.
As with the branching ratio tests, the result can be compared with the 
top mass, to check how much of $m_t$ the~$H$ at hand generates.
At $\sqrt{s}=500$~GeV there are not many events.
At $\sqrt{s}=800$~GeV a simulation has been carried out recently for
$m_H=120$~GeV, with realistic treatment of backgrounds and detector
performance~\cite{Juste:1999af}.
Assuming $1000~{\rm fb}^{-1}$ of integrated luminosity the study finds a
systematic (statistical) uncertainty of 5.5\% (4.2\%).

Finally, the self-coupling~$\lambda$ can be measured from the rate of
$ZHH$ events~\cite{Miller:1999ji}.
The cross section is small, so luminosity is especially valuable here.
A study assuming $\int{\cal L}dt=500~{\rm fb}^{-1}$
and $\sqrt{s}=500$~GeV has been started.
A measurement at the ten percent level seems feasible, once the
\nopagebreak
procedure has been optimized~\cite{Gay:2000}.

\subsection{$J^{PC}$ of the Higgs Particle}
\label{sec:spin}

In this section we examine how to determine the $J^{PC}$ quantum 
numbers of a (putative) Higgs boson.
The Higgs boson of the Standard Model has, by construction, 
$J^{PC}=0^{++}$.
The strengths and space-time structure of its
couplings to vector bosons 
are uniquely determined, as they come from the covariant 
derivative $(D_{\mu}\phi)(D^{\mu}\phi)$ terms in the Lagrangian.
Models with more Higgs doublets have additional neutral particles.
For example, with two doublets there are three neutral
particles, $h^{0}$, $H^0$, and $A^{0}$.
Neglecting $CP$ violation, the first two are $CP$-even and the last is 
$CP$-odd.
In general the mass eigenstates will be $CP$ mixtures, and with 
$CP$ violation in the Higgs sector the mixing could be significant.

Once one or more new states have been observed at the Tevatron, LHC 
and/or LC, a determination of their quantum numbers and the nature of 
their couplings to vector bosons will be a crucial step toward 
understanding electroweak symmetry breaking.
It is hard to unravel $CP$ mixtures, so we defer discussion of the 
difficulties and some prospects based on interference measurements to
Sec.~\ref{subsec:CP}.
There are several straightforward ways to obtain information on the 
spin: decay to $\gamma\gamma$ implies that the parent state 
cannot have spin~one (Sec.~\ref{subsec:VV}); the rise of the $ZH$ 
threshold depends on the spin of the~$H$ (Sec.~\ref{subsec:prod}); 
and angular distributions of $ZH$ production and decay are diagnostic 
of spin (Sec.~\ref{subsec:ang}).
The first is accessible to the LHC, the second and third are
possible only with a lepton collider.
This section focuses on Higgs bosons, but these experimental 
characteristics can be extended to other bosons as well.

\subsubsection{Decays of a putative $H$ boson to $\gamma\gamma$}
\label{subsec:VV}

Observation of the decay $H \to \gamma \gamma$ places 
restrictions on the possible spin
of a putative Higgs boson.
For the light Higgs, the decay $H \to \gamma\gamma$ is a discovery 
mode at the LHC.
Then, the Landau-Yang theorem~\cite{LandauYang} ensures that $H$ cannot 
have spin~one.
Sakurai~\cite{Sakurai} gives a proof of this in the context of $\pi^0 
\to \gamma\gamma$, and given the importance of this argument, we 
remind the reader of Sakurai's version of its proof.

Let us assume $H\to VV$ decay, where $VV$ are identical massless vector 
bosons.
In momentum space the final state wave function can be constructed 
from the polarization vectors $\vek{\epsilon}_1$ and 
$\vek{\epsilon}_2$ and the relative momentum three-vector $\vek{k}$.
This wave function must be linear in $\vek{\epsilon}_1$ and 
$\vek{\epsilon}_2$, and if the $H$ has spin~one it must transform like 
a vector under rotations.
By Bose symmetry, the wave function must be symmetric under the
interchange $\vek{\epsilon}_{1}\leftrightarrow \vek{\epsilon}_{2}$ and 
$\vek{k}\leftrightarrow -\vek{k}$.
The possible combinations
$\vek{\epsilon}_1 \times \vek{\epsilon}_2$ and 
$(\vek{\epsilon}_1\cdot\vek{\epsilon}_2)\,\vek{k}$ 
for the $H$ transformation rule are ruled out
because they are antisymmetric under the
interchange.
The combination
$\vek{k}\times(\vek{\epsilon}_1 \times \vek{\epsilon}_2)$
is symmetric, but
$\vek{k}\times(\vek{\epsilon}_1 \times \vek{\epsilon}_2) =%
\vek{\epsilon}_1\,(\vek{k}\cdot\vek{\epsilon}_{2}) -%
\vek{\epsilon}_2\,(\vek{k}\cdot\vek{\epsilon}_{1}) =0$,
because $\vek{k}\cdot\vek{\epsilon}_i=0$ for massless vector bosons.
Thus if the decay $H \to \gamma \gamma$ is observed, one can immediately
rule out the possibility that $H$ is spin~one.

A similar conclusion holds when $V$ is a stable, massive vector boson,
but not for the $Z$ boson.
The $Z$'s non-zero width means that it is essentially always off shell.

\subsubsection{Production process}
\label{subsec:prod}

A useful reaction for studies of the Higgs quantum numbers
is $e^+e^-\to ZH$ with the subsequent leptonic decay of the $Z$. 
The cross section is
\begin{equation}
	\sigma(ZH) = \frac{G_{F}^{2}m_{Z}^{4}}{96\pi s}
		\left(a_{e}^{2}+v_{e}^{2}\right) \beta
		\frac{\beta^{2}+12m_Z^2/s}{\left(1-m_Z^2/s \right)^2} 
\end{equation}
where $a_e$ is the axial and $v_e$ the vector coupling of the electron.
(In the standard model, $a_e=-1$ and $v_e=-1+4\sin^2\theta_W$.)
The factor
\begin{equation}
	\beta ^{2} = \frac{4|\vek{p}_Z|^2}{s} = 
		\left(1 - \frac{(m_H+m_Z)^{2}}{s}\right)
		\left(1 - \frac{(m_H-m_Z)^{2}}{s}\right) 
\end{equation}
arises from two-particle phase space.
The single power of~$\beta$ is characteristic of scalar-vector
production, so the threshold behavior is already diagnostic
(see Sec.~\ref{subsec:thresh}).

Events of the associated $ZH$ production can be selected independent of
the Higgs decay modes, by requiring the missing mass relative to the $Z$ 
boson to be near the known Higgs mass.
The background will come from
processes with $Z$ bosons in the final state, principally 
$e^{+}e^{-}\to ZZ$, $e^{+}e^{-}\to Z\gamma^*$, and 
$e^{+}e^{-}\to \gamma^*\gamma^*$. 
In the case of $m_{H}=120$~GeV and $\sqrt{s}=500$ GeV,
the expected numbers of events for 500~${\rm fb}^{-1}$
are shown in Table~\ref{3:tab:events}.
\begin{table}[t]
\centering
\caption[3:tab:events]{Cross sections and event rates for
500~${\rm fb}^{-1}$.}
\label{3:tab:events}
\vskip 6pt
\begin{tabular}{crrr}
\hline\hline
Process & $\sigma$ (fb) & \# of  events & \# of $Z\to e,\mu$ \\
\hline
$ZH$ & 66 & 33000 &2040 \\ 
$ZZ$ & 660 & 330000 & 44000 \\
$Z\gamma^{*}+\gamma^{*}\gamma^{*}$ & 9500 & 475000 &158000000\\ 
\hline\hline
\end{tabular}
\end{table}
It is expected that a clean sample of $ZH$ events can be identified
despite the potentially large background. This is especially true
for a light Higgs boson with a significant branching fraction for
$H\rightarrow b\bar{b}$ decays.  In this case a relatively modest
vertex detector will reduce the background in the $ZH$ events sample
to a level below a few percent.

\subsubsection{Constraining $J^{PC}$ from the cross section at threshold}
\label{subsec:thresh}

As mentioned in Sec.~\ref{subsec:prod}, the behavior of the $ZH$ production
cross section at threshold allows one to constrain the possible values
of $J^{PC}$ of the putative Higgs boson.
In Ref.~\cite{Miller:2001bi} it is shown that the dependence of the
cross section at threshold on $\beta$ 
distinguishes between different spin-parity properties of a putative
Higgs boson produced in Higgsstrahlung.  In particular, if the cross
section grows like $\beta$ at threshold then $H$ must be a $CP$-even scalar,
a $CP$-even vector ({\it i.e.}\ a pseudo-vector), or a $CP$-even spin-2 object.
For all other spin-parity assignments, including the $CP$-odd scalar,
$CP$-odd vector, $CP$-odd spin-2 object, and spins higher than two
with either parity, the cross section at threshold grows like $\beta^3$
or higher powers.

To distinguish between the three spin-parities that give a linear
rise with $\beta$ in the Higgsstrahlung cross section at threshold,
other techniques must be used.
The observation of the decay $H \to \gamma\gamma$, discussed above
in Sec.~\ref{subsec:VV}, immediately rules out spin one.
Angular distributions, discussed below in
Sec.~\ref{subsec:ang}, allow one to distinguish between the possible
spin-parity assignments.
Finally, with enough integrated luminosity doubly-differential angular
distributions can remove any ambiguities remaining for the threshold
behavior and singly-differential angular distributions.

\subsubsection{Determination of $J^{PC}$ from angular distributions}
\label{subsec:ang}

The reaction $e^+e^-\to ZX$ provides a very clean test for the quantum
numbers of the particle~$X$.
In a general model, a scalar can couple to $VV$ through a dimension-3
operator $\phi_S V^{\mu} V_{\mu}$.
A~pseudoscalar can couple to $VV$ through a dimension-5 operator,
$\phi_P F^{\mu \nu} \tilde{F}_{\mu \nu}$.
The pseudoscalar coupling to vector pairs can be large, for example in
topcolor models where $\phi_P$ denotes the topcolor pion.
Several observables are sensitive to the spin-parity of the~$X$:
\begin{itemize}
	\item the angular distribution of the $X$ decay products in its
		center-of-mass system, which depends, in general,
		on the $X$ decay mode, and so differs for $X\to f\bar{f}$
		and $X\to VV$. 
	\item the distribution in $\cos\theta$,   where $\theta$   is the angle 
		between the $e^{-}$ and the~$Z$.
	\item the distribution in $\cos\theta^*$, where $\theta^*$ is the angle
		between the final-state $e^-$ or $\mu^-$ and the $Z$ direction
		of motion, in the $Z$ center-of-mass system.
	\item the distribution in $\varphi^*$, where $\varphi^*$ is the
		angle between the $Z$ production plane and the $Z$ decay plane.     
\end{itemize} 
In the case of $J^{PC}=0^{++}$ (\emph{i.e.}, $X=h,\,H$) the differential
cross section is~\cite{Barger:1994wt}
\begin{equation}
\begin{split}
\frac{d^3\sigma (ZH)}{d\cos\theta d\cos\theta^*d\varphi^*} \sim &
	\,\sin^2\theta \sin^2\theta^* -
	\frac{1}{2\gamma}\sin2\theta \sin2\theta^*\cos\varphi^*   \\
    + & \;\frac{1}{2\gamma^2}\left[ (1+\cos^2\theta )(1+\cos^2\theta^*)
		+ \sin^2\theta \sin^2\theta^*\cos 2\varphi^*\right]   \\
    - & \;\frac{2v_{e}a_{e}}{v_{e}^2+a_{e}^2}
		\frac{2v_{f}a_{f}}{v_{f}^2+a_{f}^2}
		\frac{2}{\gamma }\left[ \sin\theta \sin\theta^*\cos\varphi^*-
			\frac{1}{\gamma }\cos \theta \cos\theta^*\right] ,
\end{split}
\label{scalar}
\end{equation}
where $\gamma$ is the Lorentz boost of the $Z$ boson.
In the  $0^{-+}$ case (\emph{i.e.}, $X=A$) the corresponding cross
section is
\begin{equation}
\begin{split}
\frac{d^3\sigma (ZA)}{d\cos\theta d\cos\theta^*d\varphi^*} \sim &
	\,1+\cos^2\theta\cos^2\theta^* -
		\frac{1}2\sin^2\theta \sin^2\theta^* \\
    - &\; \frac{1}{2}\sin^2\theta \sin^2\theta^*\cos 2\varphi^*
	  +2\frac{2v_{e}a_{e}}{v_{e}^2+a_{e}^2}
	  	\frac{2v_{f}a_{f}}{v_{f}^2+a_{f}^2}\cos\theta \cos\theta^* .
\end{split}
\label{pseudoscalar}
\end{equation}
The distinctive difference in these angular distributions
provides a basis for distinguishing between the two cases.

\begin{figure}[b]
\centering
\includegraphics[height=9cm]{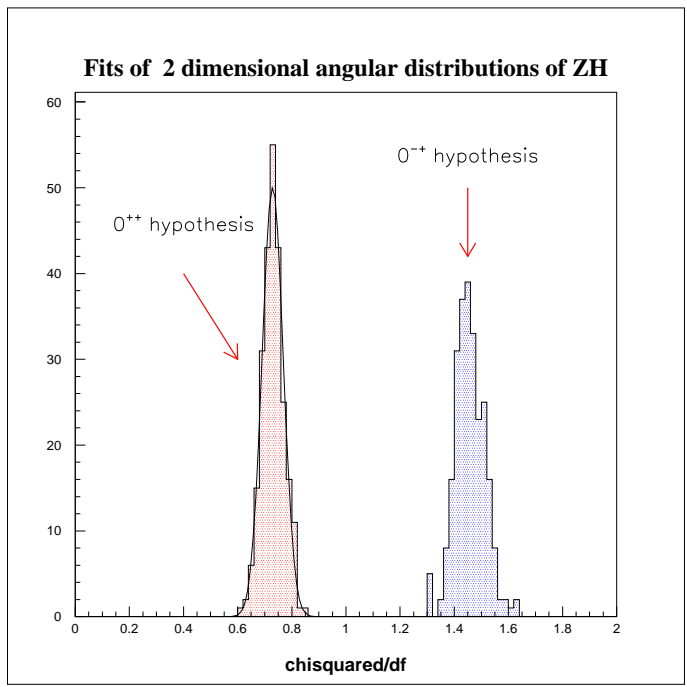}
\caption{Fits of two-dimensional angular distributions of samples of 
$ZH$ events with  the scalar and pseudoscalar hypothesis.}
\label{2dim}
\end{figure}

The triply-differential angular distributions of Eqs.~(\ref{scalar}) 
and~(\ref{pseudoscalar}) contain non-trivial correlations between the 
production angle $\theta$ and the decay angles $\theta^*$ and 
$\varphi^*$.
Most of these correlations do not contribute to doubly- 
or singly-differential distributions.
A~fit to doubly-differential distribution
$d^2\sigma(ZA)/d\cos\theta d\cos\theta^*$ of the corresponding
event samples yields a 14$\sigma$ separation between the $0^{++}$
and the $0^{-+}$ cases, see Fig.~\ref{2dim}.

\begin{figure}[b]
\centering
	\includegraphics[height=11cm]{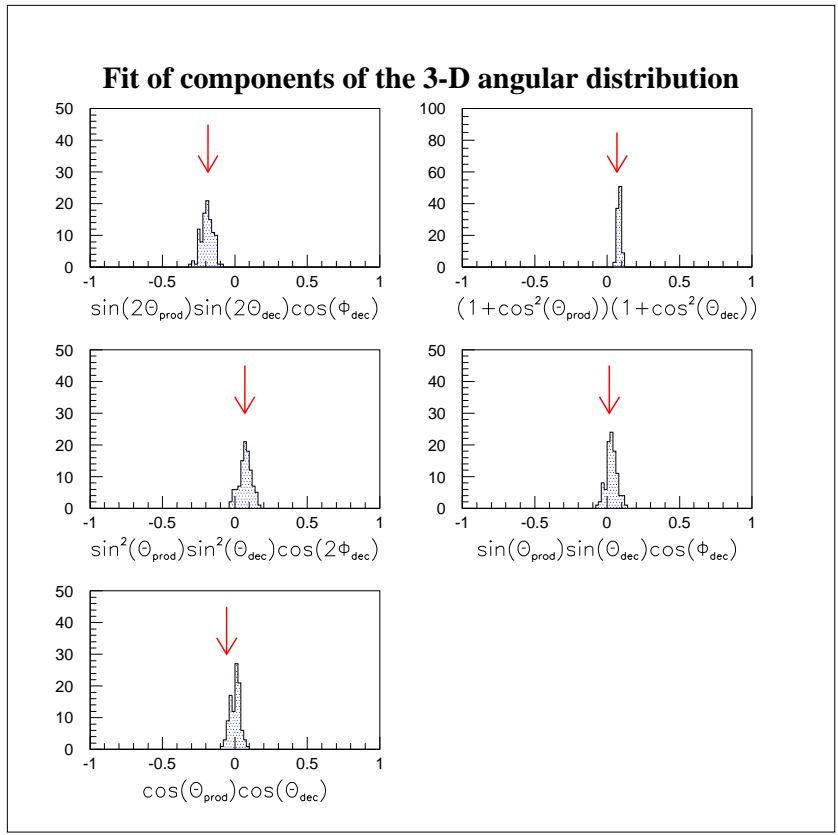}
	\caption[terms]{Fits of separate terms of the three-dimensional
		angular distributions, Eq.~(\ref{scalar}), of samples of $ZH$
		events. The relative weights of the separate terms, with respect
		to $\sin^2(\theta)\sin^2(\theta^{*})$ expected from
		Eq.~(\ref{scalar}) are indicated by the arrows.}
	\label{terms}
\end{figure}

A powerful test of the spin-parity of the Higgs boson consists of
determination of the contribution of separate terms of the
angular distributions, Eq.~(\ref{scalar}). Besides a strong confirmation 
of the expected $J^{PC}$ assignment, such decomposition can provide
limits on a non-standard $ZZH$ coupling, even for the $0^{++}$ case.
The statistical power of such an analysis of a sample of events 
corresponding to a large integrated luminosity of 1250~fb$^{-1}$ is 
shown in Table~\ref{3:tab:term} and illustrated in Fig.~\ref{terms}.
\begin{table}[t]
\centering
\caption[3:tab:term]{Comparison of expected and measured contributions 
 to the 3-D angular distribution.}
\label{3:tab:term}
\vskip 6pt
\begin{tabular}{lr@{.}lr@{$\pm$}l}
\hline\hline
\multicolumn{1}{c}{Term} & 
\multicolumn{2}{c}{Expected, Eq.~(\ref{scalar})} & 
\multicolumn{2}{c}{Measured}  \\
\hline
$\sin2\theta \sin2\theta^{*}\cos\varphi^{*}$ & \quad \quad
	$-0$&185 & $-0$.189 & 0.037 \\
$(1+\cos^{2}\theta)(1+\cos^{2}\theta^{*})$ &
	  0&0682 &  0.082 & 0.010 \\
$\sin\theta \sin\theta^{*} \cos\varphi^{*}$ &
	  0&0682 &  0.080 & 0.014 \\
$\sin^{2}\theta \sin^{2}\theta^{*} \cos2\varphi^{*}$ &
	  0&015  &  0.030 & 0.074 \\
$\cos\theta \cos\theta^{*}$ &
	  0&006  &  0.010 & 0.071 \\
\hline\hline
\end{tabular}
\end{table}

\subsubsection{Measuring $CP$ properties of a scalar boson}
\label{subsec:CP}

As described in the previous section, to distinguish models with large 
scalar or pseudoscalar couplings to vector boson pairs, it is useful 
to determine which operator the coupling comes from, by measuring angular 
distributions in $e^+e^- \to ZX$.
These and other methods can also be used to determine the $CP$ properties
of a mixed-$CP$ state.
One would first like to distinguish a $CP$-even Higgs boson from a
$CP$-odd Higgs boson, and, second, to be able to determine whether
the observed Higgs boson is a $CP$ mixture and, if so, measure the odd
and even components.

The angular dependence of the $e^+e^-\to ZH$ cross section depends
upon whether the Higgs boson $H$ is $CP$-even, $CP$-odd, or a 
mixture~\cite{Miller:2001bi,Barger:1994wt,Hagiwara:1994sw,Han:2001mi}.
Following Ref.~\cite{Han:2001mi} we parametrize the $ZZH$ vertex as
\begin{equation}
    \Gamma_{\mu\nu}(k_1,k_2) = 
	a g_{\mu\nu} +
	b \frac{k_{1\mu} k_{2\nu} - g_{\mu\nu}k_1\cdot k_2}{m_Z^2}  +
	\tilde{b}\frac{\varepsilon_{\mu\nu\alpha\beta} k_1^\alpha k_2^\beta}{m_Z^2}
    \label{eq:zzh}
\end{equation}
where $k_1$ and $k_2$ are the momenta of the two $Z$s.
The first term arises from a standard-model-like $ZZH$ coupling, and the
last two from effective interactions that could be induced by high-mass
virtual particles.
The coupling $\tilde{b}$ violates $CP$, but the other two conserve $CP$.
With this vertex the Higgsstrahlung cross section becomes
\begin{equation}
\frac{d\sigma}{d\cos\theta_Z} \propto 1
    + \frac{p_Z^2}{m_Z^2} \sin^2\theta_Z
    - 4\, {\rm Im}\left[\tilde{b}\over\tilde{a}\right]
      \frac{v_ea_e}{v_e^2+a_e^2}
      \frac{p_z\sqrt{s}}{m_Z^2} \cos\theta_Z
    +\left|\tilde{b}\over \tilde{a}\right|^2
      \frac{p^2_zs}{2m_Z^4} (1+\cos^2\theta_Z)
    \label{eq:zzhX}
\end{equation}
where $\tilde{a}=a-bE_Z\sqrt{s}/m_Z^2$; $\theta_Z$, $p_Z$, and $E_Z$
are the scattering angle, momentum, and energy of the final-state $Z$
boson; and $v_e$ and $a_e$ are the vector and axial-vector couplings
at the $e^+e^-Z$ vertex.
The term in Eq.~(\ref{eq:zzhX}) proportional to $\cos\theta_Z$ arises
from interference between the $CP$-even and $CP$-odd couplings in
Eq.~(\ref{eq:zzh}).
If the $CP$-odd coupling~$\tilde{b}$ is large enough, it can be extracted
from the forward-backward asymmetry.
Ref.~\cite{Han:2001mi} studied whether the couplings $a$, $b$ and
$\tilde{b}$ in Eq.~(\ref{eq:zzh}) could be extracted, using
$e^+e^-\to f\bar{f}H$ from Higgsstrahlung and $ZZ$ fusion, and found
the real and imaginary components of the three couplings could be 
determined from asymmetries.
A~more general analysis of the $ZZH$ \emph{and} $Z\gamma H$
couplings, using the so-called ``optimal observable'' method, found
that the $ZZH$ couplings can be well constrained with or without
beam polarization, while the $Z\gamma H$ couplings do require beam
polarization~\cite{Hagiwara:2000tk}.

It is important to note that any measurement of $CP$ violating
observables in Higgs boson production or decay is a measurement of $CP$
violation in the Higgs boson {\it couplings} to the particular initial
or final state.
It is not, however, a direct measurement of the $CP$ content of the
Higgs mass eigenstate(s).
This has not often been pointed out in the literature.
For example, an MSSM Higgs boson could be a half-and-half mixture of
the $CP$-even state $H$ and the $CP$-odd state~$A$.
If one is not too far into the decoupling regime, then the tree-level
$CP$-even coupling of $H$ to $Z$ boson pairs (denoted above by $a$)
will be non-negligible, while the loop-induced dimension-5 couplings
of $H$ and $A$ to $Z$ pairs ($b$ and $\tilde b$, respectively) will
be very small.
A measurement of the couplings $a$, $b$ and $\tilde b$ would indicate
correctly that the $CP$ violation in the Higgs couplings to vector
boson pairs is very small, because $a \gg \tilde b$.
This does not, however, indicate that the $CP$ mixing in the Higgs
mass eigenstate is small.

With this in mind let us first assume that there is no mixing of the
$CP$-odd and $CP$-even Higgs bosons.
If there is no $CP$ violation in the Higgs sector itself, \emph{i.e.}\ if
all the Higgs fields have real vacuum expectation values, then only
the $CP$-even Higgs fields couple to $ZZ$ and $W^+ W^-$ at tree level.
The dimension-5 couplings of the $CP$-even and $CP$-odd Higgs bosons
to vector pairs are induced at loop level, and so are loop suppressed.  
These loop induced couplings are very small: for example, 
the loop-induced production process $e^+e^- \to ZA$ has been 
studied in the MSSM for a 500~GeV LC, and the cross section was found
to be below 0.1~fb \cite{Akeroyd:1999gu}.
Because the $e^+e^- \to ZA$ cross section is likely to be very 
small in a general Higgs sector, 
it may be impractical to measure angular distributions 
in $ZA$ production.
Likewise, in a general Higgs sector the $CP$-odd Higgs
branching ratio to $VV$ is likely to be very small, making it difficult
to measure angular distributions in decay; for example, in the MSSM the 
branching ratios of $A \to VV$ are typically well below $10^{-2}$ 
\cite{Gunion:1992cw}.

As noted above, a mixed-$CP$ Higgs boson is well-motivated in the
MSSM~\cite{Pilaftsis:1998pe}.
Mixing of $CP$ eigenstates can be induced at the 1-loop level by soft
$CP$-violating trilinear couplings between the Higgs bosons and top
and bottom squarks.
Unfortunately, it is difficult to study the $CP$ mixing through the
Higgs couplings to vector boson pairs because the dimension-5 couplings
$b$ and $\tilde b$ described above are likely to be very small.
The mixed-$CP$ Higgs then couples to $VV$ primarily through its
$CP$-even component.
Thus, $a \gg b, \tilde b$, and the coupling is predominantly that of the
standard Higgs, suppressed by a mixing angle (since only the $CP$-even
component contributes to $a$), and the effects of the dimension-5
couplings $b$ and $\tilde b$ on the angular distributions will be small.
The mixing-angle suppression is \emph{not} diagnostic of $CP$ mixing,
because the same suppression can arise in any multi-Higgs-doublet
model (such as the MSSM) where the $VV$ couplings of $h^0$ and $H^0$
are suppressed by $\sin(\beta-\alpha)$ and $\cos(\beta-\alpha)$,
respectively.

To probe $CP$ mixing in the Higgs mass eigenstate, $CP$-violating
observables in which the $CP$-even and $CP$-odd couplings are both
large are desirable.
The couplings to photon or fermion pairs of $CP$-even Higgs bosons
are comparable to those of $CP$-odd Higgs bosons.
Three methods making use of these couplings are described in the
literature; they employ $s$-channel Higgs production at a photon or
muon collider, and none of them is possible at an $e^+e^-$~LC.
At a photon collider with transversely polarized photon beams,
the initial state is pure $CP$-even if the polarizations of the beams 
are parallel, and pure $CP$-odd if the polarizations of the beams
are perpendicular.  One can then turn on or off the $CP$-even and odd
components of a Higgs resonance by changing the orientation of the 
polarization of the photons.  Combining the observables from linearly
polarized photons with those from circularly polarized photons, 
one can disentangle the $CP$-even and $CP$-odd couplings of a Higgs
resonance to photon pairs~\cite{Gunion:1994wy}.

At a photon collider one can also take advantage of the interference 
between a Higgs-mediated
process and a process with the same final states mediated by something
else.
This is familiar from the measurement of $CP$ violation in the $B$
and $K$ meson systems.
In Ref.~\cite{Asakawa:2000jy} the process $\gamma\gamma\to t\bar{t}$
is studied near the Higgs resonance.
In this process a Higgs-mediated component of the amplitude interferes
with the continuum amplitude enabling one to determine the $CP$
properties of the Higgs.

Finally, one can use fermion polarization to measure the mixed-$CP$ 
Higgs Yukawa couplings to fermions.
In Ref.~\cite{Grzadkowski:2000hm} a muon collider with polarized beams
is proposed, to produce Higgs bosons in the $s$~channel through the
mixed-$CP$ muon Yukawa coupling, $\bar \mu (a + ib\gamma_5) \mu$.
The $CP$-even and $CP$-odd components of the Higgs coupling to muon 
pairs can be disentangled using transversely polarized muon beams.
Similarly, Ref.~\cite{Asakawa:2000es} studies the process
$\mu^+\mu^-\to H\to f\bar{f}$, where $f\bar{f}$ are third-generation
fermions.
Helicity observables in $\mu^+\mu^-$ and $f \bar f$ allow one to
measure the $CP$-even and $CP$-odd components of the Higgs couplings.

\subsection{Expected Mass of the Higgs Boson}
\label{sec:bounds}

As discussed at the beginning of this section, the physics of a
standard-model-like Higgs boson depends critically on whether the
decay to an on-shell $WW$ pair is kinematically allowed or not.
It is therefore crucial to review what is known today about the mass 
of the Higgs boson from indirect measurements and theory.
It is impossible to interpret the measurements without some recourse
to theory, but we shall do so with as few assumptions about the Higgs
sector as possible.
In particular, we do not assume that the standard model is a 
fundamental theory---we know that it is not---but rather an effective 
field theory, valid up to some scale, denoted here 
as~$\Lambda_{\rm SM}$.
This is a very mild assumption, and it weakens well-known bounds based 
on precise electroweak data, which assume $\Lambda_{\rm SM}\to\infty$.

The most powerful constraints are the precise measurements of the 
$Z^0$ line-shape from SLD and the four LEP experiments, combined with 
measurements of $\sin^2\theta_W$ from deeply inelastic scattering, of 
parity violation in cesium and thallium atoms, of the top quark mass 
in pair production at the Tevatron, and of the $W$ mass from 
the Tevatron and LEP experiments.
In general, particles beyond each experiment's kinematic reach 
contribute to the observables virtually: either in 
loop processes or as off-shell propagators.
For the Higgs boson, and other manifestations of symmetry breaking, 
the most sensitive contributions are through loops in
the $W$, $Z$, and $\gamma$ propagators.
These are often called oblique corrections, and it has become customary
to summarize them with two parameters, $S$ and~$T$, describing the weak
isopin-conserving and -violating contributions~\cite{Peskin:1990zt}.
(A third quantity~$U$, which parametrizes energy-dependent 
isospin violating effects, can be neglected at $Z$ energies.)

\begin{figure}[bp]
	\centering
 	\includegraphics[width=0.48\textwidth]{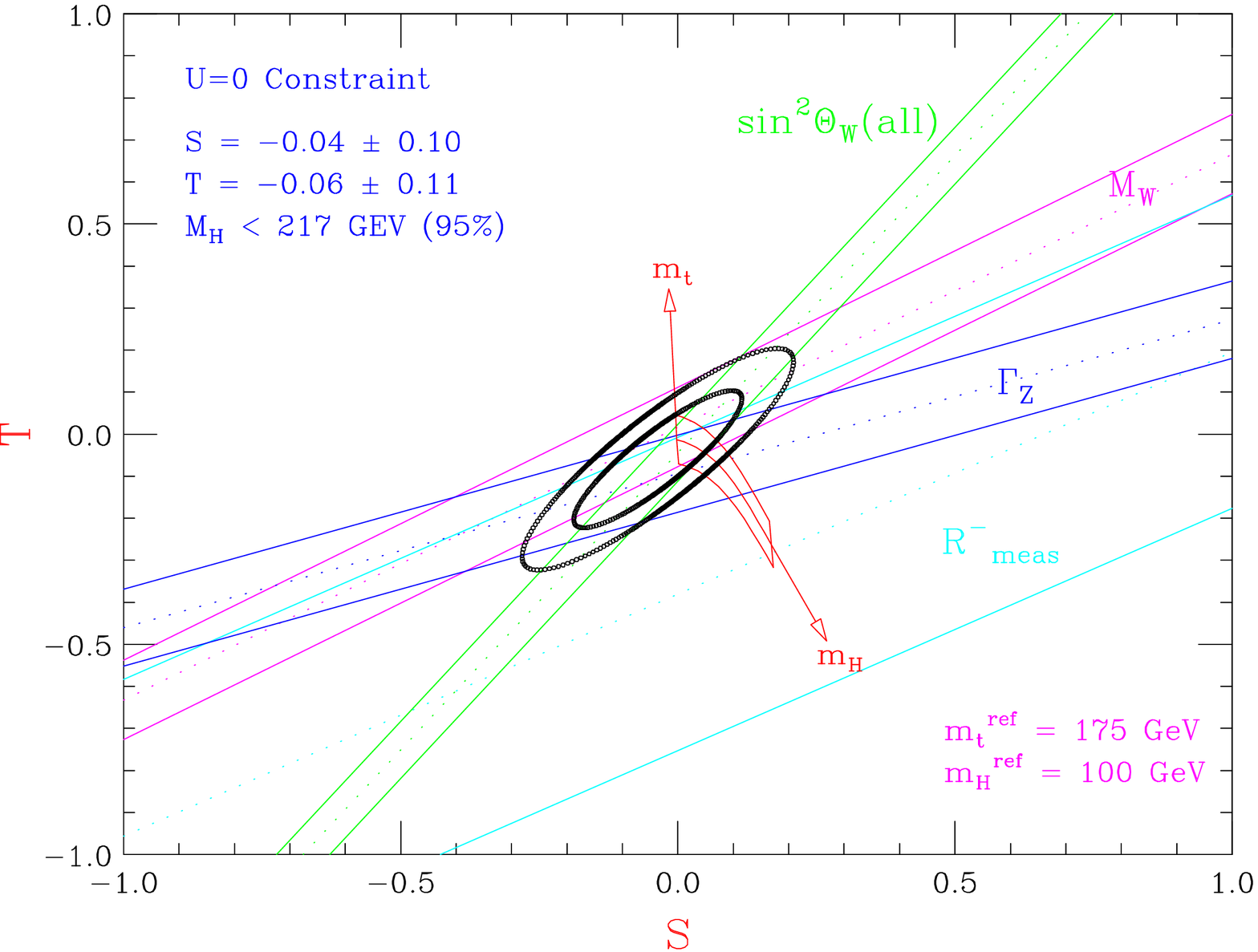} \hfill
 	\includegraphics[width=0.47\textwidth]{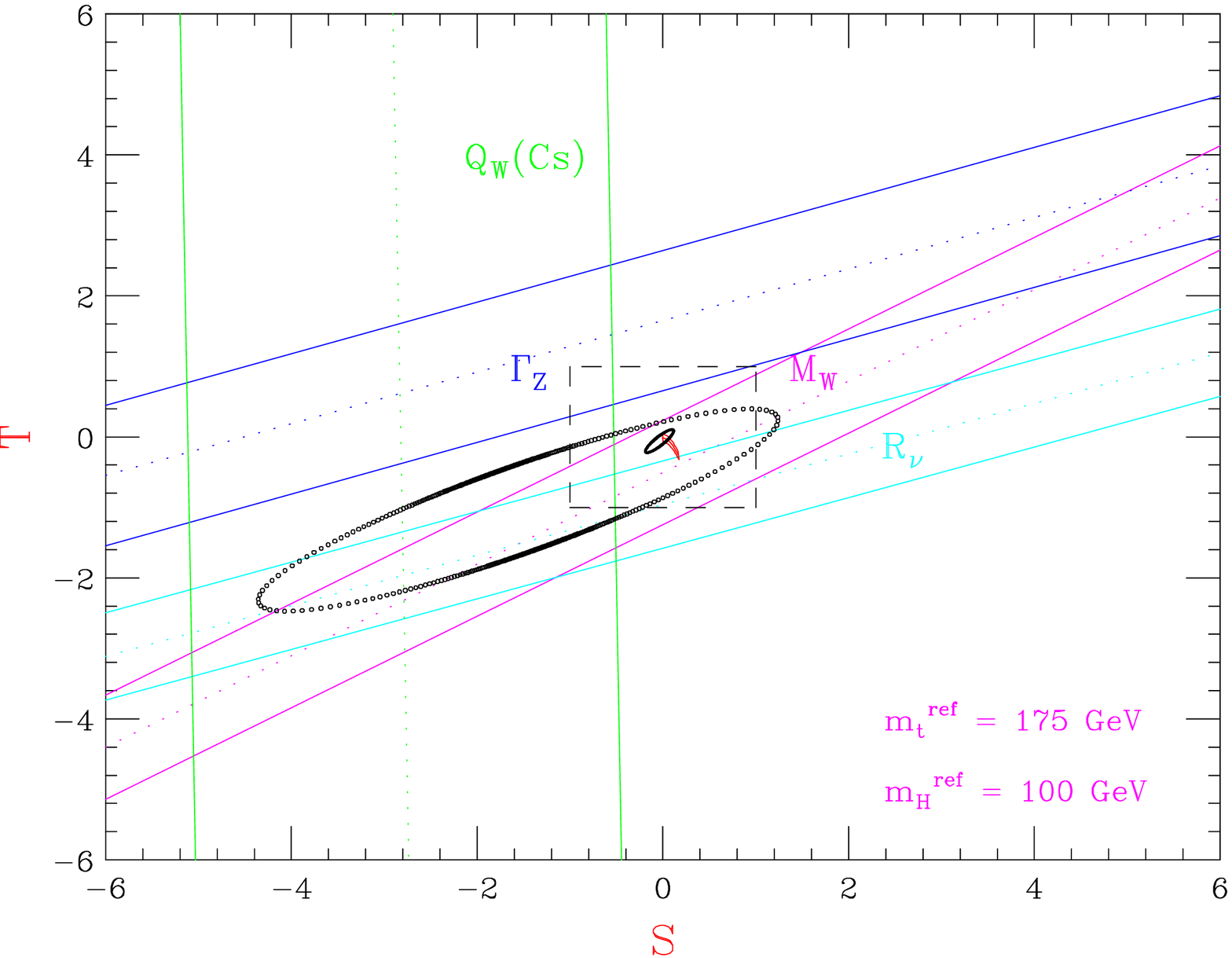}
	\caption[3:fig:ST]{Fits of precisely measured electroweak
	observables to $S$ and $T$, including 68\% confidence bands of the
	most precise ones~\cite{Swartz:1999xv}.
	(a) Fit to 14 observables from August 1999.
	(b) Fit from 1989, with (a) as inset.}
	\label{3:fig:ST}
\end{figure}

A recent fit to $S$ and $T$ of the precisely measured electroweak 
observables is shown in Fig.~\ref{3:fig:ST}(a)~\cite{Swartz:1999xv}.
Varying $m_t$ within the uncertainty of the direct measurement and~$m_H$ 
from 100~GeV to 1~TeV traces out the crescent-shaped region.
The uncertainty of other standard model parameters, apart from
$\alpha_s(m_Z)$, on this region is unimportant.
To obtain the crescent the standard model is treated as a fundamental 
field theory.

The agreement between the crescent for the standard model prediction and 
the experimentally favored ellipse is not guaranteed.
The parameters $S$ and $T$ are defined independently of unknown 
short-distance physics and are determined from the data with only 
well-substantiated aspects of the electroweak theory.
One may think of the ellipse, therefore, as an experimental measurement.
The crescent, on the other hand, is a theoretical prediction of a
particular model of the Higgs sector, namely, the standard one with
one doublet.
The overlap of the two regions demonstrates that the one-doublet model 
is a very good description of the data.

Other models trace out different regions.
For example, the MSSM's region overlaps with the ellipse in the
decoupling limit, when all superpartners are heavy.
The MSSM also agrees with the fit when only squarks and sleptons
are heavy, but charginos and neutralinos are light---as light as
100~GeV~\cite{Pierce:1997zz}.
Early models of technicolor, which do not possess a decoupling limit, 
trace out regions at significantly larger~$S$.
Models with composite Higgs bosons trace out regions connected to
the standard crescent, but extending to somewhat larger $T$: for them
the data allow an intermediate-mass or even heavy Higgs.
In this way, the constraints of the data are incisive in deciding what 
extensions of the standard model should be taken seriously:
any model with a decoupling limit naturally agrees with the data just
as well as the standard model.

\begin{figure}[bp]
	\centering
 	\includegraphics[width=0.5\textwidth]{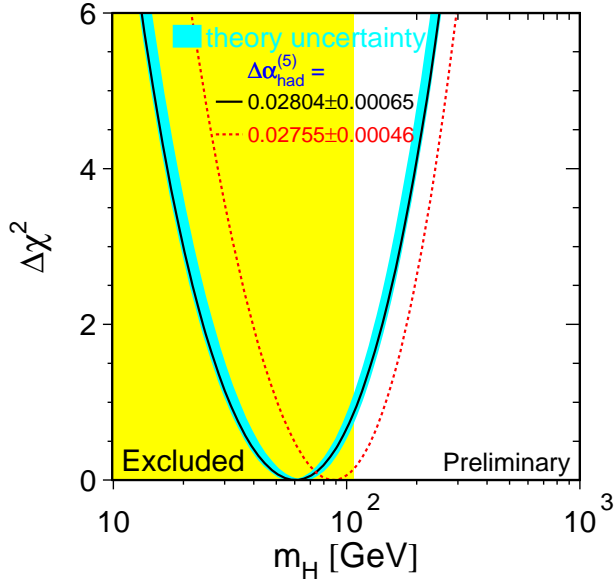}
	\caption[3:fig:blue]{Constraint on $m_H$ from the precision 
	observables, treating the standard model as a fundamental field
	theory.  Status as of August 2000~\cite{LEPEWWG}.}
	\label{3:fig:blue}
\end{figure}

It is common to distill the fit of Fig.~\ref{3:fig:ST} into a 
constraint on $\log(m_H/m_Z)$.
This process yields the well-known ``blue band'' plot of the LEP 
Electroweak Working Group~\cite{LEPEWWG}, shown in
Fig.~\ref{3:fig:blue}, and equivalent results from other groups.
As with any fit there are assumptions behind it.
An important assumption behind Fig.~\ref{3:fig:blue} is to restrict 
the free parameters to the renormalizable couplings of the 
one-doublet Higgs model, which is equivalent to assuming
$\Lambda_{\rm SM}\to\infty$.
The bound on the Higgs mass suggested by the blue-band 
fit is now $m_H<170$~GeV at 95\% (one-sided) confidence
level~\cite{Pietrzyk:2000oj}
(and $m_H<270$~GeV at 99\% CL).
At high confidence, the fit would put the standard-model Higgs in
the golden region with many measurable branching ratios.
The combination of Figs.~\ref{3:fig:ST} and~\ref{3:fig:blue}, which 
seem to imply that the real world is very like the standard model and 
that the Higgs is light, is sometimes used to argue that Higgs physics at 
the LC is nearly guaranteed to be extremely compelling.

There are two reasons to be careful about such an argument.
First, the bound is brittle, because it is really a bound on
$\log(m_H/m_Z)$.
Recent measurements from BES of the cross section for 
$e^+e^-\to{\rm hadrons}$, at $\sqrt{s}$ above and below the $\psi$ 
resonances, require a change in the treatment of the running of the 
electromagnetic coupling.
After re-fitting, the bound on the Higgs mass appears to be several 
tens of GeV higher~\cite{Pietrzyk:2000oj}.
In a more qualitative vein,
a few years ago the preferred ellipse was at somewhat lower 
$S$ and $T$.
At that time the data required non-standard models with heavy Higgs 
to have a small, negative shift in~$S$.
Now the data allow also non-standard models with a small, positive shift
in~$T$.

A second, deeper reason to be suspicious of Fig.~\ref{3:fig:blue}
is the omission of non-renormalizable interactions.
From a modern understanding of field theory, the standard model is 
an effective field theory, valid up to a scale $\Lambda_{\rm SM}$.
At energies above $\Lambda_{\rm SM}$, nature should be explained by a 
more profound field theory or, perhaps, string theory.
At present there is no experimental information on $\Lambda_{\rm SM}$,
although there are several competing theoretical ideas with 
$\Lambda_{\rm SM}$ in the range 0.25--5~TeV.
(Examples of $\Lambda_{\rm SM}$ are the typical mass of the
lowest-lying superpartners, or the scale at which composite structure
of the Higgs is evident.)

It is worth emphasizing that the scale $\Lambda_{\rm SM}$ must be 
finite, and not simply because model-building theorists believe in 
grand unification or string theory.
In the mid-to-late '80s there was great interest in the high-energy
limit of scalar field theories, such as the Higgs sector of the
standard model.
This is not an easy problem, because as the energy probed becomes 
higher, the self-couplings of scalar fields grow, and the problem 
becomes non-perturbative.
The best work was done by those working at the interface of 
particle physics and mathematical physics~\cite{Dashen:1983ts}.
To make a long story short, no way was found to take 
$\Lambda_{\rm SM}\to\infty$, unless the renormalized self-coupling
vanishes in the limit.
(This is the so-called ``triviality'' of scalar field theory, because 
there is no interaction at finite, physical energies.)
On the other hand, a phenomenologically viable theory, with 
non-vanishing self-interaction and $m_H$, is obtained for 
finite~$\Lambda_{\rm SM}$.

Once one accepts that $\Lambda_{\rm SM}$ is finite and unknown, the 
fits leading to the blue band must be redone, allowing 
higher-dimension (or non-renormalizable) interactions to float in the 
fit~\cite{Chivukula:1999az}.
These contributions are suppressed by a factor of 
$(v/\Lambda_{\rm SM})^2$, or a higher power, where $v=246$~GeV is
the Higgs field's vacuum expectation value.
Unless $\Lambda_{\rm SM}$ is close to $v$, these contributions 
are small, but today's data are precise enough to notice them even 
if~$\Lambda_{\rm SM}$ is as high as several~TeV.
It is easy to understand why they have been omitted for so
long: when precision fits were first carried out, as seen in
Fig.~\ref{3:fig:ST}(b), the data just began to constrain radiative
effects; power-suppressed contributions were in the noise.
Now, however, the precision of the data is good enough to be sensitive
to power corrections as well. 

\begin{figure}[bp]
	\centering
 	\includegraphics[width=0.8\textwidth]{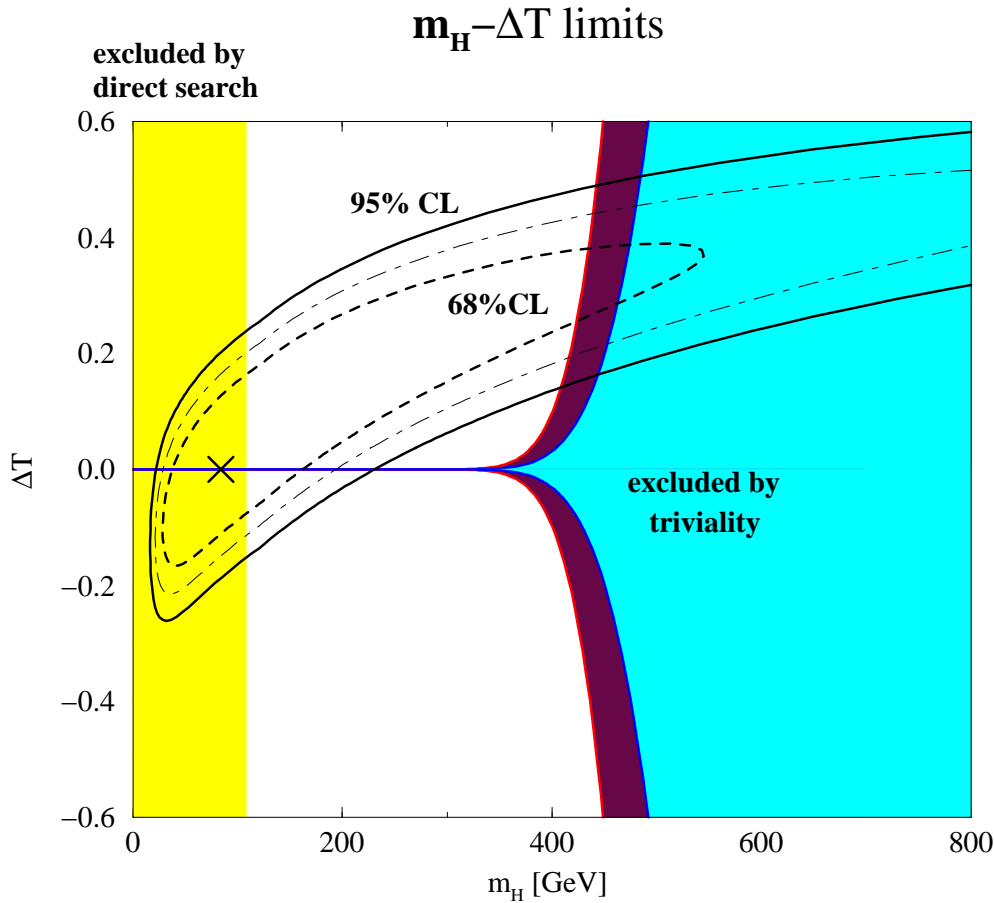}
	\caption[3:fig:mHT]{Constraint on $m_H$ and $T$.
	From Ref.~\cite{Chivukula:2000px}.
	The contours show the fit to precisely measured electroweak
	observables.
	The exclusion for low mass is from non-observation at LEP.
	In the region ``excluded by triviality'' many observables,
	not just those affecting $T$, would be expected to show
	non-standard effects at the per-mil level or higher, as discussed
	in the text.}
	\label{3:fig:mHT}
\end{figure}

The omission of the higher-dimension interactions has been restored,
in an essentially model-independent way, in at least three papers.
Hall and Kolda~\cite{Hall:1999fe} considered operators that would
be induced by TeV-scale quantum gravity, and find that the bound is
removed for $4~{\rm TeV}\lesssim\Lambda_{\rm SM}\lesssim 11~{\rm TeV}$.
Bagger, Falk, and Swartz~\cite{Bagger:2000te} considered an effective
field theory without a propagating Higgs field, which applies to
models with Higgs mass also of order~$\Lambda_{\rm SM}$ or with no
Higgs boson at all.
They found that the data are consistent with 
$\Lambda_{\rm SM}\lesssim3$~TeV.
(The precise definition of $\Lambda_{\rm SM}$ differs in the two
frameworks, so there is no conflict between the inequalities.)
Chivukula, H\"olbling, and Evans~\cite{Chivukula:2000px} considered the
renormalizable interactions of the standard model, plus interactions
that contribute to~$T$.
The results of their fits are shown in Fig.~\ref{3:fig:mHT}.
One sees that for very high scales, the usual blue-band fit is 
recovered.
If, however, $\Lambda_{\rm SM}$ is a few TeV,
the data allow the Higgs mass to be large.

There is a simple model that exploits fully the weaker bounds that
arise when treating the standard model as an effective field theory.
If one adds to the standard model fields a vector-like quark
with the $SU(3)_C\times SU(2)_W\times U(1)_Y$ quantum numbers
of the right-handed top quark and a mass of a few TeV, the Higgs
may have a mass significantly higher than the one inferred from
Fig.~\ref{3:fig:blue}~\cite{Collins:2000rz}, as high as 1~TeV.
This model is well motivated because it is an intermediate
effective theory of the top-quark seesaw model~\cite{Dobrescu:1998nm},
an explicit model of dynamically broken $SU(2)_W\times U(1)_Y$.
Another example is provided by models with extra dimensions, in which
the standard-model gauge bosons propagate in extra dimensions, 
while the fermions and Higgs boson are confined to four-dimensions.
Then the fit to the electroweak data allows a Higgs mass of up to
500~GeV~\cite{Rizzo:2000br}.
Finally, the radion, a particle that arises with warped extra 
dimensions, can cancel the Higgs's contribution to~$T$, again 
loosening the bounds~\cite{Csaki:2000zn}.

When treating the standard model as an effective theory the 
experimental bounds seem not much tighter than theoretical bounds,
based on triviality.
The triviality bound is an extension of the unitarity bound.
The latter anticipates either a physical Higgs resonance or a 
breakdown of tree-level unitarity in the $WW$ scattering amplitude 
below an energy of around 1~TeV~\cite{Lee:1977yc}.
Of course, in a real quantum field theory (and in nature) unitarity
does not break down.
The apparent breakdown at the tree level stems from a large Higgs
self-coupling, so one should handle
the full theory non-perturbatively.
As mentioned above, this possibility has been studied
extensively~\cite{Dashen:1983ts}.
This analysis finds that either the Higgs mass is less than
approximately 700~GeV, \emph{or} decay vertices or scattering
amplitudes exhibit deviations from the $\Lambda_{\rm SM}\to\infty$
limit at the percent level.
The nuance of the result leads to other, equivalent conclusions:
Fig.~\ref{3:fig:mHT}, for example, draws the bound on $m_H$ where
the deviations would be at the per-mil level.
On the other hand, the deviations could be of order one, but without
any new resonances, yet $m_H\approx1$~TeV.
The present fits of data to the standard model with
finite~$\Lambda_{\rm SM}$ yield similar numbers, but the fits are
stronger, because they close the loophole of ``deviations at some
level.''

The data-driven bound on the Higgs mass depends greatly on the 
scale $\Lambda_{\rm SM}$.
The only insights into the value of this scale are theoretical, and
the principal one is the fine-tuning problem.
In general there are radiative corrections to the parameters of the 
standard model from virtual processes between the weak scale~$v$
and~$\Lambda_{\rm SM}$.
In particular, the Higgs potential has a mass parameter $\mu^2<0$.
The parameter in the effective theory is a sum,
\begin{equation}
	\mu^2 = \mu_0^2 + c \Lambda_{\rm SM}^2 \sum_i (\pm)_i g_i^2 ,
	\label{3:eq:mu2}
\end{equation}
of the tree-level $\mu_0^2$ plus contributions from loop processes.
The constant $c$ depends on the underlying theory, the $g_i^2$ are
couplings of the Higgs to particle~$i$, and the sign is plus (minus) for
bosons (fermions).
If $g\Lambda_{\rm SM}$ is much larger than $\mu$,
there is an unnatural fine-tuning problem.
This is a serious issue, because some obvious choices for $\Lambda_{\rm
SM}$ are the scale of gravity ($M_{\rm Planck}$) or of
gauge-coupling unification ($\Lambda_{\rm GUT}$).
Consequently, it is believed that $\Lambda_{\rm SM}$ is smaller:
a few TeV, at most~\cite{Kolda:2000wi}.
The best ideas for physics between $\Lambda_{\rm SM}$ and
$\Lambda_{\rm GUT}$ (or $M_{\rm Planck}$) solve their own
fine-tuning problem by some other means.
For example, in supersymmetric models the supersymmetry requires the
terms in the sum to cancel, for energies above the susy scale.
With extra dimensions the unification scales need not be so high after
all, 10-100~TeV, so the hierarchy of scales presents no problem.
In any case, we note that the standard model's fine-tuning problem
is least severe when the scale~$\Lambda_{\rm SM}$ is relatively low,
but then the Higgs mass may be in
the intermediate-mass or heavy regions.

\subsection{Intermediate-mass Higgs Measurements}
\label{sec:inter}

In the intermediate Higgs mass range, defined in this report to be
the range from $2m_W$ to $2m_t$, the dominant decay is $\HtoWW$.
The branching fraction for this decay is a function of the weak
coupling constant $g$ and kinematic constraints for on-shell $W$
bosons.
Above $2m_Z$, the BR($\HtoZZ$) also becomes large, though smaller
than BR($\HtoWW$).
Therefore, the most significant tests one can make in most of this
mass range is whether the measured couplings of the Higgs boson to
weak gauge bosons is in agreement with the Standard Model prediction.
For the low end of this mass range ($\Mhiggs\lesssim 170$~GeV),
there is also a possibility of measuring the $b\bar{b}$ coupling.

Our approach is a simple one.  We assume that the detector design is
adequate to identify decays of the type $Z^0\rightarrow \epem$ and
$Z^0\rightarrow \mpmm$ with 80\% efficiency.  We then calculate the
number of predicted $\HtoWW$, $\HtoZZ$, $\Htobb$ decays.  For the
measurements described below, we make the following assumptions:
\begin{itemize}
\item 250~fb$^{-1}$ of delivered luminosity
\item $\sqrt{s} = 500$~GeV
\item Associated production of Higgs via the
process $\epem \rightarrow Z^0 H$, followed by $Z^0\rightarrow
\epem$ or $Z^0 \rightarrow \mpmm$
\item Identification of the Higgs
events through the missing mass technique
\end{itemize}
We have constrained the sample to associated production to give
a direct measurement of the branching ratios by measuring the
number of Higgs events and the number of decay events in the same
dataset, independent of luminosity measurements and cross section
calculations.  Extrapolations for other values of the luminosity
should be straightforward.

\subsubsection{Estimates of statistical uncertainties}

\begin{figure}[bp]
	\centering
 	\includegraphics[width=0.48\textwidth]{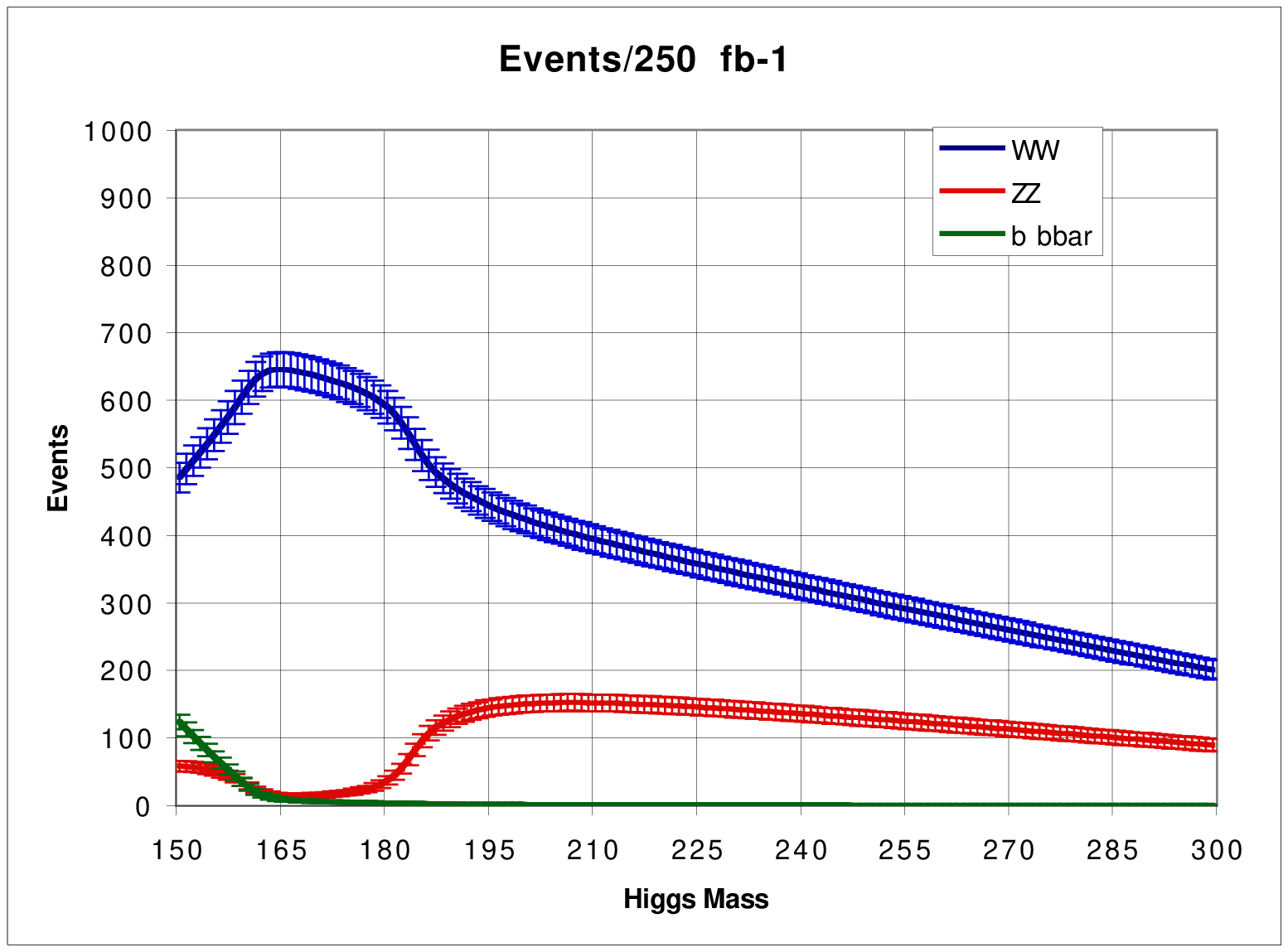}\hfill
 	\includegraphics[width=0.48\textwidth]{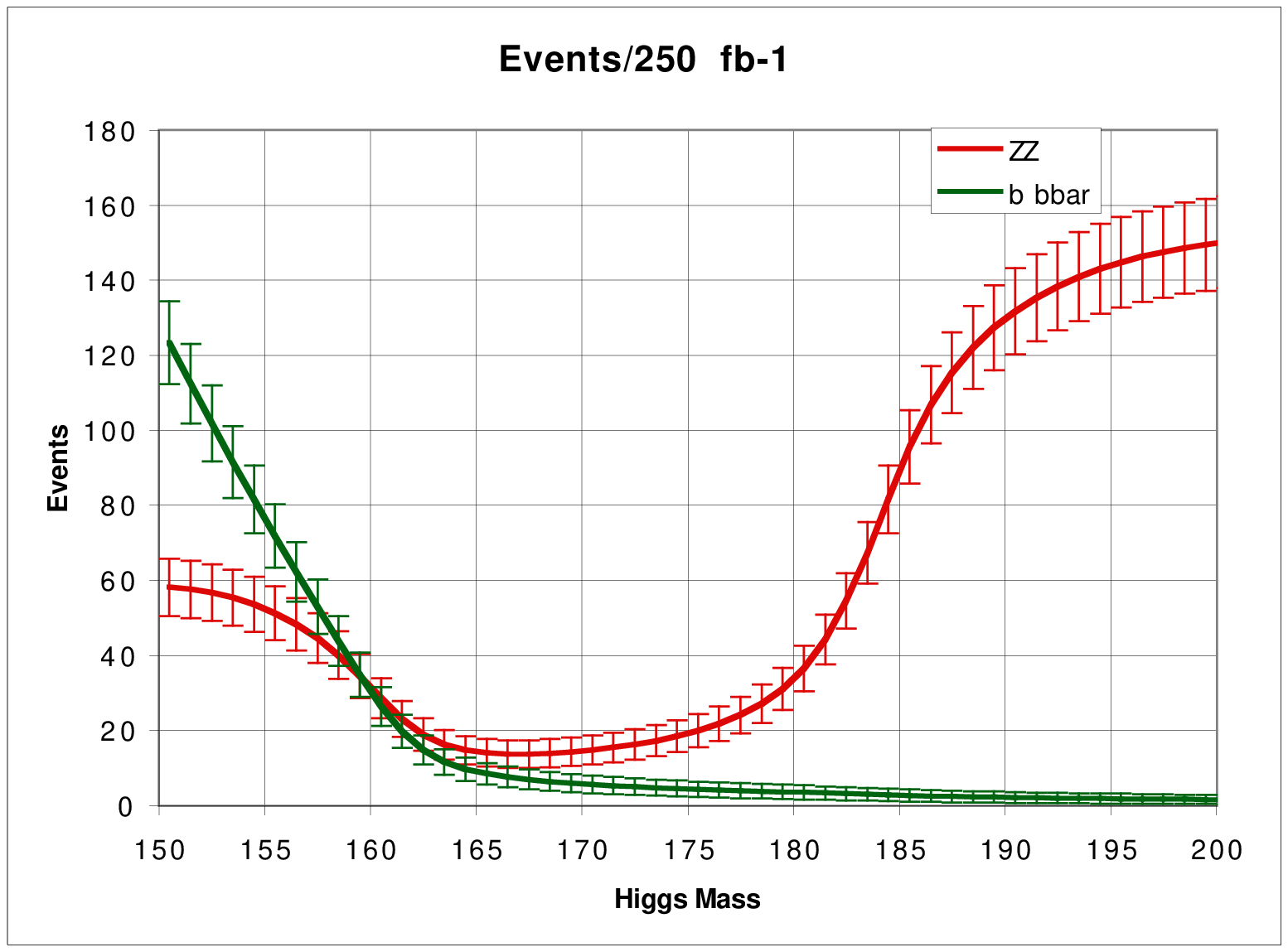}
	\caption[fig-nevents]{Number of events predicted for associated Higgs 
	production with a center of mass energy of 500~GeV, 250~fb$^{-1}$ 
	delivered luminosity, where the associated $Z^0$ is identified by 
	decays into $\epem$ or $\mpmm$.
	(a) Number of $\HtoWW$, $\HtoZZ$, and $\Htobb$.
	(b) Blow-up of $\HtoZZ$ and $\Htobb$ only.}
	\label{fig-nevents}
\end{figure}

In Fig.~\ref{fig-nevents}, we present the number of predicted
$\HtoWW$, $\HtoZZ$, and $\Htobb$ events in 250~fb$^{-1}$ at
$\sqrt{s}=500$~GeV.
The cross section calculation for associated
production comes from reference~\cite{Kilian:1996iy} and the branching
ratios from the HDECAY program~\cite{Djouadi:1998yw}.
We have not included any additional decay branching fractions or
identification efficiencies at this stage.

For reasonable expectations of $W$ identification efficiencies (50\%
or better)~\cite{Barate:2000gi}, we would expect to identify
$\gtrsim$ 100 events over the entire mass range, with
significantly more for $\Mhiggs$ in the region 150~GeV to
200~GeV.  As a result, the statistical uncertainty on the
measurement of BR($\HtoWW$) will be $\lesssim$ 10\%.

The $Z^0$ and $b$ measurements are more problematic.  In
Fig.~\ref{fig-nevents}(b), we focus on the predicted number of events for
$\HtoZZ$ and $\Htobb$.  Unless one is able to distinguish  hadronic
$Z^0$ decays  from hadronic $W$ decays, it will be difficult to have
adequate statistics to measure BR($\HtoZZ$) as the total number of
$Z^0Z^0$ events is $\lesssim$ 150 over the entire mass range.
If one could identify one of the two $Z^0$'s in the Higgs decays
(through leptons or $b\bar{b}$) 40\% of the time, at best
the statistical uncertainty of BR($\HtoZZ$) would be $\sim$ 11\%
for $\Mhiggs\sim$  210~GeV.  For $\Mhiggs < 180$~GeV,
the statistical uncertainties would be larger than 25\%.

With this approach, the measurement of BR($\Htobb$) with precision
better than 25\% will only be possible for $\Mhiggs\lesssim 160$~GeV.
For larger masses, the branching fraction into $b\bar{b}$ is too small.

\subsubsection{Different strategies for $Z\to b\bar{b}$}

\begin{figure}[bp]
	\centering
 	\includegraphics[height=0.42\textheight]{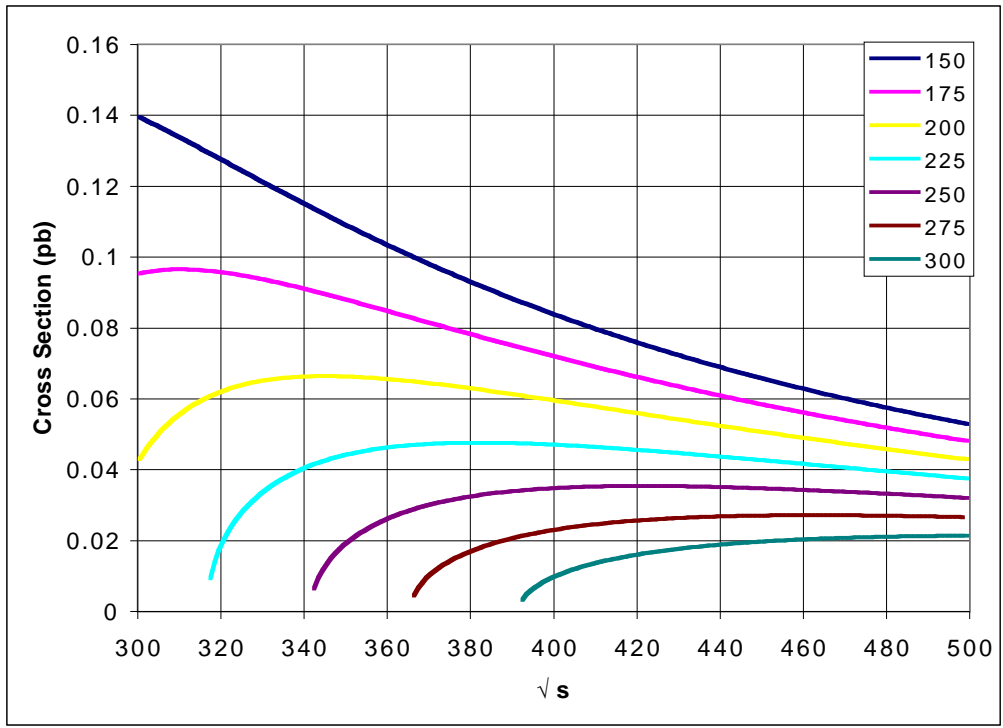}
	\caption[fig-sigma]{The cross section in pb$^{-1}$ for associated
	Higgs production vs.\ center of mass energy.  The seven curves are
	for Higgs masses from 150~GeV to 300~GeV.}
\label{fig-sigma}
\end{figure}

A center of mass energy of 500~GeV is not optimal for all Higgs
masses in the range 150~GeV to 300~GeV.
As can be seen in Fig.~\ref{fig-sigma}, the cross section for
associated Higgs
production depends upon both the Higgs mass and the center of
mass energy.
The peak value of the production cross section occurs at center of
mass energy near $\Mhiggs+m_Z+50$~GeV.
For the low end of the mass range ($\lesssim175$~GeV), the production
cross section is 2--2.5 times higher at $\sqrt{s}=300$~GeV than at
the nominal energy $\sqrt{s}=500$~GeV.

\begin{figure}[bp]
	\centering
	\includegraphics[width=0.47\textwidth]{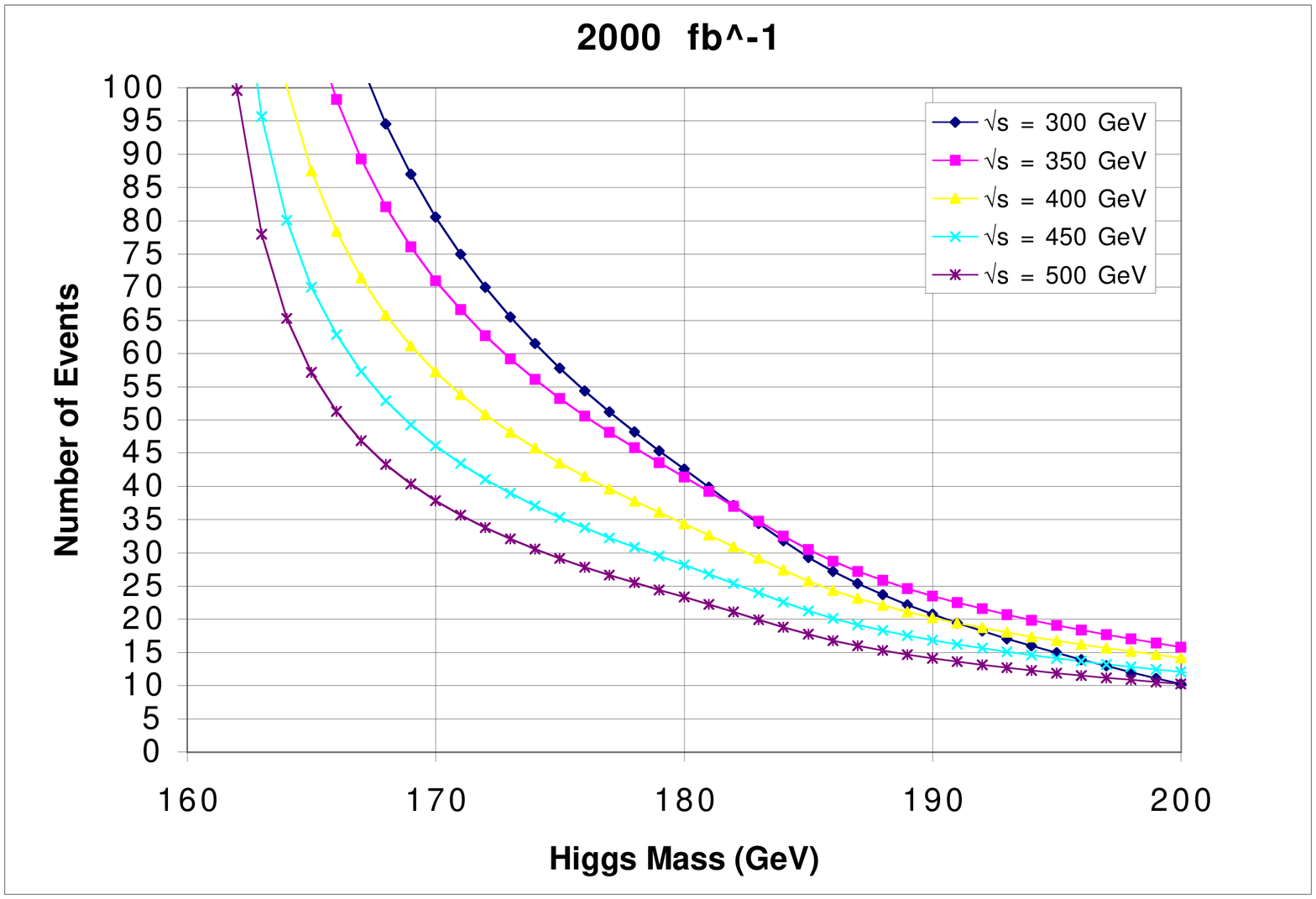}\hfill
	\includegraphics[width=0.47\textwidth]{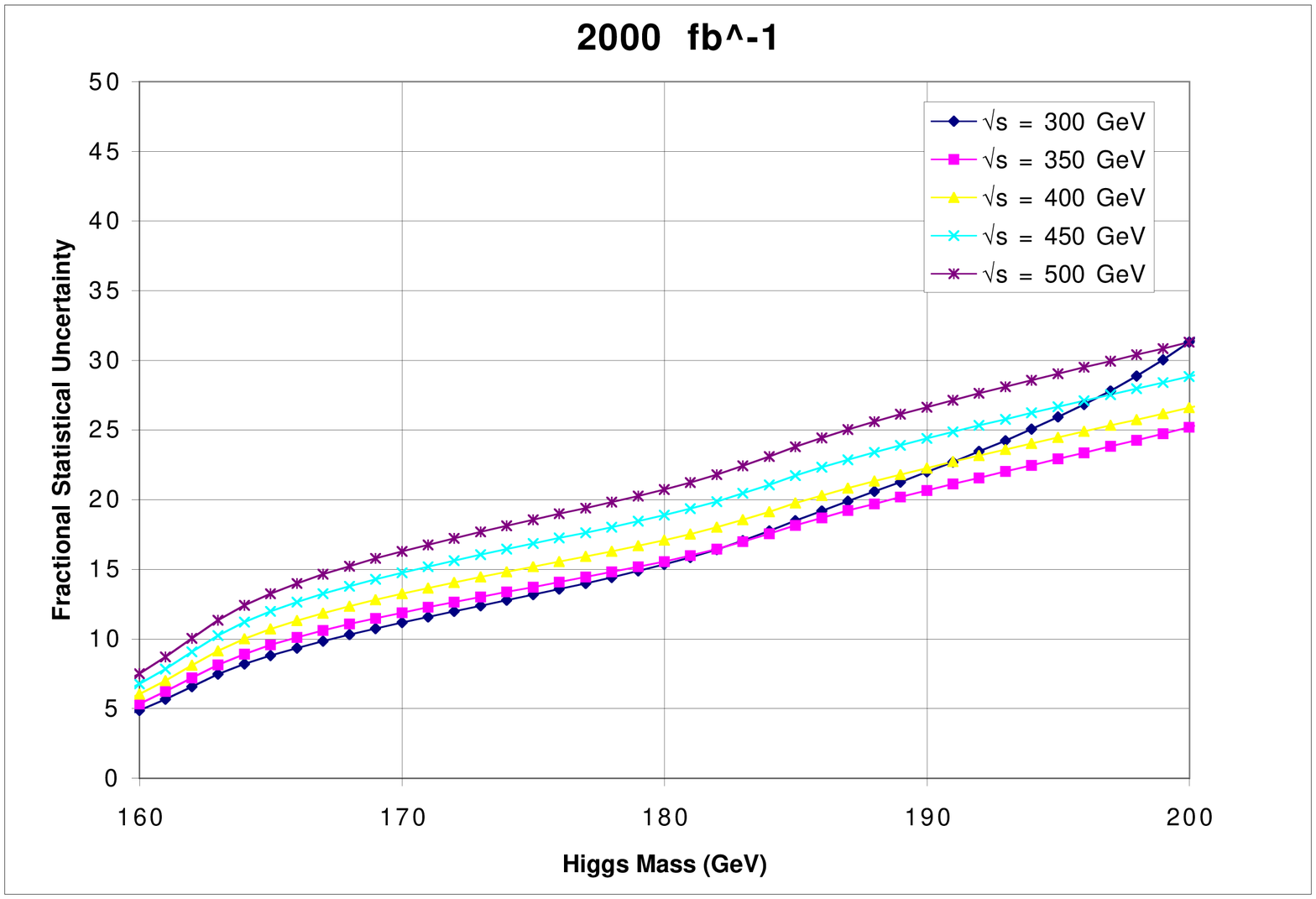}
	\caption[bb2000]{$ZH$ events in 2000~fb$^{-1}$ with 
	$Z\to l^+l^-$ ($l=e,\mu$) and $\Htobb$ vs.\ $m_H$.
	(a) Number of events. (b) Statistical error on BR($\Htobb$).
	The five curves are for $\sqrt{s}$ from 300~GeV to 500~GeV.}
\label{bb2000}
\end{figure}

As stated above, the measurement of $\Htobb$ is statistics limited for
the entire mass range.
A possible approach to increase the statistics in this sample is
to look for the $\nu\bar\nu$ decays of the associated $Z^0$, with the
experimental signature being two tagged $b$s with mass consistent
with the Higgs, along with significant missing energy from the two
neutrinos.
Since the BR($Z^0\to\nu\bar{\nu})$ is three times that of $\epem$
plus $\mpmm$, the gain extends the reach with 25\% statistical
uncertainty to $\Mhiggs \sim 165$~GeV.
Another possibility would be to consider hadronic decays of the~$Z^0$.
This would require more detailed study, because mass reconstruction for
both $Z$ and $H$ would have to be simulated.
Finally, one should acknowledge that a measurement of a rare decay would
come at the end of the Higgs phase of the LC program.
Figure~\ref{bb2000} shows the number of $ZH,\,Z\to l^+l^-,\,\Htobb$ 
events, and the resulting statistical error, in a long run of 
2000~fb$^{-1}$, as a function of $m_H$ for several different possible 
running energies.
Combining these strategies should, presumably, improve the prospects 
for this measurement.

\subsection{Heavy Higgs Measurements}

In this section we consider the contribution of an~LC if
the Higgs is heavy, $m_H>350$~GeV, and has standard-model
couplings.
As discussed above, such a heavy Higgs would require the existence of
a non-standard effect; however,
that effect could exist at a mass scale of several TeV, so that
the heavy Higgs could possess couplings close to those of the
standard model.
For example, Ref.~\cite{Collins:2000rz} found that the top-seesaw
model can satisfy the electroweak constraints with a Higgs boson in
this region, while the additional vector-like quark has mass above 1~TeV.

In this discussion, we assume that experiments at the LHC would
discover this heavy Higgs, since a signal greater than 5$\sigma$ is
claimed by both CMS~\cite{CMS} and ATLAS~\cite{ATLAS} for 30~fb$^{-1}$
for $350~{\rm GeV}<m_H\lesssim 1$~TeV.
We ask what measurements an~LC could contribute to better the
understanding of a heavy Higgs boson.

We consider the specific case for $m_H~=500$~GeV and the Higgs
has standard couplings.
Then the standard-model width is calculated to be 70~GeV, and the
branching ratios into the dominant decay modes are: 55\% to $W^+W^-$,
25\% to $Z^0 Z^0$, and 20\% to $t\bar{t}$.

At the LHC, the production cross section is 4~pb for
$m_H~=500$~GeV. The decay of the Higgs into pairs of $Z^0$'s
and the subsequent decay of the $Z^0$'s into either $e^+ e^-$ or $\mu^+ \mu^-$
gives a cross section times branching ratio into the four lepton final
state, ``$4\ell$'' (the {\it golden} mode), of 3.2~fb.
In 300~fb$^{-1}$, and assuming acceptance times efficiency to be 40\%, the
ATLAS TDR~\cite{ATLAS} states that 390 events in the golden mode can be used
to measure the mass to a relative error better than 0.3\%, the width to
6\%, and the product of production cross section times branching ratio to
12\%.
(The last assumes a 10\% uncertainty on the luminosity determination.)
The LHC should be able to make a precision measurement on the ratio
of branching ratios ${\rm BR}(H\to WW)/{\rm BR}(H\to ZZ)$.
Other measurements, such as the $CP$ nature of the Higgs,
are not discussed here.

Several measurements could potentially benefit from an~LC.
There is no obvious physics case that would require improving on LHC's
mass measurement.
The machine with the largest usable statistics should be able to 
measure the width more precisely.
At LHC, the measurement of the width comes from a direct fit of the 
data to the $4\ell$ lineshape, and its uncertainty is dominated by the 
statistical precision of the fit.
The LC has a better chance of using the hadronic and neutrino decays 
of the two $Z$ bosons.
The LHC appears unlikely to be able to make a precise measurement
of $WWH$ and $ZZH$ couplings, due to both uncertainties in the
normalization of luminosity and the lack of a sufficiently accurate
value of the production cross section.

At an~LC, the production of a very heavy Higgs requires both
high energy and high luminosity. We consider 
running at $\sqrt{s}=800$~GeV with an integrated luminosity of
2500~fb$^{-1}$.
At this energy, the production cross section
of $H\nu\bar\nu$ dominates at 10~fb, but the production cross section of
$HZ$ also contributes significantly at 6~fb. In all, approximately 40,000
Higgs would be produced. Even with such a glorious data set, the number of
expected $4\ell$ candidates is only 46 events assuming 100\% 
acceptance and efficiency and no backgrounds. We assume that in the case
of an~LC, the golden mode will not contribute to a
high precision measurement due to limited statistics.

The expected sample that would be produced for Higgs decaying into
$W^+W^-$ where either one or both $W$'s decay leptonically is
7,700~events.
For Higgs decaying into $Z^0 Z^0$, with at least one $Z^0$ decaying into
a lepton pair ($e$ or $\mu$) and the other $Z^0$ decaying either into
neutrinos or $b\bar{b}$, one expects 480 events. Other decay modes of the
vector bosons appear to either have large backgrounds or have great
difficulties in their reconstruction ($c$-tagging could help, for instance).
Unlike the LHC, it is possible that the~LC could observe and
measure the branching ratio of Higgs into $t\bar{t}$. Final state modes 
that involve at least one lepton and two $b$-jets would provide 
a sample of 2750 events. Assuming that 10\% of this sample is usable (within
the acceptance, found efficiently over background, correctly $b$-tagged),
the relative error on the branching ratio could be 6\%. For a very heavy
standard model-like Higgs, assuming that a sufficiently large data sample 
could be obtained, the~LC could be the best place to measure
the coupling of the Higgs to fermion pairs and, thus, explore the
connection between fermion masses and electroweak symmetry breaking.

The last point is important, because a heavy Higgs with nearly
standard couplings is almost certainly part of a larger sector,
with other particles at higher mass.
Then, while planning or awaiting an upgrade of the LC energy, it would
be attractive to explore the underlying physics by running the~LC in a
``Giga$Z$'' configuration~\cite{Erler:2000jg}.
By going beyond the current precision for the $Z$ lineshape and $W$
and top masses, one can pinpoint the allowable region in the $S$-$T$
plane and, thus, constrain theoretical models while gaining insight
into the appropriate energy scale for new physics.


\section{Extensions of the Standard Model}

The fundamental particles observed so far are the gauge bosons
of the $SU(3)_C \times SU(2)_W \times U(1)_Y$ symmetry group, the
longitudinal degrees of freedom of the $W^\pm$ and $Z^0$, and three
generations of quarks and leptons.
Remarkably enough, there is a strong theoretical argument for the
existence of further physical phenomena~\cite{Lee:1977yc}:
if the particles observed so far were the only existing ones, then 
the $WW$ scattering cross section would violate perturbative unitarity 
at a scale of order 1~TeV.
Therefore, either the $W$ and $Z$ must have strongly coupled
self-interactions at the TeV scale, or new fundamental
degrees of freedom, beyond those observed so far, must exist.
The scale of the new phenomena is within the reach of future collider
experiments, and searching for them is a main
goal of high-energy physics.
In this section we discuss the known alternatives for these new
phenomena, and their implications for experiments at an~LC.

As discussed in Sec.~\ref{sec:higgs}, models with a Higgs boson,
whether fundamental or composite, give a good description of all
available data.
Nevertheless, one should ask if there are phenomenologically viable
models without a Higgs boson.
At present it is not possible to reject such alternatives completely,
because, as mentioned above, with no Higgs boson the self-interactions
of the $W$ and $Z$ must become strong at the TeV scale.
In that case one cannot use a loop expansion to compute the physical
quantities relevant for comparing with the electroweak data.
With our currently limited understanding of strongly-coupled field
theories, it is hard to decide whether a theory without a
Higgs boson can be viable.

Technicolor is a well-known class of theories of electroweak symmetry
breaking without a Higgs boson~\cite{technicolor}.
Its defining feature is that the longitudinal degrees of freedom of
the $W$ and $Z$ are composite states of some new fermions (called
techni-fermions) bound by an asymptotically-free gauge interaction
that confines below the TeV scale.
If this interaction is similar to QCD, then one may use
experimental data for hadrons to derive predictions.
It turns out that the electroweak data impose a strict upper bound
on the number of techni-fermions~\cite{Peskin:1990zt,Holdom:1990tc}.
Even for the minimal number of techni-fermions, it would be necessary
that some additional interactions cancel in part the contributions of
the techni-fermions to~$S$.
Therefore, QCD-like technicolor with many techni-fermions is an
example of a theory without a Higgs boson that is ruled out.
On the other hand, if the technicolor interaction is significantly
different than QCD ({\it e.g.}, if the scale-dependence of the
technicolor coupling is very mild, a paradigm called ``walking
technicolor''), then one does not know how to test predictions
of the theory against the data.
Assuming that such a theory turns out to be viable and, indeed,
realized in nature, its most important experimental test would be $WW$
scattering at energies of order 1~TeV.
Such an experiment would reveal the strongly-coupled gauge boson
interactions, potentially leading to the production of vector
resonances or other bound states.
The capability of an~LC in this channel has been studied for several
distinct models in the report for Snowmass~'96~\cite{Barklow:1997nf}.
Some technicolor models also allow the existence of light pseudo-scalar
particles, which would be interesting to study at an~LC.

Even though it may be conceivable that the Higgs boson does not exist, 
the impressive fit of the standard model to the experimental data may
be taken as an indication for the existence of a Higgs boson.
In the rest of this section we will assume that a Higgs boson indeed
exists, the effective theory below a scale of a TeV being the standard
model with the possible addition of other light states.

It is important to emphasize that even if the standard model proves
to be the correct description of nature at energies up to a TeV
or so, the standard model is at best a low-energy effective theory
valid only up to some higher energy scale $\Lambda_{\rm SM}$.
This statement is supported by a variety of arguments.
In addition to the breakdown of scalar field theories, discussed
in Sec.~\ref{sec:bounds}, there are two other robust arguments.
The first is that the ${\rm U}(1)_Y$ gauge coupling also increases with 
energy, so this sector presumably also breaks down.
The other is that the standard model does not include gravity. 
These problems could be resolved if $\Lambda_{\rm SM}$ is as large
as the Planck scale ($M_{\rm Planck} \approx 2 \times 10^{19}$~GeV),
but it appears more likely that $\Lambda_{\rm SM}$ is in the TeV
range or below.
The reasoning is phrased in the literature in various forms, and it
is often called the hierarchy problem. In its simplest form it
boils down to the question of why the weak interactions
are so much stronger than the gravitational interactions or,
equivalently, why the electroweak scale $v = 246$~GeV is so much
smaller than the Planck scale.
A theory which includes both the standard model and gravitational
interactions should address this problem, together with the associated
fine-tuning problem, Eq.~(\ref{3:eq:mu2}).
Broadly speaking, there are three classes of theories which
attempt to solve the hierarchy problem:
\begin{itemize}
  \item supersymmetric extensions of the standard model;
  \item theories of Higgs compositeness;
  \item theories with extra dimensions.
\end{itemize}
In each one of these classes of models there are new particles
and interactions at a scale in the TeV range or below.
In what follows we discuss these theories in turn.

\subsection{Supersymmetric Extensions of the Standard Model}
\label{sec:susy}

Supersymmetry (susy) is a space-time symmetry connecting fermions and
bosons.
In a certain well-defined sense, it is the largest space-time symmetry
consistent with a unitary $S$-matrix~\cite{Haag:1975qh}.
It also arises in string theory, which is the leading candidate to unify
gauge and gravitational forces.
For these reasons, it offers an attractive theoretical framework for
particle physics, whether or not it is directly relevant at the TeV
scale.

Susy also has several features that make it the most popular extension of
the standard model of electroweak symmetry breaking:
\begin{itemize}
\item Susy solves the fine-tuning aspect of hierarchy problem.
The symmetry between particles of different spin (superpartners)
leads to exact relations between diagrams containing loops of
the respective superpartners.
As a consequence, the quadratic radiative corrections to~$\mu^2$ in 
Eq.~(\ref{3:eq:mu2}) cancel.
\item The minimal supersymmetric extension of the standard model
(MSSM) offers tantalizing evidence that the gauge couplings of
$SU(3)_C\times SU(2)_W\times U(1)_Y$ unify
at a very high energy scale $\Lambda_{\rm GUT}$,
on the order of $2\times10^{16}$~GeV~\cite{Dimopoulos:1981yj}.
(This is often summarized as a prediction of $\sin^2\theta_W$.)

The possibility that a simpler structure lurks
behind the known gauge symmetries is extremely intriguing,
with important implications, for example in proton decay
experiments and in neutrino physics.
Perhaps more importantly for collider phenomenology, gauge
coupling unification in the MSSM works only if the spectrum of the
new superpartners is within an order of magnitude of the TeV
scale~\cite{Bagger:1995bw}.
\item In most phenomenologically viable models of low-energy supersymmetry
one introduces a new conserved quantum number, $R$-parity, whose sole
purpose is to forbid too rapid proton decay mediated by superpartners.
In $R$-parity conserving theories the lightest superpartner (LSP) is 
stable and may be a suitable dark matter candidate. Its relic abundance
varies greatly over parameter space, but
nevertheless there are significant regions where the LSP relic
density is of just the right size~\cite{Feng:2000gh,Ellis:2000we}.
\end{itemize}

These arguments only represent circumstantial evidence for
supersymmetry.
On the other hand, the simplest susy extensions have
some troublesome features that one should bear in mind:
\begin{itemize}
\item The $\mu$ problem.
Although susy protects the weak scale from large radiative corrections,
there is still a potentially dangerous {\em tree-level} contribution to 
Eq.~(\ref{3:eq:mu2}), which is allowed by all symmetries, including
supersymmetry. The term $\mu_0^2$
on the right-hand side of Eq.~(\ref{3:eq:mu2}) now consists
of two contributions---a susy-breaking piece whose natural size is
of the order of a TeV, and a susy-preserving piece, which is expected
to be on the order of the fundamental new physics scale, \emph{e.g.}, the
Planck scale. Hence, an uncomfortably large hierarchy remains.
Various solutions to the $\mu$ problem have been claimed in the
literature~\cite{Polonsky:1999qd}, but none seems to be very
compelling.
\item The susy flavor problem. 
Generic models suffer from severe problems associated with
large flavor-changing effects. Without a dedicated suppression 
mechanism, the superpartners of the first two generations' quarks and
leptons must have masses as large as
100~TeV or higher~\cite{Gabbiani:1996hi}.
\item The proton decay problem. 
With the most general Yukawa-type couplings between quarks,
leptons and their superpartners,
the proton would decay too rapidly through susy-mediated processes.
The simplest and most elegant solution to this problem is the
imposition of $R$-parity (see above).
\end{itemize}

There is also a pragmatic reason to consider susy in detail from
an experimental point of view.
It offers a predictive, renormalizable and perturbatively
calculable model with a wide variety of possible phenomenology.
Study of the numerous possible signatures in susy models leads to a
detector design that is well-suited to the observation of new physics,
no matter what its origin.
In the same sense, susy models are useful for comparing the physics
capabilities of various future colliders, as they illustrate generic,
yet disparate scenarios for new physics.
More details can be found in the extensive literature---see,
\emph{e.g.}, the whitepaper~\cite{Bagger:2000iu} or well-known
reviews~\cite{Murayama:1996ec}, and references therein.

\subsubsection{Classification of susy models}

A truly supersymmetric theory of particle physics would have only a
few more parameters in addition to the standard ones.
The interactions with gauge bosons and their superpartners
(gauginos) are fixed by gauge symmetry and supersymmetry.
Likewise, the couplings to Higgs bosons and their superpartners
(higgsinos) are fixed by the observed fermion mass spectrum
and supersymmetry.
Of the few remaining parameters allowed by susy, some may be eliminated
by other considerations, \emph{e.g.}, by imposing $R$-parity.

If exact, supersymmetry guarantees the degeneracy of the standard
particles and their superpartners.
Since this is not observed in nature, supersymmetry must be broken,
and a model of TeV-scale physics must also have susy-breaking parameters.
Unfortunately, there are too many parameters in general, and a completely
systematic study is impractical.
Hence, to make progress, one must make additional assumptions.
One is on the field content.
The minimal field content which includes the standard model fields is a
model with two Higgs doublets, called the minimal supersymmetric
standard model (MSSM).
Consistent with supersymmetry and its theoretical motivations, it is,
however, possible to have more fields, leading to non-minimal
extensions (NMSSMs).

The number of free parameters can be reduced by assuming a mechanism
for susy breaking.
Typically this is achieved by postulating two sectors, a hidden sector,
in which dynamical susy breaking takes place, and the sector in which
we live.
One then requires some mechanism to mediate the breaking of susy from
the hidden sector to ours.
The different options are known as
\begin{itemize}
\item gravity mediation, or supergravity (SUGRA)~\cite{Chamseddine:1982jx};
\item gauge mediation, via standard~\cite{Dine:1993yw} and/or
	non-standard~\cite{Mohapatra:1997kv} gauge interactions;
\item anomaly mediation~\cite{Randall:1999uk};
\item gaugino mediation~\cite{Kaplan:1999ac};
\item Scherk-Schwarz susy breaking~\cite{Scherk:1979zr}.
\end{itemize}
Rather than review the detailed features of the spectrum in each
of these classes of models, we shall mention a few of the 
important distinctions below, when comparing
the LHC and the LC physics reach and capabilities.

\subsubsection{How does susy justify a linear collider?}

In this subsection we present some arguments that make the case for
an LC, in case low energy susy is connected to electroweak
symmetry breaking. For this purpose, we outline the
physics program needed to verify that susy has indeed been observed,
and the specific ways in which an LC would help elucidate not
only electroweak symmetry breaking but also supersymmetry breaking.
We also contrast with the information gleaned from the LHC experiments. 

A discovery of susy would mark the beginning of a golden era for 
particle physics. All known particles should have
superpartners, with interactions and properties of their own.
To confirm the picture experimentally one must~\cite{Feng:1995zd}
\begin{enumerate}
\item prove that the new particles have the correct
quantum numbers to be superpartners;
\item show that the couplings to gauge bosons and gauginos
are equal (up to radiative corrections~\cite{Hikasa:1996bw}),
as predicted by supersymmetry;
\item show also that the couplings to Higgs bosons and higgsinos
are related;
\item correctly identify the flavors of the squark and slepton resonances;
\item measure precisely the superpartners' mass spectrum;
\item measure mixing angles of squarks, sleptons, charginos, and
neutralinos;
\item combine the masses and mixing angles to extract the fundamental
parameters of the Lagrangian.
\end{enumerate}
If the superpartners in question are within the kinematic reach of
an LC, then, on each of these seven counts, it can
provide the most incisive information.
In principle, an LC with high enough energy can produce the
whole susy spectrum and can, thus, fully reconstruct the susy model.


The conventional wisdom is that the LHC is a discovery machine,
while linear colliders are for precision studies.
What this usually means is that the LHC can discover {\em more}
particles, while the LC can measure their properties
(masses, couplings, widths, quantum numbers etc.)~{\em better}.
Because of its higher parton center-of-mass energy, the LHC can
certainly {\em produce} more species of particles than a LC in the
1~TeV range.
But it is not at all certain that all of them can be {\em observed}
at the LHC, because of the larger backgrounds.
For example, observation of first and especially second generation
squarks appears to be rather challenging at the LHC.
On the one hand, one has to extract the squark signal under a much
larger gluino signal.
On the other hand, having extracted a squark signal, it appears rather
difficult to show how many squark flavors are present.
These questions have not been studied in great detail to date and
deserve attention.

It is important to keep in mind that even if the LHC is able to
observe more superpartners than the LC, the two sets do not necessarily
coincide.
Generically, the LHC does better for strongly interacting superpartners,
while the LC does better for sleptons, charginos and neutralinos.
In this way, the two machines complement each other.

It is often said that one should abandon susy as a key to electroweak
physics, if the LHC does not observe it.%
\footnote{The notion is based on the most widely studied models:
SUGRA and minimal gauge mediation.
Then, squarks and gluinos are among the heaviest superpartners.
They decay to gauginos and hard jets, and subsequent decays of
the gauginos may produce hard leptons.
The combination of hard jets, leptons, \emph{and} missing energy is
such a distinctive signature that it is practically impossible to miss
at the LHC.}
But there are realizations of supersymmetry for which such a
conclusion is unwarranted.
It is easy to construct models, where discovering susy at the LHC is
not at all straightforward.
We give two examples for illustration.
First, consider the following hierarchy of masses
\begin{equation}
	m_{\tilde g}\gg m_{\tilde q} \sim m_{\tilde \ell} > m_{\tilde h}
\end{equation}
where $m_{\tilde g}$, $m_{\tilde q}$, $m_{\tilde \ell}$ and
$m_{\tilde h}$ are the masses of the gauginos, squarks, sleptons and
higgsinos, respectively.
Such a hierarchy can be typical of models based on Scherk-Schwarz susy
breaking, with gauge fields in the bulk and chiral fields on the
wall~\cite{Antoniadis:1999sd,HCKM}.
If, furthermore, the gauginos are all heavy enough to be beyond the 
reach of the LHC, while all scalars are nearly degenerate, then the 
only observable signatures of susy come from slepton and squark pair 
production with subsequent two-body decays to higgsinos. 
The resulting dilepton plus missing energy signature has been 
analyzed~\cite{CMS} and the reach extends up to 
$m_{\tilde\ell}\sim 350$~GeV at most.
A dijet signature, even if observable, is far from diagnostic of 
supersymmetry, and its appearance would elicit a number of alternative 
explanations.
Second, there are also well-motivated susy models in which the dominant
discovery signatures have many $\tau$ leptons.
At the~LC calorimetry can be used to detect hadronic decays of
the~$\tau$, whereas at at hadron machines the underlying event would
tend to obscure them.
Thus, in either of these scenarios the key discoveries could be left
for an~LC.

\subsubsection{Contrast with LHC}

Let us review some of the expectations for susy at the LHC.%
\footnote{We are grateful to Frank Paige for comments.}
As long as the colored superpartners (gluino and squarks) are lighter
than about $1$~TeV, strong production dominates and is not small
compared to the QCD background. For the scenarios
that have been under investigation, all with $M_{\rm susy}\le 1$~TeV,
the signal can be separated from the standard model backgrounds with
simple cuts, yielding evidence for new particles
and a rough estimate of the mass scale~\cite{Tovey:2000wk}.
Next, one would examine the susy sample for special features.
Depending on the particular model, one may be able to isolate specific
decay chains and then use kinematic end points to determine combinations
of masses, for example
$\tilde \chi_2^0\to\tilde \ell\to \tilde\chi^0_1\to\tilde G$
in gauge-mediated models~\cite{Hinchliffe:1999ys}, or
$\tilde q\to \tilde\chi^0_2\to\tilde\ell\to\tilde\chi^0_1$
in gravity mediated~\cite{Hinchliffe:1997iu},
anomaly-mediated~\cite{Paige:1999ui}
or string-inspired~\cite{Allanach:2000kt} models.

In summary, partial reconstruction of susy decay chains has been
demonstrated for several sample cases, and this technique should
extend to more generic cases.
Moreover, experience from the Tevatron suggests that one should
expect the LHC collaborations to achieve more with real data than
in simulation.
Nevertheless, a feature of the decay-chain method is that it yields
mass differences, relative to the LSP mass, but the LSP mass itself
is not well measured.

If an~LC can produce at least one visible particle, it can be
detected, and its decay can be used to measure the LSP mass.
Then one could combine the LSP mass from the LC with the mass
differences from the LHC to determine the spectrum with a precision
that LHC alone would not have, and a kinematic reach that an LC (with
$\sqrt{s}$ not larger than 1~TeV) would not have.
One should note that in reconstruction of electroweak superpartners 
at an LC one can measure their masses precisely~\cite{Martyn:1999tc}.
Even for squark mass measurements, an LC has an advantage 
because of the fixed center of mass energy, and squark masses can be 
measured with a precision of up to a few GeV~\cite{Feng:1994sd}.

It is easy to identify other advantages of linear colliders.
First, consider the question of measuring the putative superpartners' 
couplings.
These can appear in two places: in production of superpartners, and in 
their subsequent decays.
However, any particle with a mass of a few hundred~GeV, and
charged under any of the standard gauge interactions,
decays promptly at the collision point. It is, thus, 
impossible to extract its couplings from a lifetime measurement.
One can also consider ratios of branching ratios, but then at least
one of the couplings has to be known by other means.
At a hadron collider it may be difficult to disentangle all the
different superpartner decays, since they are all simultaneously
present.
Therefore, the cleaner and more straightforward way to measure the
couplings is through the production mechanism, \emph{i.e.}, from a
cross-section measurement of direct sparticle production.

Even then, there are several complications. First, one
has to extract events where the superpartners are being directly
pair-produced, rather than appearing at the end of other 
particles' decay chains. At the LHC, this typically reduces the
discriminating power against standard-model backgrounds, and leads to
smaller signal rates. For example, almost any susy model has
a region of parameter space where the sleptons are lighter 
than the squarks or gluinos, and can be produced in cascade
decays of those (colored) superpartners. Unfortunately, because of the
relatively small slepton production cross section,
the LHC can detect {\em direct} production of sleptons only up to
$\sim 350$~GeV. The slepton reach gets significantly higher if
indirect production from squark and gluino decays is also included,
but this is of course not helpful for a slepton cross-section
measurement.

There are several kinds of measurements that are much better suited to
lepton than hadron machines.
One arises when there is significant mixing
between the scalar partners of the left- and right-handed
fermions. In that case, to extract the coupling to
the weak eigenstate, one must measure the mixing angle as well.
Here, polarized beams give a unique advantage to lepton colliders.
In many processes there is more than one intermediate state, 
and sometimes some of the
virtual particles (usually $t$-channel) are either unknown or poorly
measured. Beam polarization helps again, as one
can enhance or suppress certain subprocesses.
Quantum numbers of putative superpartners are most easily determined 
from the threshold behavior of sparticle pair production or angular 
distributions of decays, as discussed for the Higgs boson in 
Sec.~\ref{sec:spin}.
The well-defined energy and clean environment of an LC provide
a clear edge over the LHC.

Another case where an LC is superior in disentangling the new physics,
is when there are several superpartners that are roughly degenerate
in mass.
Some examples can be readily found in the existing susy models:
\begin{itemize}
\item The two heavy neutral Higgs boson states of the MSSM,
the $CP$-even $H^0$ and the $CP$-odd $A^0$,
can have a mass splitting below
the experimental resolution of the LHC detectors.
\item The mass splittings of higgsinos
are typically on the order of a few GeV, because they
are induced only by mixing
effects in the chargino and neutralino mass matrices.
Therefore, the decay products in
higgsino transitions are extremely soft and unobservable at the LHC.
But by carefully scanning through higgsino thresholds, an
LC can measure the masses precisely.
\item Sleptons and squarks from the first two generations must
be nearly degenerate, to satisfy flavor-changing constraints.
\end{itemize}

A final piece of physics which can only be done at an LC is to elucidate
the nature of a stable, neutral LSP. LSP pair production by itself
does not give an observable signature, so one must either 
tag with an initial state photon, or consider associate production 
of the LSP and next-to-lightest neutralino. For an analysis as 
subtle as this one, the clean environment of an LC is crucial.

We should point out that experimental tests of the relations between
the Higgs and higgsino couplings are extremely difficult at both the
LHC and an LC.
First, these tests can only be carried out for third-generation
sfermions.
Second, those couplings rarely appear in production processes, and their
extraction from ratios of branching ratios appears to be difficult.
Furthermore, the necessary decay channels must be open, which need not
always be the case.

Let us comment on the optimum energy and luminosity.
For the superpartners, energy is clearly a premium, because with
higher energy a greater fraction of the spectrum is likely to
be accessible.
Until a superpartner is observed in an experiment, one can only
guess the appropriate starting energy.
In the next subsection, we shall discuss information from a variety
of running or approved future experiments, both in particle physics
and astrophysics, that may soon see some new signal.
Nevertheless, it is likely that the full elucidation of susy will
eventually require lepton colliders with $\sqrt{s}$ of a few~TeV.

Once even a few superpartners are within its kinematic reach, an LC
will provide physics information which is inaccessible at the LHC.
High luminosity is then needed for better precision in measuring the
fundamental parameters of the susy Lagrangian.
In particular, with LC measurements one can start to unravel the mechanism
of susy breaking.
To do so, one needs the most important masses as precisely as possible.
An important feature of an LC is its ability
to perform threshold scans in energy.
This leads to very precise determinations of the
masses~\cite{Martyn:1999tc}, and it becomes
possible, without model assumptions, to distinguish among the susy 
breaking mechanisms listed above~\cite{Blair:2000gy}.

\subsubsection{A glimpse into the future}

If susy exists, how long do we have to wait to see the
first hints of new physics?
This question is especially important for LC planning, since the 
answer depends on how heavy the susy spectrum is and, therefore, has 
bearing on the required collider energy.
With no experimental guidance, one has to rely on theoretical prejudice.
The traditional avenue is to place upper limits on the superpartner
masses based on naturalness arguments~\cite{Barbieri:1988fn}.

A more predictive, and somewhat better motivated, estimate can be made 
if one assumes that the LSP explains the missing non-baryonic 
dark matter in the universe.
The LSP is then neutral and, hence, a mixture of an electroweak gaugino 
and a higgsino.
Until recently, it was thought that
only a gaugino-like LSP could be a good dark matter candidate.
In that case one can derive model-dependent upper
limits on the superpartner spectrum~\cite{Ellis:2000yp},
with obvious implications for an LC.
Recent work~\cite{Feng:2000gh,Feng:2000zu} shows, however, that 
throughout a large portion of the (previously unexplored) parameter 
space, a mixed gaugino-higgsino neutralino state is also a perfectly 
good dark matter candidate.
The possibility of a significant higgsino fraction has far-reaching
consequences for
current and near-future astrophysical searches, which look for traces
of dark matter annihilation at the center of the Earth, Sun or our galaxy.
As it turns out, {\em before the LHC commences}, the combination
of all collider searches and both direct and indirect dark matter
detection probes will map out all of the cosmologically preferred
parameter space that is accessible to an~LC with $\sqrt{s}=0.5$~TeV.

\begin{figure}[t]
	\centering
	\includegraphics[height=3.4in]{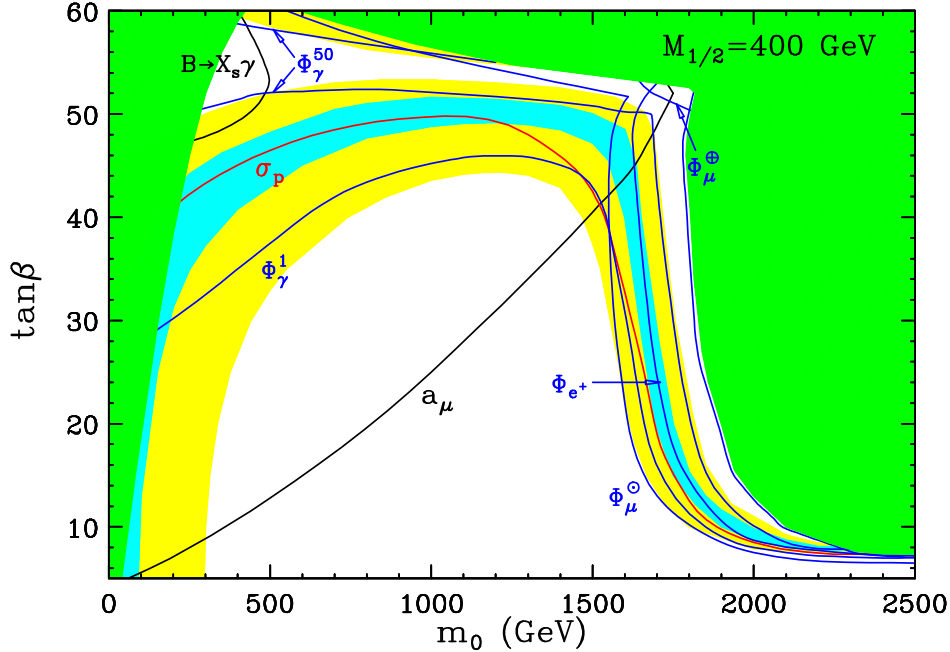}
	\caption{Estimated reach of various high-energy
		collider and low-energy precision searches (black),
		direct dark matter searches (red), 
		and indirect dark matter searches (blue)
		before the LHC begins operation, for 
		fixed $M_{1/2}=400$~GeV, $A_0 = 0$, and $\mu > 0$.
		The signals considered are listed in the text.
		The projected sensitivities used for each experiment
		are given in Ref.~\cite{Feng:2000zu}. 
		The regions probed extend the
		curves toward the forbidden, green region.
		The regions with preferred (acceptable) LSP relic
		density are shaded in light blue (yellow).}
	\label{LCreach}
\end{figure}

Figure~\ref{LCreach} shows how this works in the case of minimal 
supergravity~\cite{Feng:2000zu}, although the conclusion remains valid 
in other susy settings as well~\cite{FM}.
The cosmologically preferred
(allowed) regions are shown in light blue (yellow).%
\footnote{On grey-scale devices, yellow in Fig.~\ref{LCreach} is
lighter than blue, and both are lighter than green.}
The green regions are excluded either by LEP or for theoretical
reasons. The curves show the sensitivity limits of various experiments
(the regions probed extend towards the excluded region). The signals
considered are upward going muon flux from neutralino annihilation
at the center of the Earth ($\Phi_{\mu}^{\oplus}$) and the Sun
($\Phi_{\mu}^{\odot}$); the continuum photon or positron flux
from neutralino annihilation in our galaxy ($\Phi_\gamma$ and 
$\Phi_{e^+}$, respectively); direct dark matter detection in tabletop
experiments ($\sigma_P$); and constraints from $B\to X_s \gamma$ and 
muon $g-2$.
For illustration the gaugino mass parameter is chosen so that the
chargino mass is roughly 250~GeV, barely within reach of a 500~GeV LC.
But for lower gaugino masses, the sensitivity of the shown experiments 
is even better.
We see that the combination of particle physics and astrophysical
searches covers all the cosmologically interesting parameter space,
and at least one is bound to observe a signal {\em before} the LHC
begins operation.%
\footnote{While this report was being written, the Muon $g-2$
Collaboration announced a measurement of the muon anomalous
magnetic moment~\cite{Brown:2001mg},
which deviates from the standard-model prediction by~$2.6\sigma$.
While this is not a proof of the existence of new physics,
the magnitude of the effect is generic in the simplest supersymmetric
models~\cite{Feng:2001tr}.
If the deviation holds up and is due to supersymmetry, it suggests
either light sleptons and charginos, or large $\tan\beta$ and, hence,
final states with $\tau$ leptons.}
Conversely, if there is no hint of supersymmetry before that time,
no superpartners will be within reach of a 0.5~TeV lepton collider
(or dark matter would require another explanation).

\subsubsection{Higgs properties in susy models}
\label{sec:susyHiggs}

Since supersymmetry has a decoupling limit, most susy models have a
light Higgs boson with nearly standard properties.
As discussed in Sec.~\ref{sec:lightHiggs}, precise measurements
of this Higgs boson's couplings could reveal its susy
nature~\cite{Battaglia:1999re,Carena:2001bg}.
In addition to this state, susy models have other, usually heavier
Higgs boson states.
By studying them one can measure the ratios of their vacuum expectation
values, for example $\tan\beta$ in models with two
doublets~\cite{Feng:1997xv}.
More details on heavy Higgs bosons can be found in the existing
literature~\cite{Djouadi:1995mr}.

A new scenario~\cite{Dobrescu:2000yn},
which was investigated during the Study, may arise in the simplest
non-minimal extension, or NMSSM.
The novel feature is the presence in the spectrum of light, mostly
singlet, $CP$-odd scalars, which are called axions~$A^0$.
Their origin is easily understood as pseudo-Nambu-Goldstone bosons of
an approximate, spontaneously broken global U(1) symmetry.
The axions are, therefore, naturally light, because their mass is generated
only by terms which explicitly break the symmetry.
One then typically
finds that the decay $h^0\to A^0A^0$, of the
standard-model-like Higgs boson to a pair of axions,
is kinematically allowed. 
A thorough study of the parameter space revealed that the branching
fraction for this non-standard decay mode is typically on the order
of $50\%$ and can compete with the conventional mode $h^0\to b\bar{b}$.
Preliminary studies showed that the resulting Higgs boson signature
$h^0\to 4b$ is marginally unobservable in Run~II of the Tevatron
(based on Run~I efficiencies).
At an LC, however, this decay should be easy to detect.

\subsection{Non-supersymmetric Alternatives}
In many discussions of physics beyond the standard model,
electroweak symmetry is assumed either to be connected to supersymmetry
or to be driven by strong dynamics, such as technicolor, without
a Higgs boson.
There is, however, a fertile middle ground of composite models.
Here, typically, a strong binding mechanism accompanies, or drives, the
breaking of electroweak symmetry, but some of the states have all the
properties of a fundamental Higgs boson.
Indeed, at sub-TeV energies these scenarios are often best described by
a (possibly extended) Higgs sector, and the strong dynamics is apparent
only above a TeV or so.

In this section, we review some features of such composite models in four
dimensions, and more recent ideas inspired by the possibility of extra
spatial dimensions.
For completeness we include here other signals of extra dimensions that
could be seen in linear colliders.

\subsubsection{Higgs compositeness in four dimensions}
\label{sec:compositeHiggs}

The triviality of the Higgs sector in the standard model
provides a new energy scale~$\Lambda_{\rm SM}$.
It may be called the compositeness
scale, because at higher scales the Higgs boson is no longer a physical
degree of freedom.
Instead, some other fundamental degrees of freedom
should become important. In the standard model with a light Higgs 
boson the compositeness scale is too high to be relevant for
experiments. However, if there are new particles at the TeV scale,
then the scale dependence of the Higgs couplings may be accelerated
and the compositeness scale may be in the TeV range. 

Composite models in which the Higgs field is made of
some new fermions~\cite{Kaplan:1984sm}, or arises as a
result of some new strong interactions involving the top
quark~\cite{Bardeen:1990ds,Cvetic:1997eb} have been extensively
studied in the past.
More recently, a class of models, called top-seesaw theory, in which a
Higgs field appears as a bound state of the top quark with a new heavy
quark, has proven phenomenologically viable and free of excessive
fine-tuning~\cite{Dobrescu:1998nm,Chivukula:1998wd}.
Furthermore, the top quark is naturally the heaviest standard
fermion in these
models, because it participates directly in the breaking of the
electroweak symmetry.
This theory has a decoupling 
limit, so at low energy it behaves as the standard model 
and, therefore, agrees with the electroweak data, as discussed in
Sec.~\ref{sec:bounds}.

The interaction responsible for binding the Higgs field is
provided by a spontaneously broken gauge symmetry, such
as topcolor~\cite{Hill:1991at}, or some flavor or family
symmetry~\cite{Burdman:1999vw}.
Note that such interaction is asymptotically free, allowing for a 
solution to the hierarchy problem.
At the same time the interaction is non-confining, and therefore with 
a very different behavior than the confining technicolor interaction 
discussed at the beginning of this section.

Typically, in the top-quark seesaw theory the Higgs boson is
heavy, with a mass of order 500~GeV~\cite{Chivukula:2000px}. 
However, the effective theory
below the compositeness scale may include an extended Higgs sector,
in which case the mixing between the $CP$-even scalars could bring the
mass of the standard-model-like Higgs boson down to 100~GeV or
so~\cite{Chivukula:1998wd,Dobrescu:1999gv}. 
One interesting possibility in this context is that there is a light 
Higgs boson with nearly standard couplings to fermions and gauge
bosons, and nevertheless its decay modes are completely non-standard.
This happens whenever, as in the NMSSM discussed in 
Sec.~\ref{sec:susyHiggs}, a $CP$-odd scalar has a mass less than half 
the Higgs mass and the coupling of the Higgs to a pair of $CP$-odd 
scalars is not suppressed.
The Higgs boson decays in this case into a pair of $CP$-odd
scalars, each of them subsequently decaying into a pair of
standard model particles, with model dependent branching
fractions~\cite{Dobrescu:2000jt}.
If the Higgs boson has standard model branching fractions, then the
capability of an LC depends on $m_H$, as discussed in
Sec.~\ref{sec:higgs}.
On the other hand, if the Higgs boson has non-standard decays, 
an $e^+ e^-$ collider may prove very useful in disentangling the
composite nature of the Higgs boson, by measuring its width and
branching fractions.

The heavy quark constituent of the Higgs has a mass of a few TeV 
while the gauge bosons associated with the strong interactions that
bind the Higgs are expected to be even heavier.
Above the compositeness scale there must be some additional 
physics that leads to the spontaneous breaking of the
gauge symmetry responsible for binding the Higgs. This may involve 
new gauge dynamics~\cite{Collins:2000rz}, or fundamental scalars and
supersymmetry.
Evidently, an $e^+ e^-$ collider could  study these
interesting strongly-interacting particles only if it operates at 
energies above a~TeV.

Other models of Higgs compositeness have been proposed
recently~\cite{Georgi:2000wt}, and more are likely to be constructed
in the future.

\subsubsection{Theories with extra dimensions}
In the last few years there has been an explosion of interest in
models with extra spatial dimensions.
We consider here several popular scenarios:
large (\emph{i.e.}, mm-sized) extra dimensions for gravity only;
TeV-sized extra dimensions accessible to some, but not all, standard particles;
universal extra dimensions accessible to all particles; and
warped extra dimensions, where the curvature of the extra dimensions 
plays a crucial role.
\vspace*{1em}

\noindent $\bullet$ {\sl Large extra dimensions} 
\nopagebreak

The recent intensive investigations of physics in extra
dimensions has been sparked by the observation that
the graviton could propagate in compact spatial dimensions
as large as a millimeter~\cite{largedim}.
As a result, the strength of gravity would be modified at short
distances, so that the fundamental scale of quantum gravity could be
in the TeV range, rather than $10^{19}$~GeV.
This observation not only changes completely the nature of the
hierarchy problem, but also leads to definite phenomenological
predictions. The main implication is that there are
Kaluza-Klein (KK) excitations of the graviton, {\it i.e.}, spin-2
particles with universal couplings to matter, but suppressed by the
Planck scale. The spectrum of KK gravitons
is very dense, with spacings between states given by the
inverse size of the extra dimensions. Although each of these states
is very weakly coupled, the large number of states gives rise
to large signals at energies comparable with the fundamental (TeV)
scale.

In collider experiments the KK gravitons have two classes of
effects: real KK graviton emission, characterized by missing
transverse energy, and virtual KK graviton exchange,
characterized by anomalous di-fermion or di-boson production at large
invariant masses~\cite{Han:1999sg}.
The current direct bounds on the scale of quantum gravity are of order
1~TeV and are set at the Tevatron~\cite{Abbott:2000zb}
and at LEP~\cite{Abbiendi:2000hh}.
More stringent bounds are set by the electroweak
data, but these are model dependent and less robust.

Both the LHC and an~LC with $\sqrt{s} =0.5$--1~TeV would be able to 
probe KK-graviton processes for fundamental scales up to several TeV.
Moreover, if the scale of quantum gravity is indeed in this range, 
then collider experiments at $\sqrt{s}$ of several TeV would find 
phenomena which are well beyond any current theoretical understanding.
\vspace*{1em}

\noindent $\bullet$ {\sl TeV-size extra dimensions}
\nopagebreak

Unlike the extra dimensions accessible only to gravity, which could be
macroscopic, extra spatial dimensions accessible to
standard model particles are constrained by Tevatron and LEP
data to be smaller than of order $(1~{\rm TeV})^{-1}$.
Nevertheless, the existence of TeV-size extra dimensions is a logical
possibility, and is motivated by various theoretical considerations,
such as the generation of hierarchical quark and lepton masses, or
the potential for gauge coupling unification at a scale in the TeV
range~\cite{Dienes:1999vg}.

The immediate consequence of this scenario is the existence
of towers of KK excitations for the particles that
propagate in the TeV-size extra dimensions.
For example, the gluons would have spin-1 color-octet excitations. 
Their masses are given by $\sqrt{j}/R$ where $R$ is the radius of
the extra dimensions and $j$ is an integer that labels the
KK level. The number of states on each level depends
on the number of extra dimensions.
Both the density of occupied levels
and the average number of states ($D_n$) on a level increase with the
number of extra dimensions, as can be seen in Table~\ref{KKlevels}.
\begin{table}
\centering
\caption{The mass, $M_n$, and number of states, $D_n$, of the $n$th
occupied KK level, for $\delta\leq 3$ extra dimensions,
compactified on a torus.
For illustration, only the first nine levels are shown
here explicitly.}
\label{KKlevels}
\vskip 6pt
\renewcommand{\arraystretch}{1.3}\small
\begin{tabular}{l|cccccccccc}
\hline \hline
                        &  $n$   & $\;1\;$ & 2 & 3 & 4 & 5 & 6 & 7 & 8 & $9$ \\
\hline
$\delta=1$      & $M_nR$ & 1 & 2 & 3 & 4 & 5 & 6 & 7 & 8 & 9 \\
                        &  $D_n$ & 2 & 2 & 2 & 2 & 2 & 2 & 2 & 2 & 2 \\
\hline
$\delta=2$      & $M_nR$ & 1 & $\sqrt{2}$ & 2 & $\sqrt{5}$ & $\sqrt{8}$
                                & 3 & $\sqrt{10}$ & $\sqrt{13}$ & 4  \\
                        & $D_n$  & 4 & 4 & 4 & 8 & 4 & 4 & 8 & 8 & 4 \\
\hline
$\delta=3$      & $M_nR$ & 1 & $\sqrt{2}$ & $\sqrt{3}$ & 2 & $\sqrt{5}$
                                & $\sqrt{6}$ & $\sqrt{8}$ & 3 & $\sqrt{10}$ \\
                        & $D_n$  & 6 & 12 & 8 & 6 & 24 & 24 & 12 & 30 & 24 \\
\hline \hline
\end{tabular}
\end{table}
The $W, Z$ and photon would have color-singlet spin-1 excitations with 
a mass spectrum similar to the KK gluons, but slightly perturbed due 
to the electroweak symmetry breaking.
In the popular case in which the quarks and leptons are localized 
on a three-dimensional domain wall (a 3-brane), the KK excitations of 
the gauge bosons have the same couplings, up to a factor of order one, 
as the corresponding standard model states.
Therefore, the $Z$ and photon KK states may be produced in the $s$ 
channel in $e^+e^-$ collisions.
At the same time, the KK excitations of the electroweak gauge bosons 
contribute at tree-level to the electroweak observables and, as a 
consequence, are constrained to lie above $\sim 4$~TeV, unless there 
are even more compensating effects.
Standard-model quarks and leptons might propagate in extra dimensions, 
and for each of these chiral fermions there is a tower of vector-like 
fermions ({\it i.e.}, four-component spinors with the left- and 
right-handed components carrying the same charges) with mass 
separations of order~$1/R$.

The KK excitations of standard model particles can be produced at the LHC 
unless the inverse radius of the extra dimensions is above $\sim 6$~TeV.
An~LC with $\sqrt{s} = 0.5$--1~TeV is unlikely
to produce directly any of these KK excitations, but is very
sensitive to their presence via virtual effects.
It could, for example, provide evidence that a 4~TeV
KK excitation of the $Z$ is accompanied by an almost 
degenerate KK excitation of the photon~\cite{Rizzo:2000en}.
One should emphasize here that the observation
of a signal in $e^+e^-$ collisions fitting the virtual effects
of certain KK excitations would not prove
that extra dimensions indeed exist. Instead, the virtual effects
may very well be due to some other new heavy particles from a
four-dimensional model (for example a collection of $Z^\prime$ bosons 
rather than KK excitations of the photon and $Z$).
On the other hand, observation of a series of KK resonances,
say in a multi-TeV LC, not only would establish the existence of extra
dimensions, but also would determine the number of extra dimensions
and their structure.

\vspace*{1em}

\noindent $\bullet$ {\sl Universal extra dimensions}
\nopagebreak

A qualitatively distinct case is that {\it all} 
standard model particles propagate in extra dimensions.
The KK number is then conserved at each vertex, 
so that the KK excitations may be produced only in groups 
of two or more. Hence, direct bounds from the Tevatron 
and LEP are significantly lower. Moreover, the KK states
do not contribute at tree level to the electroweak observables. 
The current mass bounds on the first KK states are as low as 300~GeV
for $\delta = 1$ universal extra dimension, and in the range 
400--800~GeV for $\delta = 2$~\cite{Appelquist:2000nn}.

These loose bounds make the universal extra dimensions particularly 
interesting for collider experiments.
At $\sqrt{s} > 600$~GeV, an LC could already pair-produce KK leptons, 
quarks and gauge bosons.  One possibility is that the KK states decay
outside the detector, so that the signal would be pairs of highly 
ionizing tracks. Alternatively, interactions that do not
conserve momentum in the extra dimensions may allow the KK states 
to decay promptly into pairs of 
standard model particles.

\vspace*{1em}

\noindent $\bullet$ {\sl Warped extra dimensions}
\nopagebreak

Another explanation for the observed weakness of gravitational
interactions is given by the possibility that there is one extra
dimension, and the graviton is localized on a 3-brane 
far from the 3-brane where our world of standard-model
particles is localized~\cite{Randall:1999ee}.
On our 3-brane the wave function of the graviton is suppressed
exponentially, and in this way the huge hierarchy between the
electroweak and Planck scales is naturally induced.
The graviton again has KK excitations, but
these now have a certain non-uniform mass spectrum in the TeV range,
and rather large couplings at the TeV scale.
These spin-2 resonances would have a striking signal at any
collider sufficiently energetic to produce them. Specific studies of
these KK gravitons at the LHC and at an~LC can be found in 
Ref.~\cite{Davoudiasl:2000jd}.

It may also be possible that some of the standard model fields
propagate in the bulk of the extra dimensions, in which case
spin-0 and 1 resonances with an interesting spectrum could be produced
at the LHC and an~LC~\cite{Davoudiasl:2000wi}.

\subsubsection{Higgs boson properties in models with extra dimensions}

The presence of extra dimensions and, more generally, of any 
new physics beyond the standard model, may affect searches for
the Higgs boson in several ways:
\begin{itemize}
\item
It may modify the rates for producing the Higgs boson in standard processes.
For example, mixing in the Higgs sector of the MSSM typically suppresses the
production cross section. 
Likewise, in models with large extra dimensions, 
or with warped extra dimensions, the Higgs boson mixes with the spin-0
component of the graviton~\cite{Giudice:2000av}.
\item It may modify the branching ratios for the standard-model-like
Higgs boson,
\emph{e.g.}, by adding new possible decay channels (such examples are
given in Sections~\ref{sec:susyHiggs} and \ref{sec:compositeHiggs}), 
or through radiative corrections 
due to new particles (see, \emph{e.g.}, Ref.~\cite{Carena:1999gk}).
\item It may yield qualitatively new Higgs production mechanisms.
For example, some flavor models in extra
dimensions~\cite{Arkani-Hamed:1998sj,Arkani-Hamed:2000yy}
couple the standard Higgs boson to a bulk scalar (called a flavon) through
non-renormalizable interactions. This may lead to the possibility of
Higgs-flavon associated production at both hadron and lepton
colliders~\cite{Cheng:1999rx}.
The Higgs boson branching fractions are nearly standard, and
the observed signature is $b\bar{b}\met$. At the Tevatron and the LHC
it is almost impossible to separate it from $Zh^0$, but a lepton
collider would be able to reveal the presence of extra-dimensional physics.
\end{itemize} 

Apart from these phenomenological implications, various new phenomena
due to the extra dimensions shed a new light on the origin of 
electroweak symmetry breaking 
and fermion mass
generation~\cite{Arkani-Hamed:1998sj,Arkani-Hamed:1999dc,Dienes:1999vg}.
An interesting example is provided by the TeV-size extra dimensions.
The KK excitations of the standard model gauge
bosons and fermions give rise to a scalar bound state with 
the quantum numbers of the standard model Higgs
doublet~\cite{Cheng:1999bg,Arkani-Hamed:2000hv}. 
The Higgs boson appears as a composite scalar with 
a combination of KK modes of the top-quark playing the
role of constituents. 
This can be easily understood given that the KK excitations
of the gluons and electroweak gauge bosons induce a strongly-coupled
attractive interaction between the left- and right-handed top-quark
fields. 

Let us concentrate on a simple scenario, where all standard model gauge
bosons and fermions (or at least the third generation of fermions)
propagate in extra dimensions~\cite{Arkani-Hamed:2000hv}.
It is remarkable that out of the many possible bound states involving
the quarks and leptons, the most deeply bound state
has the quantum numbers of the Higgs doublet.
Indeed, this state acquires a vacuum expectation value and, thus,
breaks electroweak symmetry!
Furthermore, this composite Higgs doublet has a large Yukawa coupling
to the top-quark, which explains the large top mass.
This is a direct consequence of the experimentally determined gauge
charges of the quarks and leptons.

The framework that emerges here is very appealing.
At a scale in the TeV range, called for convenience the string scale,
$\Lambda_s$, the only degrees of freedom 
are the $SU(3)_C \times SU(2)_W \times U(1)_Y$ gauge bosons and the
three generations of quarks and leptons, all of them propagating 
in a higher-dimensional spacetime.
(At even higher energy scales, the fundamental degrees of freedom of a
theory incorporating quantum gravity are expected to become relevant,
such as the winding modes of string theory.)
Below the scale $\Lambda_s$, fermion--anti-fermion pairs bind via
the $SU(3)_C \times SU(2)_W \times U(1)_Y$ interactions.
Then, below the scale $1/R < \Lambda_s$ that sets the
size of the extra dimensions, the physics is 
described by an effective four-dimensional theory.
This effective theory is the usual standard model,
with the possible addition of a few
other scalars, such as the heavy states of a two-Higgs-doublet model.
(Interestingly enough, in the case of four extra dimensions there is
also a potentially light bound-state with the quantum numbers of a
bottom squark~\cite{Arkani-Hamed:2000hv}.)

As well as being a simple model consistent with all the experimental
data, this framework is also predictive. The top mass is predicted 
with an uncertainty of about 20\% and agrees well with the experimental
value. The Higgs mass also is determined theoretically, and the result 
is a narrow range 
\begin{equation}
	165~{\rm GeV} < m_H < 230~{\rm GeV}. 
\label{prediction}
\end{equation}
The correct prediction of the top mass and, more importantly, the 
prediction of the Higgs quantum numbers are unmatched by the 
standard model or its supersymmetric extensions.
This model does suffer from some problems, which are, however,
shared by the MSSM.
It accommodates, but does not explain, the light quark and lepton masses.
Also, a moderate tuning is required to keep the electroweak
scale below $1/R$. 
A fuller comparison of the MSSM with this model of composite
Higgs from extra dimensions can be found in Sec.~6 of
Ref.~\cite{Arkani-Hamed:2000hv}.
Here we only add that in the case of universal extra dimensions
the fine-tuning is no longer an issue because of 
the rather loose bound on~$R$  ($1/R > 400$--800~GeV for two
extra dimensions).

A dedicated study of the signatures of the composite Higgs model from
extra dimensions at the LHC and an~LC has not yet
been pursued.
The prediction in Eq.~(\ref{prediction}) places the Higgs boson in the
intermediate mass range, discussed in Sec.~\ref{sec:inter}.
More striking than the Higgs physics would be the non-standard phenomena
discussed above.
Eventually multi-TeV lepton collisions would be desired, to verify the
layers of physics at the electroweak, compactification, and string
scales.

\section{Conclusions and Recommendations}

Despite the success of the standard model of particle physics, we do not
expect it to give a complete explanation of nature, at the
very least because it does not incorporate gravity.
At an aesthetic level, theorists expect gauge forces to unify with
each other and with gravity, somewhere below the Planck mass.
At a more concrete level, solid theoretical investigation of the
``triviality problem'' provides strong evidence that the scalar sector
breaks down when applied to arbitrarily high energies.
Consequently, the scalar (Higgs) sector is useful only if it is
treated as an effective field theory, valid only up to some finite
energy scale, $\Lambda_{\rm SM}$.
Indeed, it is fair to call the $SU(3)_C \times SU(2)_W \times U(1)_Y$
gauge symmetry and the quarks' and leptons' quantum numbers, which
all are here to stay, laws of nature.
The scalar sector and interactions with it, on the other hand, merely
represent a working model.

Above the scale $\Lambda_{\rm SM}$ new forms of matter are at play.
The masses of quarks, charged leptons, and electroweak gauge bosons
are generated by interactions with this matter.
(These masses are otherwise forbidden by the electroweak gauge
symmetry.)
These interactions also lead to flavor and $CP$ violation,
at least in the quark sector.
Understanding this matter is thus a fundamental and central problem
for physics.
Despite the essential information that will come from hadronic
collisions at the Tevatron and the LHC, it will take $e^+e^-$
(or~$\mu^+\mu^-$) collisions at this scale to comprehend it fully.
The most plausible scale for $\Lambda_{\rm SM}$ is around a~TeV.
This is not much higher than the scale of electroweak symmetry breaking,
set by $v=(\sqrt{2}G_F)^{-1/2}=246$~GeV.
If $\Lambda_{\rm SM}$ were to be much higher, then $v$ would
be unnaturally small compared to the mass scale of the agents of
symmetry breaking.

Our ignorance of $\Lambda_{\rm SM}$ and of the form of the new matter
make it both easy and difficult to argue in favor of a linear $e^+e^-$
collider with $\sqrt{s}=0.5$--1~TeV.
The easy part is as follows.
If~$\Lambda_{\rm SM}$ is in the natural range, such an LC would start
to explore many new states.
The precise and model-independent mode of experimentation makes it
likely that LC measurements would be able to diagnose the origin
of electroweak symmetry breaking.
If, alternatively, $\Lambda_{\rm SM}$ is (unnaturally) high, fits to
precisely measured electroweak observables indicate that the lowest
lying $CP$-even scalar---the Higgs boson---is light.
Indeed, it is light enough for LC measurements to test the main
features of mass generation in the standard model.

The hard part of the argument also stems from recognizing that the
standard Higgs sector is, at best, an effective theory.
Any list of measurements and, thus, their luminosity and energy
requirements hinges on a scenario for the underlying physics.
Many scenarios in the literature lead to an exciting program.
Inevitably, others are less exciting.
Nevertheless, the LC affords an excellent opportunity to complement the
LHC in elucidating what breaks electroweak gauge symmetry.

For the Higgs boson, this report concentrates on scenarios that are
close to the standard model.
This is motivated mostly by the precise electroweak data.
Models with a decoupling limit, by definition, can support
a light Higgs with new states at high masses.
They automatically satisfy constraints of the data in this and,
typically, an adjacent neighborhood of parameter space.
Within this neighborhood the phenomenology can change, and in some
models the change may be drastic.
More often than not, however, the potential of an~LC becomes more
interesting.

Another reason to consider nearly-standard Higgs bosons is that it
gives two simple and reasonable scenarios for the Higgs program at~LC.
If the Higgs mass satisfies $m_H<2m_W$, then a detailed profile of
the Higgs is measurable~\cite{Battaglia:2001jb}.
High luminosity is essential, because precisely measured branching
ratios are sensitive to (possible) higher-mass Higgs bosons.
Such a program should have an impact comparable to measurements
of $Z$ boson properties at LEP and SLC.
If $m_H>2m_W$ Higgs physics becomes more difficult:
the decay to $b\bar{b}$ is rare and to $\tau^+\tau^-$ and
$c\bar{c}$ are very rare.  
It is not yet clear how much integrated luminosity is needed for
the rare modes, and without them the Higgs program is not quite as
compelling.
(If $m_H>2m_t$ the branching ratio to $t\bar{t}$ is large
and measurable.)
In non-supersymmetric extensions of the standard model such values of
the Higgs mass are consistent with the precisely measured observables,
because the Higgs boson's contribution to them is small and could be
compensated by similarly small contributions from TeV-scale particles.
If the Higgs is indeed so heavy, then the flexibility of an~LC to
carry out the Giga$Z$ option (as well as improved measurements in
$m_W$ and $m_t$) is attractive, because the indirect insights might
help guide us to the next energy scale.

We also considered several extensions of the standard model---%
supersymmetry, compositeness and extra spatial dimensions---%
in which new physics resides at or below 1 TeV.
Even with only a handful of accessible superpartners, an LC can perform
many measurements complementary or superior to the LHC.
Some of the most important are as follows:
the precision on the mass of the lightest superpartner is expected to be
only 10\% at the LHC, but 1\% at the LC.
Direct studies of the slepton sector are a challenge at the LHC,
but an LC can measure the masses, couplings and mixing parameters
of the sleptons rather well.
The well-defined center of mass energy of an LC is extremely helpful
in resolving nearby resonances.
We have even identified a case where the observation of supersymmetry
at the LHC would remain ambiguous, and await confirmation elucidation
at an LC.
Moreover, linear colliders are also able to test definitively various
supersymmetric relations and mass sum rules.
In all cases, precision is needed for gaining insight into the
underlying physics of supersymmetry breaking, and an LC of appropriate
energy will be an essential tool to reach this goal.

There are several scenarios with extra spatial dimensions.
The LC can help discriminate between them.
For example, measurements of the mass spectrum of Kaluza-Klein (KK)
modes can be used to determine the number of extra dimensions and
their geometric structure.
Even when the energy is not high enough to produce KK excitations
directly, they can be exchanged as virtual particles.
Thus, some information on the KK spectrum can be obtained by measuring
scattering cross sections---in a way reminiscent of studying the $W$
boson in charged-current interactions, though not as much as in direct
production.

These and other scenarios point to several features that a future
$e^+e^-$ facility should have:
\begin{enumerate}
    \item High (integrated) luminosity.
	Precision is desired for the couplings of the Higgs boson to
	the known particles.
	Indeed, in some cases, such as the Higgs self-coupling, or the
	$b\bar{b}$ coupling for $m_H>2m_W$, one must rely on rare processes.
	In susy and extra-dimension scenarios, there are also many other
	measurements to make, some requiring certain energies or particular
	combinations of beam polarization.

	\item Polarization.
	At the electroweak scale and above, left- and right-handed
	fermions are fundamentally different.
	Choosing the right initial state, in response to the data,
	could prove vital.
	In susy, for example, polarization helps to pin down the quantum
	numbers of the superpartners, and it is the key that allows full
	exploration of the charginos and neutralinos.

	\item Flexibility.
	Scans of the most important superpartner thresholds are valuable
	for deducing how supersymmetry is broken.
	In some other scenarios the Giga$Z$ program~\cite{Erler:2000jg}
	is not merely interesting, but helpful for pointing to
	accelerators beyond the LHC and a 1~TeV~LC.
	The $\gamma\gamma$ option can contribute to Higgs studies, and
	the $e^-e^-$ option is a good tool for studying the selectron,
	through $e^-e^-\to\tilde{e}^-\tilde{e}^-$.

	\item Energy above 1~TeV.
	The TeV scale does not stop at $\sqrt{s}=1$~TeV, so energy
	upgrades are certainly crucial.
	The heavier superpartners may require $\sqrt{s}>1$~TeV;
	direct production of several resonances in a KK tower certainly
	does.
	It would be useful to understand how much of a sub-TeV layout could
	be reused for a multi-TeV~LC.
\end{enumerate}
The first three features appear in the NLC/JLC-X and TESLA designs,
both of which are relatively mature.
Higher energies appear to be upgrades (at least in length) or to await
development of different technology, such as two-beam acceleration.

In the course of our study we addressed the experimental possibilities
for a nearly-standard Higgs boson in the intermediate-mass and heavy
regions.
We also studied how angular distributions can be used to determine
particles' spin, particularly for the Higgs boson.
On the theoretical side, we examined the connection of the LSP in susy
to dark matter, and surveyed non-supersymmetric extensions of the
standard model.
There are a few questions that our studies have pointed to:
\begin{itemize}
	\item Measurement of BR($H \to b\bar{b}$) for $m_H > 2m_W$.
		We have attempted a first examination, using only $ZH$
		events and then only leptonic $Z$ decays.
		How well, as function of $m_H$, can one do, when neutrino
		and hadronic $Z$ decays are added?
		When $\bar{\nu}H\nu$ production is included?

	\item Further simulations of spin determination.
		We have studied the Higgs and plan refinements.
		It would also be interesting to look at superpartners' spins.

	\item Using light Higgs branching ratios to distinguish
		the standard model from non-standard models.
		Current studies have focussed on the MSSM and have
		omitted $CP$-violation.

	\item The decay $h/H\to AA\to b\bar{b}b\bar{b}$,
		which is allowed in an NMSSM and in certain composite
		models~\cite{Dobrescu:2000jt}.
		The LC looks promising, but a detailed study has not been done.

	\item Studies of benchmark scenarios in which susy remains obscure
		even in the light of LHC data.
		We have found examples either based on Scherk-Schwarz susy breaking
		or with $\tau$-rich final states.
\end{itemize}
Further issues will arise, as new theoretical ideas for physics at the
TeV scale turn into testable scenarios, and as experiments yield more
constraints and new surprises.

\section*{Acknowledgments}

We would like to thank many colleagues who participated in parts of 
this study, in particular, Marcela Carena, Tom Dombeck, James Done,
Gene Fisk, Norman Graf, Paul Grannis, Chris Hill, Teruki Kamon,
Boaz Klima, Chris Quigg, Alvin Tollestrup, Andre Turcot,
Rick Van Kooten, Avi Yagil, and G.~P.~Yeh.
Over the nine-month course of the study we were lucky to learn 
about LC physics and technology from Marco Battaglia, Richard DuBois,
Norbert Holtkamp, Stanis\l aw Jadach, Tony Johnson, Wolfgang Kilian,
Michael Peskin, Mike Ronan, and Frank Zimmerman.
We thank Paul Grannis, Kaoru Hagiwara, David J.~``the~Red'' Miller,
Hugh Montgomery, Frank Paige, Michael Peskin, Thomas Rizzo, and
Peter Zerwas for many perceptive comments on the manuscript.
Finally, we thank Mike Shaevitz and Mike Witherell for commissioning
this study and for their support during its course.

\end{document}